\documentclass[12pt]{iopart}
\usepackage{slashed}
\usepackage{amsfonts}
\usepackage{enumerate}
\usepackage{graphicx}

\newcommand{\Ord}{\mathcal{O}}
 
\newcommand{\nn}{\nonumber \\}
\newcommand{\el}{\nonumber \\ &&}

\newcommand{\elale}{\nonumber \\ &=&}

\newcommand{\M}{\mathcal{M}}
\newcommand{\T}{\theta}
\newenvironment{changemargin}[2]{%
  \begin{list}{}{%
    \setlength{\topsep}{0pt}%
    \setlength{\leftmargin}{#1}%
    \setlength{\rightmargin}{#2}%
    \setlength{\listparindent}{\parindent}%
    \setlength{\itemindent}{\parindent}%
    \setlength{\parsep}{\parskip}%
  }%
  \item[]}{\end{list}}

\def\eqn#1{eq.~(\ref{#1})}

\newcommand{\be}{\begin{equation}} 
\newcommand{\ee}{\end{equation}} 
\newcommand{\cmb}{\begin{changemargin}}
\newcommand{\cme}{\end{changemargin}}
\newcommand{\bea}{\begin{eqnarray}} 
\newcommand{\eea}{\end{eqnarray}} 
\newcommand{\D}{\delta}
\newcommand{\g}{\gamma}
\newcommand{\A}{\alpha}
\newcommand{\B}{\beta}
\newcommand{\G}{\Gamma}

\newcommand{\mc}{\mathcal}
\newcommand{\mb}{\mathbb}
\def\Tr{\mathop{\rm Tr}\nolimits}

\def\spa#1.#2{\langle#1\,#2\rangle}
\def\spb#1.#2{[#1\,#2]}
\def\sandmm#1.#2.#3{%
\left\langle\smash{#1}{\vphantom1}\right|{#2}%
\left|\smash{#3}{\vphantom1}\right]}
\def\spab#1.#2.#3{\sandmm#1.#2.#3}
\def\spba#1.#2.#3{\sandpp#1.#2.#3}
\def\spaa#1.#2.#3.#4{\sandmp#1.{#2#3}.#4}
\def\spbb#1.#2.#3.#4{\sandpm#1.{#2#3}.#4}
\def\spash#1.#2{\spa{\smash{#1}}.{\smash{#2}}}
\def\spbsh#1.#2{\spb{\smash{#1}}.{\smash{#2}}}

\def\ksl{\not{\hbox{\kern-2.3pt $k$}}}

\def\e{\epsilon}

\def\Ord{{\cal O}}
\def\a{{\cal A}}

\def\Nc{N_c}
\def\Nsym{{\cal N}=4}

\def\li#1{{\mathop{\rm Li}\nolimits}_#1}

\def\pol{\varepsilon}

\begin{document}
\title[One-loop $\mathcal{N}=4$ SYM amplitudes in $d$ dimensions]{One-loop $\mathcal{N}=4$ super Yang-Mills scattering amplitudes in $d$ dimensions, relation to open strings and polygonal Wilson loops}
\author{R M Schabinger}
\address{Instituto de F\'{i}sica Te\'{o}rica UAM/CSIC,
	Universidad Aut\'{o}noma de Madrid,
        Cantoblanco, E-28049 Madrid, Espa\~{n}a}
\ead{robert.schabinger@uam.es}        
\begin{abstract}
In this review we discuss some recent developments related to one-loop $\mathcal{N}=4$ super Yang-Mills scattering amplitudes calculated to all orders in $\e$. It is often the case that one-loop gauge theory computations are carried out to $\Ord(\e^0)$, since higher order in $\e$ contributions vanish in the $\e \rightarrow 0$ limit. We will show, however, that the higher order contributions are actually quite useful. In the context of maximally supersymmetric Yang-Mills, we consider two examples in detail to illustrate our point. First we concentrate on computations with gluonic external states and argue that $\mathcal{N}=4$ supersymmetry implies a simple relation between all-orders-in-$\e$ one-loop $\mathcal{N}=4$ super Yang-Mills amplitudes and the first and second stringy corrections to analogous tree-level superstring amplitudes. For our second example we will derive a new result for the all-orders-in-$\e$ one-loop superamplitude for planar six-particle NMHV scattering, an object which allows one to easily obtain six-point NMHV amplitudes with arbitrary external states. We will then discuss the relevance of this computation to the evaluation of the ratio of the planar two-loop six-point NMHV superamplitude to the planar two-loop six-point MHV superamplitude, a quantity which is expected to have remarkable properties and has been the subject of much recent investigation.
\end{abstract}
\maketitle

\tableofcontents

\section{Overview}
In recent years, tremendous progress has been made towards a more complete understanding of the scattering amplitudes in $\mathcal{N}=4$ super Yang-Mills theory~\cite{origN=4} (hereafter simply $\Nsym$). $\Nsym$ is special in that it has more symmetry than any other gauge theory, especially in its so-called planar limit~\cite{origtHooft,origMald}. Although the theory's S-matrix has been under investigation for nearly 30 years~\cite{origGreenSchwBrink}, the last five have been particularly exciting. Numerous ground-breaking discoveries have been made (like the application of the AdS/CFT correspondence~\cite{origMald} to gluon scattering at strong coupling~\cite{origAldayMald}, a hidden dual superconformal symmetry of the planar theory~\cite{DHKSdualconf}, and a dual description of the leading singularities of the S-matrix as integrals over periods in a Grassmann manifold~\cite{dualSmat} to name just a few) and there is no reason to believe that we have learned everything $\Nsym$ has to teach us. 

Recently, existing tools for the calculation of one-loop $\Nsym$ amplitudes to all orders in the dimensional regularization~\cite{origtHooftVelt} parameter have been developed further~\cite{myfifth} and this is one of the main themes discussed in this review. This regularization parameter, $\e$, is introduced to cut off the IR divergences that appear in massless gauge theory calculations (we encourage readers less familiar with the structure of IR divergences in gauge theory to peruse \ref{dimregs}). We will illustrate our refined methods by considering examples where our results find useful application. At times we will develop aspects of $\Nsym$ S-matrix theory that appear to be of purely academic interest, but, in fact, a significant part of the computational machinery discussed in this review can be applied to calculations in any quantum field theory. When techniques are applicable only in $\Nsym$ we will try to emphasize this. Before delving into the details of the problems we want to solve, we present the Lagrange density of the model and discuss some of its salient features.

The field content of the $\mathcal{N}=4$ model consists of a gauge field $A_{\mu}$, four Majorana fermions $\psi_i$, three real scalars $X_p$, and three real pseudo-scalars $Y_q$. All fields are in the adjoint representation of a compact gauge group, $G$. The Lagrange  density of $\mathcal{N}=4$ is given by~\cite{myfirst}\footnote{Here we use the conventions of Peskin and Schroeder~\cite{PeskinSchroeder} for the Lagrange density, which differ somewhat from the conventions of~\cite{myfirst}. Throughout this review, when not explicitly defined, the reader may assume that our conventions for perturbation theory coincide with those of Peskin and Schroeder.} 
\bea
\label{LagDens}
\mathcal{L} &&= \textrm{tr} \bigg\{ -\frac{1}{2} F_{\mu \nu} F^{\mu \nu} + \bar{\psi_i} \slashed{D}  \psi_i + D^{\mu} X_p D_{\mu} X_p + D^{\mu} Y_q D_{\mu} Y_q \\ \nonumber
&& - i g \bar{\psi_i} \alpha^p_{i j}[X_p,\psi_j]+g \bar{\psi}_i \gamma_5 \beta^q_{i j} [Y_q,\psi_j] \\ \nonumber
&& +\frac{g^2}{2}\bigg([X_l,X_k][X_l,X_k]+[Y_l,Y_k][Y_l,Y_k]+2[X_l,Y_k][X_l,Y_k]\bigg )\bigg\},
\eea
where the $4 \times 4$ matrices $\alpha^p$ and $\beta^q$ are given by  \footnote{$\sigma_0$ is the $2 \times 2$ identity matrix.}
\bea
&& \alpha^1 = \left(\begin{array}{cc} i \sigma_2 & 0 \\ 0 & i \sigma_2 \end{array}\right),~~\alpha^2 = \left(\begin{array}{cc} 0 & - \sigma_1 \\ \sigma_1 & 0 \end{array}\right),~~\alpha^3 = \left(\begin{array}{cc} 0 & \sigma_3 \\ -\sigma_3 & 0 \end{array}\right),  \\ \nonumber
&& \beta^1 = \left(\begin{array}{cc} -i \sigma_2 & 0 \\ 0 & i \sigma_2 \end{array}\right),~~\beta^2 = \left(\begin{array}{cc} 0 & -i \sigma_2 \\ -i \sigma_2 & 0 \end{array}\right),~~\beta^3 = \left(\begin{array}{cc} 0 & \sigma_0 \\ -\sigma_0 & 0 \end{array}\right).
\eea
Once the gauge group and coupling constant $g$ are fixed, the theory is uniquely specified. It turns out that in scattering amplitude calculations it is somewhat more typical to pair up the scalars and pseudoscalars and work with three complex scalar fields. The presence of four supercharges means that there is an $SU(4)$ R-symmetry acting on the fields.  This symmetry acts on the state space as well and dictates selection rules for $\Nsym$ scattering amplitudes.

One of the first remarkable discoveries made about the $\Nsym$ model is that the
scale invariance of the classical Lagrange density (\ref{LagDens}) remains a symmetry at
the quantum level~\cite{SohniusWest}, implying that the $\beta$ function vanishes to
all orders
in perturbation theory. It follows~\cite{HoweStelleWest,HoweST,BrinkLN1,BrinkLN2,Lemes} that the
theory is UV finite in
perturbation theory (it turns out that the $\beta$ function remains
zero non-perturbatively
as well, but this is trickier to prove~\cite{Seiberg}). The classical superconformal
invariance
of the classical Lagrange density (\ref{LagDens}) (see B for a brief
discussion of
the $\Nsym$ superconformal group) continues to be a quantum mechanical
symmetry of
all correlation functions of gauge-invariant operators.

Most of the work on $\Nsym$ scattering amplitudes focuses on the massless, superconformal $\Nsym$ model described above but we note in passing that it is also possible to construct an $\Nsym$ model with both massive and massless fields~\cite{Fayet}. One can give some of the scalar fields in (\ref{LagDens}) vacuum expectation values  (VEVs) at the cost of superconformal invariance and some of the generators of $G$. Formally, the fact that the six scalar fields can acquire VEVs without breaking supersymmetry implies that the theory has a six-dimensional moduli space of vacua. The $\Nsym$ model where some, but not all, of the gauge group generators are broken by scalar VEVs is called the Coulomb phase of the theory. While most of the literature has focused on the massless, conformal phase of the theory, the Coulomb phase is also quite interesting and is starting to attract the attention it deserves~\cite{myfirst,Higgsreg1,Higgsreg2,Higgsreg3,Higgsreg4}. Unfortunately, a proper discussion of the Coulomb phase is beyond the scope of this review article (however, see reference~\cite{JHennRev}) and we focus our attention exclusively on the S-matrix in the conformal phase of the theory, using dimensional regularization to regulate the IR divergences.

Given the assertion that $\Nsym$ scattering amplitudes are free of UV divergences, we might guess that, say, the one-loop four-gluon amplitude in $\Nsym$ is built out $D = 4 - 2\e$ triangles and boxes, but not $D = 4 - 2\e$ bubbles, since only one-loop bubbles are UV divergent. Remarkably, this turns out not to be the case; the one-loop four-gluon $\Nsym$ amplitude is built out of box integrals only. What is even more remarkable is that, with the caveat that we drop contributions $\Ord(\e)$ and higher, this conclusion holds~\cite{1001lessons} for $n$-gluon scattering amplitudes\footnote{As will be made clear later, it is now known that this conclusion holds for one-loop $\Nsym$ scattering amplitudes with arbitrary external states.}.

For our purposes, the result in the above paragraph will not suffice; we are interested in studying $\Nsym$ amplitudes to all orders in $\e$ and we therefore need to modify the integral basis. Actually, this is not too hard. It has been clear at least since the work of~\cite{VermNeer} that all one has to do is add scalar pentagon integrals to the basis. Then one can express any one-loop $\Nsym$ scattering amplitude in terms of pentagons and boxes to all orders in $\e$. These ideas will be explained in much more detail in Section \ref{revcomp} after the necessary framework has been reviewed. 

However, before continuing, it is important to understand precisely what is meant by ``to all orders in $\e$,'' as it is language that occurs very frequently in what follows. In this work we focus on establishing relations that would allow one to calculate certain one-loop $\Nsym$ scattering amplitudes to $\mathcal{O}(\e^n)$ provided that one is able to calculate the relevant box and pentagon integrals to $\mathcal{O}(\e^n)$. This necessitates a brief discussion of the calculational status of the box and pentagons themselves. The calculation of scalar Feynman integrals to higher orders in epsilon has been studied recently by a number of different authors ({\it e.g.} \cite{KTn,LSSn,DDGSn}), but it is unfortunately not yet possible to straightforwardly calculate a given scalar Feynman integral to $\mathcal{O}(\e^n)$. If one is interested in making the epsilon expansion of one of the one-loop amplitudes discussed later on in this review explicit, references~\cite{KTn}, \cite{LSSn}, and \cite{DDGSn} together with Smirnov's textbook\footnote{This textbook discusses many generally applicable calculational techniques and treats the epsilon expansion of several specific box integrals in detail.} on the evaluation of Feynman integrals~\cite{Sn} should suffice to determine the one-loop Feynman integrals that enter into our calculations through $\mathcal{O}(\e^2)$. As we shall see, this is good enough for those applications within the scope of the present paper.

Another major theme of~\cite{myfifth} reviewed here is a novel relation between one-loop scattering amplitudes in $\Nsym$ gauge theory and tree-level scattering amplitudes in open superstring theory\footnote{Tree-level amplitudes of massless particles in open superstring constructions compactified to four dimensions have a universal form~\cite{ST1}.}. With a bit of inspiration, the relationships to be discussed can be derived from the existing string theory literature. To the best of our knowledge, however, they are unknown at the time of this writing. Although, we are in possession of a simple and general derivation of the relations presented in Section \ref{gsrel}, we shall also test them explicitly in the simplest non-trivial case to establish confidence that there are no gaps in our logic. 

What do we mean by ``the simplest non-trivial case?'' We proceed in the spirit of~\cite{DHKSdualconf,SUSYBCFW,origElvang} and represent states of definite helicity in a way that makes their $SU(4)_R$ transformation properties manifest. In what follows $g^\pm(p_i)$ is a positive or negative helicity gluon of momentum $p_i$, $\phi_a^{\pm}(p_i)$ is a positive or negative helicity fermion of flavor $a$ and momentum $p_i$, and $S^\pm_a(p_i)$ is a complex scalar of flavor $a$ and momentum $p_i$. A scalar has no helicity so the assignment of ``$+$'' and ``$-$'' is a bit arbitrary. One could, for example, assign the $+$ label to holomorphic scalars and the $-$ label to anti-holomorphic scalars. 
\cmb{-.5 in}{0 in}
\bea
&&g^+(p_i)\leftrightarrow p_i 
\nn
\phi_1^+(p_i) \leftrightarrow p_i^1 ~~~~~~~~~ \phi_2^+(p_i) &&\leftrightarrow p_i^2 ~~~~~~~~~ \phi_3^+(p_i) \leftrightarrow p_i^3 ~~~~~~~~~ \phi_4^+(p_i) \leftrightarrow p_i^4
\nn
S_1^+(p_i) \leftrightarrow p_i^{12}  ~~~~~~~~~~~~~~ &&S_2^+(p_i) \leftrightarrow p_i^{23} ~~~~~~~~~~~~~ S_3^+(p_i) \leftrightarrow p_i^{13}
\nn
S_1^-(p_i) \leftrightarrow p_i^{34} ~~~~~~~~~~~~~~ &&S_2^-(p_i) \leftrightarrow p_i^{14} ~~~~~~~~~~~~~ S_3^-(p_i) \leftrightarrow  p_i^{24}
\nn
\phi_1^-(p_i) \leftrightarrow p_i^{234} ~~~~~~~~~ \phi_2^-(p_i) &&\leftrightarrow p_i^{134} ~~~~~~~~~\phi_3^-(p_i) \leftrightarrow p_i^{124} ~~~~~~~~~ \phi_4^-(p_i) \leftrightarrow p_i^{123} 
\nn
&& g^-(p_i) \leftrightarrow p_i^{1234}
\label{operatormap}
\eea
\cme
The only {\it a priori} non-zero scattering amplitudes are those that respect the R-symmetry; it must be possible to collect $k$ complete copies of $\{1,2,3,4\}$ ($k$ is of course a non-negative integer) or the helicity amplitude under consideration is identically zero. For instance, $A(k_1^1,k_2^2,k_3,k_4^3)$ must vanish whereas $A(k_1^{1234},k_2,k_3,k_4)$ is {\it a priori} non-zero (with $k = 1$). Actually, it turns out that supersymmetry forces all scattering amplitudes with $k = 0~ {\rm or}~1$ to be equal to zero. Consequently, the first non-zero amplitudes have $k = 2$. Such amplitudes are called MHV amplitudes for historical reasons\footnote{MHV stands for maximally helicity violating. The $n$-point MHV amplitude describes, for example, a scattering experiment where two negative helicity gluons go in and $n-4$ positive helicity gluons and 2 negative helicity gluons come out. Such an outcome violates helicity as much as is possible at tree-level in QCD.}. The notation $N^{k - 2} MHV$ is a standard and convenient way to describe how close to MHV the helicity configuration of a given scattering amplitude is.  
 
A few years ago, Stieberger and Taylor~\cite{ST1} discovered a relation between the one-loop gluon $\Nsym$ MHV amplitudes and the tree-level gluon open superstring MHV amplitudes for which they had no explanation. Our work explains the relation they found and generalizes it as much as possible. Since Stieberger and Taylor showed explicitly that all MHV amplitudes satisfy the relation, it is of some interest to look at the simplest uncalculated example as an explicit test of our proposed generalization of the simpler Stieberger-Taylor relation. In other words, we ought to calculate the all-orders-in-$\e$ one-loop six-gluon\footnote{It is straightforward to check that at least six external particles need to participate in order to get an NMHV amplitude.} next-to-MHV (NMHV) amplitudes in $\Nsym$. Fortunately, Stieberger and Taylor have already tabulated all independent six-gluon NMHV amplitudes in open superstring theory~\cite{ST4} compactified to four dimensions. The existence of these results will make it significantly easier to check our proposed relations. Furthermore, our relations shed some light in a non-obvious way on an old result in pure Yang-Mills. In a nutshell, we are able to explain why $A^{1-{\rm loop}}_{1;\,\mathcal{N}=0}(k_1,k_2,\cdots,k_n)$ \footnote{The subscript ``$1$'' in $A^{1-{\rm loop}}_{1;\,\mathcal{N}=0}(k_1,k_2,\cdots,k_n)$ indicates that only the planar part of the scattering amplitude is retained.}vanishes when $n > 4$ and three of the gluons are replaced by photons. The precise statement of our relations between gauge and string theory is somewhat technical and we postpone further discussion of it to Section \ref{gsrel}. Suffice it to say that the gauge theory side of our relation requires one-loop $\Nsym$ amplitudes calculated to all orders in $\e$.

Before we can discuss our results for all-order one-loop $\Nsym$ amplitudes, we have to remind the reader of several exciting recent developments in the theory of $\Nsym$ scattering amplitudes. We now specialize to the planar limit (this is crucial for what follows) and discuss some of the remarkable features of the $\Nsym$ S-matrix in this limit. Particularly exciting is the fact that, in the planar limit, it is possible to completely solve the perturbative S-matrix (up to momentum independent but coupling constant dependent pieces) for the scattering of either four gluons or five gluons (and, by $\Nsym$ supersymmetry, all four- and five-point amplitudes). Starting from the work of~\cite{ABDK}, Bern, Dixon, and Smirnov~\cite{BDS} made an all-loop, all-multiplicity proposal for the finite part of the MHV amplitudes in $\Nsym$. In this paper~\cite{BDS}, BDS explicitly demonstrated that their ansatz was valid for the four-point amplitude through three loops. Subsequent work demonstrated that the BDS ansatz holds for the five-point amplitude through two-loops~\cite{TwoLoopFive} and that the strong coupling form of the four-point amplitude calculated via the AdS/CFT correspondence has precisely the form predicted by BDS~\cite{origAldayMald}.  

In fact,~\cite{origAldayMald} sparked a significant parallel development. Motivated by the fact that the strong coupling calculation proceeded by relating the four-point gluon amplitude to a particular four-sided light-like Wilson loop, the authors of~\cite{DKS4pt} were able to show that the finite part of the four-point light-like Wilson loop at one-loop matches the finite part of the planar one-loop four-gluon scattering amplitude. The focus of~\cite{DKS4pt} was on the planar four-gluon MHV amplitude, but it was shown in~\cite{BHTWLn} that this MHV amplitude/light-like Wilson loop correspondence holds for all one-loop MHV amplitudes  in $\Nsym$. As will be made clear in Section \ref{supercomp}, an arbitrary $n$-gluon light-like Wilson loop should be conformally invariant in position space.\footnote{Strictly speaking, the conformal symmetry is anomalous due to the presence of divergences at the cusps in the Wilson loop. If one regulates these divergences and subtracts the conformal anomaly, then the finite part of what remains will be conformally invariant.} What was not at all obvious before the discovery of the amplitude/Wilson loop correspondence is that $\Nsym$ scattering amplitudes must be (dual) conformally invariant in {\it momentum} space. 

It turns out that this hidden symmetry (referred to hereafter as dual conformal invariance) has non-trivial consequences for the $\Nsym$ S-matrix. Assuming that the MHV amplitude/light-like Wilson loop correspondence holds to all loop orders, the authors of~\cite{DHKSward} were able to prove that dual conformal invariance  fixes the form of all the four- and five- point gluon helicity amplitudes (recall that non-MHV amplitudes first enter at the six-point level) to all orders in planar $\Nsym$ perturbation theory. Up to trivial factors, they showed that the functional form of the (dual) conformal anomaly coincides with that of the BDS ansatz. Subsequently, work was done at strong coupling~\cite{BM,BRTW} that provides evidence for the assumption made in~\cite{DHKSward} that the MHV amplitude/light-like Wilson loop correspondence holds to all orders in perturbation theory. Quite recently, the symmetry responsible for the correspondence was understood from a perspective that bears on the results seen at weak coupling as well~\cite{DrummondFerro}.

The idea is that, due to the fact that non-trivial conformally invariant cross-ratios can first be formed at the six-point level, one would na\"{i}vely expect the four- and five-point amplitudes to be momentum-independent constants to all orders in perturbation theory. It is well-known, however, that gluon loop amplitudes have IR divergences. These IR divergences explicitly break the dual conformal symmetry and it is precisely this breaking which allows four- and five- gluon loop amplitudes to have non-trivial momentum dependence. In fact, the arguments of~\cite{DHKSward} allowed the authors to predict the precise form that the answer should take and they found (up to trivial constants) complete agreement with the BDS ansatz to all orders in perturbation theory.

At this stage, it was unclear whether the appropriate hexagon Wilson loop would still be dual to the six-point MHV amplitude at the two-loop level. This question was decisively settled in the affirmative by the work of~\cite{BDKRSVV} on the scattering amplitude side and~\cite{DHKS2nd2L6pt} on the Wilson loop side. Another issue settled by the authors of~\cite{BDKRSVV} and~\cite{DHKS2nd2L6pt} was the question of whether the BDS ansatz fails at two loops and six points. It had already been suggested by Alday and Maldacena in~\cite{AMlargen} that the BDS ansatz must fail to describe the analytic form of the $L$-loop $n$-gluon MHV amplitude for sufficiently large $L$ and $n$, but it had not yet been conclusively proven until the appearance of~\cite{BDKRSVV} and~\cite{DHKS2nd2L6pt} that $L = 2$ and $n = 6$ was the simplest possible example of BDS ansatz violation. The difference between the full answer and the BDS ansatz is called the remainder function and it is invariant under the dual conformal symmetry. 

Since full two-loop six-point calculations are extremely arduous, one might hope that there is a smoother route to proving the above fact. In fact, Bartels, Lipatov, and Sabio Vera~\cite{BLSV1,BLSV2} derived an approximate formula for the imaginary part of the two-loop remainder function in a particular region of phase-space and multi-Regge kinematics. For some time, this formula was the subject of controversy, due to subtleties associated with analytical continuation of two-loop amplitudes. In~\cite{mysecond} the present author confirmed the controversial result\footnote{To appreciate the subtlety here the reader may wish to read the discussion at the end of~\cite{mysecond,BNST} as it relates to the erratum at the end of~\cite{DDG}.} of BLSV for the imaginary part of the remainder by explicitly continuing the full results of~\cite{DHKS2nd2L6pt} into the Minkowski region of phase-space in question.

Drummond, Henn, Korchemsky, and Sokatchev (DHKS)~\cite{DHKS2nd2L6pt} recently discovered an even larger symmetry of the planar S-matrix. DHKS~\cite{DHKSdualconf} found that there is actually a full dual $\Nsym$ superconformal symmetry acting in momentum space, which they appropriately christened dual superconformal symmetry. One of the main ideas utilized in ~\cite{DHKSdualconf} is that all of the scattering amplitudes with the same value of $n$ are unified into a bigger object called an on-shell $\Nsym$ superamplitude. This superamplitude can be further expanded into $k$-charge sectors and we will often refer to the $k$-charge sectors of a given superamplitude as superamplitudes as well. For example, the $n = 6$, $k = 2$ superamplitude would contain component amplitudes like $\mathcal{A}\left(p_1^{1234},p_2^{1234},p_3,p_4,p_5,p_6\right)$ and $\mathcal{A}\left(p_1^{1},p_2^{1234},p_3^{234},p_4,p_5,p_6\right)$ among others. 

DHKS~\cite{DHKSdualconf}  made an intriguing conjecture for the ratio of the $k = 3$ and $k = 2$ six-point superamplitudes. They argued that the $k = 3$ superamplitude is naturally written as the $k = 2$ superamplitude times a function invariant under the dual superconformal symmetry. DHKS explicitly demonstrated that their proposal holds in the one-loop approximation. This ratio function has recently been the subject of intensive investigation and there are strong arguments in favor of it~\cite{BHMPYangian2}. Nevertheless, it would be nice to see explicitly that the two-loop ratio function is invariant under dual superconformal symmetry and this was done quite recently by Kosower, Roiban, and Vergu for an appropriately defined even part of the ratio function~\cite{KRV}. It turns out that the all-orders-in-$\e$ one-loop six-point $\Nsym$ NMHV superamplitude is necessary to explicitly test the dual superconformal invariance of the ratio function at two loops in dimensional regularization. In Section \ref{supercomp} we review some of the developments that led to the discovery of dual superconformal symmetry and summarize its important features. We then explain how the all-orders-in-$\e$ one-loop formulas we present in Section \ref{gluoncomp} for purely gluonic amplitudes can be supersymmetrized in a way that manifests this hidden symmetry as much as possible.

To summarize, the structure of this review is as follows. In Section \ref{revcomp} we briefly review some well-known results used later on in the review and fix some notation. In Section \ref{gluoncomp} we discuss a new, efficient approach to the calculation of all-orders-in-$\e$ one-loop $\Nsym$ amplitudes, with the one-loop six-point gluon NMHV amplitude as our main non-trivial example. In Section \ref{gsrel} we discuss a novel relation between one-loop $\Nsym$ gauge theory and tree-level open superstring theory and illustrate its usefulness by solving an old puzzle in pure Yang-Mills. In Section \ref{supercomp} we begin by reviewing $\Nsym$ on-shell supersymmetry, the BDS ansatz, and few other important related results. We then elaborate on the light-like Wilson loop/MHV amplitude correspondence and dual superconformal invariance. We show that, with a modest amount of additional effort, the gluonic pentagon coefficents derived in Section \ref{gluoncomp} can be supersymmetrized and written in a way that meshes well with the dual superconformal symmetry. Finally, we provide an alternative supersymmetrized form which was used by Kosower, Roiban, and Vergu in their recent test of the dual superconformal invariance of the two-loop NMHV ratio function~\cite{KRV}. We conclude in Section \ref{sum} where we recapitulate the main results reviewed. In addition, we provide several appendices where we discuss important topics that deserve some attention but would be awkward to include in the main text. In \ref{dimregs} we discuss some technical aspects of dimensional regularization and the structure of the IR divergences in planar $\Nsym$ gauge theory at the one-loop level. Finally, in \ref{sconf}, we give a brief introduction to the $\Nsym$ superconformal group and present in considerable detail the $\Nsym$ superconformal and dual superconformal algebras. \ref{sconf} contains the complete conformal and dual superconformal algebras and corrects various misprints existing in the literature.
\section{Brief Summary of Computational Techniques}
\label{revcomp}
In this section we remind the reader of the tools that make state-of-the-art gauge theory computations possible. In \ref{planar} we define the planar limit of Yang-Mills theory and explain how working in this limit dramatically simplifies the color structure of scattering amplitudes at the multi-loop level. In \ref{SH} we establish our spinor helicity conventions. In \ref{GUD} we introduce the $D$ dimensional generalized unitarity method at the one-loop level in the context of $\Nsym$ and discuss the integral basis, valid to all orders in $\e$, needed to use it. 
\subsection{The Planar Limit and Color Decomposition}
\label{planar}
Long ago, 't Hooft observed that non-Abelian gauge theories simplify dramatically~\cite{origtHooft} in a particular limit, in which one fixes the combination $\lambda = 2 \Nc ~g^2$, eliminates $g$ in favor of $\Nc$ and $\lambda$, and then takes $\Nc$ to infinity ($\lambda$ is referred to as the 't Hooft coupling in his honor). One thing 't Hooft conjectured was that large $\Nc$ gauge theory ought to have a stringy description. This idea was given new life by Maldacena in his ground-breaking work~\cite{origMald} on the near-horizon geometry of $AdS_5 \times S_5$. In brief, Maldacena showed that type IIB superstring theory in an $AdS_5 \times S_5$ background is dual to a $\Nsym$ SYM gauge theory. Maldacena's duality was incredibly novel because it related planar $\Nsym$ at strong coupling to classical type IIB superstring theory at weak coupling. In this review, we will see that unexpected simplicity also emerges in the planar limit of {\it weakly} coupled $\Nsym$.  

For our purposes, the advantage of working in the planar limit is that the well-known tree-level color decomposition formula 
\begin{changemargin}{-.8 in}{0 in}
\bea
\label{finloopcol}
&& \a^{1-{\rm loop}}_{\Nsym}\left(k_1^{h_1},~\cdots,~k_n^{h_n}\right) \\
&&\quad
= g^{n-2}{g^2 \Nc \mu^{2\e} e^{-\gamma_E \e}\over (4 \pi)^{(2-\e)}} \sum_{\sigma \in S_n/\mathcal{Z}_n} {\rm Tr}[T^{a_{\sigma(1)}}~\cdots T^{a_{\sigma(n)}} ]  A^{1-{\rm loop}}_{1;\,\Nsym}\left(k_{\sigma(1)}^{h_{\sigma(1)}},~\cdots,~k_{\sigma(n)}^{h_{\sigma(n)}}\right)
\nonumber\\&&\quad
+ g^{n-2}{g^2 \mu^{2\e} e^{-\gamma_E \e} \over (4 \pi)^{2-\e}} \sum_{m = 2}^{[{n \over 2}]+1}\Bigg( \sum_{\sigma \in~ S_n/(\mathcal{Z}_{m-1}\times\mathcal{Z}_{n-m+1})} {\rm Tr}[T^{a_{\sigma(1)}}~\cdots T^{a_{\sigma(m-1)}}]\times
\el \qquad\qquad\qquad\qquad\qquad\quad
\times{\rm Tr}[T^{a_{\sigma(m)}}~\cdots T^{a_{\sigma(n)}}] A^{1-{\rm loop}}_{2;\,\Nsym}\left(k_{\sigma(1)}^{h_{\sigma(1)}},~\cdots,~k_{\sigma(n)}^{h_{\sigma(n)}}\right)\Bigg)\,{\rm ,}\nonumber
\eea
\end{changemargin}
the single-trace\footnote[1]{Our normalization conventions for $SU(\Nc)$ differ from those of Peskin and Schroeder; we use ${\rm Tr}[T^a T^b] = \D^{a b}$ as opposed to ${\rm Tr}[T^a T^b] = {\D^{a b}\over 2}$.} color structures have an explicit factor of $\Nc$ out front that the double-trace structures do not. It follows that the single-trace structures dominate in the large $\Nc$ limit. To be explicit, the planar $L$-loop color decomposition formula is
\cmb{-.6 in}{0 in}
\bea
&&
{\cal A}_1^{L-{\rm loop}}\left(k_1^{h_{\sigma(1)}}, k_2^{h_{\sigma(2)}}, ~\ldots, k_n^{h_{\sigma(n)}}\right) = 
g^{n-2}\left({g^2 \Nc \mu^{2\e} e^{-\gamma_E \e} \over (4 \pi)^{2-\e}}\right)^L
\el
\qquad\quad
\times \sum_{\sigma \in~ S_n/Z_n}
 \, \Tr[ T^{a_{\sigma(1)}} T^{a_{\sigma(2)}}
   \ldots T^{a_{\sigma(n)}} ]
      A_{1}^{L-{\rm loop}}\left(k_1^{h_{\sigma(1)}}, k_2^{h_{\sigma(2)}}, ~\ldots, k_n^{h_{\sigma(n)}}\right)\,{\rm .} \nn
\label{LLoopDecomposition}
\eea
\cme
Clearly, this is going to be much easier to work with than a full $L$-loop color decomposition.

Although $\Nsym$ supersymmetry by itself is very powerful and puts highly non-trivial constraints on the perturbative S-matrix, $\Nsym$ supersymmetry together with the planar limit is even more powerful. In Section \ref{supercomp} we will discuss a so-called hidden symmetry of the $\Nsym$ S-matrix that emerges in the large $\lambda$ limit. This symmetry, called dual superconformal invariance is like a copy of the ordinary superconformal invariance of the $\Nsym$ model that acts in {\it momentum} space\footnote{The Lagrange density of $\Nsym$ is manifestly superconformally invariant in position space. We encourage the reader unfamiliar with superconformal symmetry to peruse \ref{sconf}.}. 
  
\subsection{Spinor Helicity Formalism}
\label{SH}
In this subsection we fix our conventions for the spinor helicity formalism~\cite{Calkul1,Calkul2,Calkul3,Calkul4,Calkul5,Calkul6}. Our underlying assumption in using this method is that the correct way to deal with the S-matrix is to compute helicity amplitudes. This approach is a useful one in practice because many helicity amplitudes are protected by supersymmetry or related by discrete symmetries (entering either from parity invariance or the color decomposition). In fact, our conventions are very nearly those of~\cite{Dixon96rev}, though we prefer to rewrite everything in more modern notation.

To this end, we define
\be
u^+(p_i) \equiv  |i~\rangle ~~~~~~ u^-(p_i) \equiv |i~] ~~~~~~ \bar{u}^+(p_i) \equiv [i~| ~~~~~~ \bar{u}^-(p_i) \equiv \langle i~| \,{\rm .}
\ee
Then Mandelstam invariants $(p_i+p_j)^2$ and gluon polarization vectors are expressed as
\bea
s_{i j} = \spa{i}.j\spb{j}.i 
\qquad
\pol^+_\mu(p_i) = {\spab{q_i}.\gamma_\mu.i \over \sqrt{2}\, \spa{q_i}.i} 
\qquad
\pol^-_\mu(p_i) = {\langle i |\, \gamma_\mu \,| q_i ] \over \sqrt{2}\, \spb{i}.{q_i}}
\label{polvecs}
\eea
Finally, for reference, we take this opportunity to define the $n$-gluon tree-level MHV amplitude (Parke-Taylor formula~\cite{ParkeTaylor}) in terms of spinor variables:
\be
A_{n;\,\spa{i}.{j}}^{\textrm{\scriptsize{MHV}}}\equiv A^{tree}\left(1,...,i^{1234},...,j^{1234},...,n\right)=i { \langle i j \rangle^4 \over \langle 1 2 \rangle \langle 2 3 \rangle...\langle n 1 \rangle} \,{\rm .}
\label{ParkeTayl}
\ee 
\subsection{Generalized Unitarity in $D$ Dimensions}
\label{GUD}
We now turn to loop-level calculations. Most of the calculations in this review are performed at the one-loop level, but the ideas reviewed in this subsection, with appropriate modifications, have been applied to multi-loop calculations as well (see~\cite{CarrascoRev} for a review). At the time of this writing, the program of {\it generalized unitarity} pioneered by Bern, Dixon, Dunbar, and Kosower~\cite{BDDKMHV,BDDKNMHV} (for a review see~\cite{BernRev}) has almost completely replaced the traditional Feynman diagram based approach to one-loop calculations. In particular, it is now possible to calculate the virtual corrections to any one-loop scattering process in the Standard Model~\cite{Garzelli}. The great thing about generalized unitarity is that it works in very general situations. In particular, generalized unitarity is compatible with dimensional regularization because dimensional regularization preserves unitarity. As was shown shortly after the seminal papers on the generalized unitarity technique were published, there is no inherent restriction to $\Ord(\e^0)$; if desired, one can compute amplitudes to all orders in $\e$ by working a little harder.~\cite{BernMorgan}

It has been known at least since the work of~\cite{VermNeer} that any one-loop planar $\Nsym$ amplitude can be written as
\be
A_1^{1-{\rm loop}}(k_1^{h_1},\cdots,k_n^{h_n}) = \sum_\alpha C_\alpha I^{(\alpha)}_4 + \sum_\beta D_\beta I^{(\beta)}_5\,{\rm ,}
\label{gen1loop}
\ee
where $\alpha$ or $\beta$ labels the specific kinematic structure of the box or pentagon integral (more on our labeling scheme below).  Much of the power of the generalized unitarity technique comes from (\ref{gen1loop}), so it is worth spending some time trying to understand it. Of course (\ref{gen1loop}) is very special to $\Nsym$~\cite{1001lessons, BDDKMHV}. Less supersymmetry ({\it i.e.} $\mathcal{N}=1$ super Yang-Mills) makes analogous relations less powerful and a little more difficult to work  with (see {\it e.g.}~\cite{Forde}). For $\mathcal{N}=0$ it becomes harder still. 

Before we start, we need a convenient way to enumerate the box and pentagon topologies for a planar $n$-particle scattering process. Consider, as usual, a regular $n$-gon with one external line attached at each vertex. In an approach based on Feynman diagrams this would be the highest-point Feynman integral that could possibly appear prior to integral reduction. There are 
\be
\left(\begin{array}{c} n \\ n-4\end{array}\right) = {n! \over (n-4)! 4!}
\label{noboxes}
\ee
ways to collapse this $n$-gon down to a box and
\be
\left(\begin{array}{c} n \\ n-5\end{array}\right) = {n! \over (n-5)! 5!}
\label{nopents}
\ee
ways to collapse it down to a pentagon. Consequently, it is natural to label each box or pentagon in the integral basis by, respectively, an $n-4$-tuple or $n-5$-tuple of integers corresponding to the internal lines that need to be erased to produce the box or pentagon in question\footnote{Our convention will be to start counting with the propagator connecting the 1st and $n$th vertices.}. In this work we will mostly be interested in $n = 6$ for which the above enumeration gives 15 boxes and 6 pentagons. 

Formula (\ref{noboxes}) gives the largest number of boxes that could possibly appear. Depending on the helicity configuration, certain classes of boxes may make no contribution to the sum in (\ref{gen1loop}). Because it will make our notation clear and because it will be nice to have them later on, we write the $C_\alpha I^{(\alpha)}_4$ terms in (\ref{gen1loop}) out explicitly for $A_1^{1-{\rm loop}}\left(k_1^{1234},k_2^{1234},k_3,k_4,k_5,k_6\right)$:\footnote{In eq. (\ref{6ptgMHV}) $s_i \equiv s_{i\,i+1}\equiv(k_i+k_{i+1})^2$ and $t_i \equiv s_{i\,i+1\,i+2} \equiv (k_i+k_{i+1}+k_{i+2})^2$, where indices are mod 6. We will frequently use this notation in our discussions of six-point scattering. The notation can, of course, be generalized to describe a basis of kinematic invariants for arbitrary $n$. For instance, at the eight-point level, invariants like $w_{1} \equiv (k_1+k_2+k_3+k_4)^2$ will enter.}
\begin{changemargin}{-.6 in}{0 in}
\bea
&&A_1^{1-{\rm loop}}\left(k_1^{1234},k_2^{1234},k_3,k_4,k_5,k_6\right) = 
\frac{1}{2}{A_{n;\,\spa{1}.{2}}^{\textrm{\scriptsize{MHV}}} }\Big(- s_3 s_4 I_{4}^{(1,2)}- {s_4 s_5} I_{4}^{(2,3)} - {s_5 s_6} I_{4}^{(3,4)} 
\el \qquad\qquad\qquad
-{s_1 s_6 }I_{4}^{(4,5)} - {s_1 s_2 }I_{4}^{(5,6)} - {s_2 s_3 } I_{4}^{(1,6)}+ (s_3 s_6 - t_2 t_3) I_{4}^{(1,4)} 
\el\qquad\qquad\qquad
+ (s_1 s_4 - t_1 t_3) I_{4}^{(2,5)} + (s_2 s_5 - t_1 t_2) I_{4}^{(3,6)} +  \Ord(\e)\Big) 
\label{6ptgMHV}
\eea
\cme
We see that in the six-point MHV amplitude all of the boxes with two adjacent external masses enter with zero coefficient. Clearly, the zero mass box will appear in (\ref{gen1loop}) only for the special case of four particle scattering. For general $n$, planar $\Nsym$ MHV amplitudes are built out of one mass and two mass easy boxes~\cite{BDDKMHV} and planar $\Nsym$ NMHV amplitudes are built out of one mass, two mass easy, two mass hard, and three mass boxes~\cite{allNMHV}; four mass boxes don't appear until the eight-point ${\rm N}^2$MHV level. In particular, the absence of two mass easy basis integrals in $A_1^{1-{\rm loop}}\left(k_1^{1234},k_2^{1234},k_3^{1234},k_4,k_5,k_6\right)$ does not generalize to higher $n$ NMHV amplitudes. The reader unfamiliar with the standard notation (``two mass easy'' and so forth) used for the box integrals may wish to consult reference~\cite{myfifth} or~\cite{BrittoRev}. 

Before going further, we need to carefully define the one-loop Feynman integrals which enter into our perturbative calculations. In general, the Feynman integrals that enter into calculations in massless gauge theories have severe IR divergences that need to be regulated. In dimensional regularization~\cite{origtHooftVelt} one regulates the IR divergences by analytically continuing the scattering amplitude under consideration from $D = 4$ to $D = 4 - 2 \e$ and then computing its Laurent expansion about $\e = 0$. We make the definition
\bea
\label{IntDef}
\!\!\!\!\!\!\!\!\!\!\!\!\!\!\!\!\!\!\!
I_n^{D=4-2 \e}
&\equiv& i (-1)^{n+1}(4\pi)^{2-\e} \int 
    {d^{4-2\e} p \over (2\pi)^{4-2\e} }
  { 1 \over p^2 \ldots
    (p-\sum_{i=1}^{n-1} K_i )^2 }
 \\   &=& i (-1)^{n+1} (4\pi)^{2-\e} \int 
    {d^{4} p \over (2\pi)^{4} }
    {d^{-2\e} \mu \over (2\pi)^{-2\e} }
  { 1 \over (p^2 - \mu^2) \ldots (
    (p-\sum_{i=1}^{n-1} K_i )^2 - \mu^2) } \,{\rm .}
    \nonumber
\eea
The prefactor $i (-1)^{n+1} (4\pi)^{2-\e}$ cancels a factor of $i (-1)^{n} (4\pi)^{\e-2}$ that always arises in the calculation of one-loop Feynman integrals and on the second line we have explicitly separated out the integrations into four dimensional and $-2\e$ dimensional pieces. This second form will be particularly useful in the context of $D$ dimensional generalized unitarity.

In this work, we will never need explicit formulas for basis integrals expanded in $\e$. By combining the principle of generalized unitarity~\cite{AnalyticSmatrix}, the fact that the integrals in (\ref{gen1loop}) form a complete basis for one-loop planar $\Nsym$ scattering amplitudes, and the fact that all box integrals can be uniquely specified by how they develop residues when viewed as contour integrals in $\mathbb{C}^4$, we can deduce all of the $C_\alpha$ coefficients in the sum of eq. (\ref{gen1loop}) for any given amplitude without explicitly evaluating a single Feynman integral~\cite{BDDKMHV,origBCF,SharpLS}. This procedure, called the leading singularity method~\cite{SharpLS}, is extremely powerful but nevertheless fails to determine some subset of the $D_\beta$'s. We will have to supplement the leading singularity method with an independent $D$ dimensional generalized unitarity calculation to derive the pentagon coefficients not fixed by it. We look at this in detail in Section \ref{gluoncomp}. Before moving on, we work through an exceptionally simple example, aspects of which will be useful later on in the review.

We  consider the amplitude $A^{1-{\rm loop}}_{1;\,\mathcal{N}=0}(k_1,k_2,k_3,k_4)$ in pure Yang-Mills theory. Following~\cite{BernMorgan}, we remind the reader of the second form for $I_4^{D = 4 - 2\e}$ where we split up the integral over the loop momentum into four dimensional and $-2\e$ dimensional pieces:
\cmb{-1.8 in}{0 in}
\be
I_4^{D=4-2 \e} = -i (4\pi)^{2-\e} \int 
    {d^{4} p \over (2\pi)^{4} }
    {d^{-2\e} \mu \over (2\pi)^{-2\e} }
  { 1 \over (p^2 - \mu^2) ((p-k_1)^2 - \mu^2) ((p-k_1-k_2)^2 - \mu^2)(
    (p+k_4)^2 - \mu^2) } \,{\rm .}\nn
\ee
\cme
If we consider an $s$-channel cut of the above zero mass box integral, we find the on-shell conditions
\be
p^2 = \mu^2 \qquad (p-k_1-k_2)^2 = \mu^2\,{\rm .}
\ee
It follows that, to reconstruct the complete one-loop integrand in $D$ dimensions using the principle of generalized unitarity, one should simply imagine that the lines of the tree amplitudes on either side of the unitarity cut(s) (external lines of the trees that have $p$-dependent momenta) have a mass $\mu$. Actually, the procedure of gluing trees together to form loops is a little more complicated in our approach because we do not have in hand an analog of the spinor helicity framework in $-2\e$ dimensions\footnote{It is possible that, with a bit of inspiration, we might be able to profitably make use of some combination of the formalisms worked out in~\cite{Dennen,CheungC,BCDHI}.} Consequently, the whole process is more closely related to traditional perturbation theory. In particular, summing over internal degrees of freedom inside the loop being reconstructed is much more labor intensive than it is in four dimensions. One trick to try and avoid tedious algebra, which works better in some situations than in others, is to perform a supersymmetric decomposition of the amplitude. For example, if we rewrite a loop of gluons in the following way:
$$A_g = (A_g + 4 A_f + 3 A_s) - 4 (A_f + A_s) + A_s$$
We see that the contribution from a loop of gluons ({\it i.e.} pure Yang-Mills theory) can be derived by summing the answer in $\Nsym$ and the contribution from a loop of complex scalars and then subtracting off the contribution from four $\mathcal{N}=1$ chiral multiplets. For the present application this works beautifully because the first two terms on the right-hand side of the above equation are protected by supersymmetry and vanish. It follows that
\be
A^{1-{\rm loop}}_{1;\,\mathcal{N}=0}(k_1,k_2,k_3,k_4) = A^{1-{\rm loop}}_{1;\,{\rm scalar}}(k_1,k_2,k_3,k_4)
\label{all+susydecomp}
\ee
and, in this particular case, we can avoid some numerator algebra by calculating $A^{1-{\rm loop}}_{1;\,{\rm scalar}}(k_1,k_2,k_3,k_4)$ instead of $A^{1-{\rm loop}}_{1;\,\mathcal{N}=0}(k_1,k_2,k_3,k_4)$. 

Generalized unitarity applied to $A^{1-{\rm loop}}_{1;\,{\rm scalar}}(k_1,k_2,k_3,k_4)$ gives\footnote{In what follows, we will very often be interested in amplitudes where some of the external states have definite helicity and some should be thought of as having any of the possible physical polarizations. We label external states with indeterminate polarization as $(q)_{x}$ where $q$ is the momentum carried by the external particle and $x$ denotes the particle type ($s$ or $\bar{s}$ for scalar states, $f$ or $\bar{f}$ for fermion states, and $g$ for gluon states).}
\cmb{-1.2 in}{0 in}
\bea
&&{1\over (4 \pi)^{2-\e}} A^{1-{\rm loop}}_{1;\,{\rm scalar}}(k_1,k_2,k_3,k_4) = \int 
    {d^{4} p \over (2\pi)^{4} }
    {d^{-2\e} \mu \over (2\pi)^{-2\e} }
  \bigg({ i \over p^2 - \mu^2} A^{tree}_{\mu^2}\left((-p)_{s},k_1,k_2,(p-k_1-k_2)_{\bar{s}}\right)
\el{i\over(p-k_1-k_2)^2 - \mu^2} A^{tree}_{\mu^2}\left((-p+k_1+k_2)_s,k_3,k_4,p_{\bar{s}}\right)
  \el+ { i \over p^2 - \mu^2} A^{tree}_{\mu^2}\left((-p)_{\bar{s}},k_1,k_2,(p-k_1-k_2)_s\right) {i\over(p-k_1-k_2)^2 - \mu^2} A^{tree}_{\mu^2}\left((-p+k_1+k_2)_{\bar{s}},k_3,k_4,p_s\right)\bigg) \,{\rm .}\nn
\eea
\cme
The massive scalar amplitudes $A^{tree}_{\mu^2}\left((-p)_s,k_1,k_2,(p-k_1-k_2)_{\bar{s}}\right)$ and $A^{tree}_{\mu^2}\left((-p)_{\bar{s}},k_1,k_2,(p-k_1-k_2)_s\right)$ are equal, as are $A^{tree}_{\mu^2}\left((-p+k_1+k_2)_s,k_3,k_4,p_{\bar{s}}\right)$ and $A^{tree}_{\mu^2}\left((-p+k_1+k_2)_{\bar{s}},k_3,k_4,p_s\right)$. Using
\bea
A^{tree}_{\mu^2}\left((-p)_s,k_1,k_2,p'_{\bar s}
\right) &=& {i \mu^2 \spb1.2 \over \spa1.2 ((p-k_1)^2-\mu^2)}~~~~{\rm and} \\
A^{tree}_{\mu^2}\left(
p'_s,k_3,k_4,p_{\bar{s}}\right) &=& {i \mu^2 \spb3.4 \over \spa3.4 ((p+k_4)^2-\mu^2)}\,{\rm ,}
\eea
which can be derived from Feynman diagrams (with $p'_s$ and $p'_{\bar s}$ determined by momentum conservation), we find 
\cmb{-1.0 in}{0 in}
\bea
&&{1\over (4 \pi)^{2-\e}} A^{1-{\rm loop}}_{1;\,{\rm scalar}}(k_1,k_2,k_3,k_4) = {1\over (4 \pi)^{2-\e}} {2 \spb1.2 \spb3.4\over \spa1.2 \spa3.4}\times
\el \times \int 
    {d^{4} p \over (2\pi)^{4} }
    {d^{-2\e} \mu \over (2\pi)^{-2\e} }
  { \mu^4 \over (p^2 - \mu^2)((p-k_1)^2-\mu^2)((p-k_1-k_2)^2 - \mu^2)((p+k_4)^2-\mu^2)} \nn
&&= {1\over (4 \pi)^{2-\e}} {2 i \spb1.2 \spb3.4\over \spa1.2 \spa3.4} I_4^{D = 4 - 2\e}[\mu^4]
\eea
\cme
That is:
\be
\!\!\!\!\!\!\!\!
A^{1-{\rm loop}}_{1;\,\mathcal{N}=0}(k_1,k_2,k_3,k_4) = A^{1-{\rm loop}}_{1;\,{\rm scalar}}(k_1,k_2,k_3,k_4) = {2 i \spb1.2 \spb3.4\over \spa1.2 \spa3.4} I_4^{D = 4 - 2\e}[\mu^4]\,{\rm .}
\ee
A basis integral with some power of $\mu^2$ inserted in the numerator is usually referred to as a $\mu$-integral and such terms will play a central role in this work. It is often convenient to rewrite $\mu$-integrals in terms of dimensionally shifted integrals. This is easily accomplished by manipulating the $-2\e$ dimensional part of the integration measure in eq. (\ref{IntDef}). For $r \in \mathbb{N}$, this analysis leads to
\be
I_n^{D = 4 - 2\e}[\mu^{2 r}] = -\e(1-\e)(2-\e)\cdots(r-1-\e)I_n^{D = 2 r + 4 - 2\e}
\label{DSmu}
\ee
relating $\mu$-integrals and dimensionally-shifted integrals. Now, a very interesting phenomenon can occur, which we illustrate by applying eq. (\ref{DSmu}) to our result for $A^{1-{\rm loop}}_{1;\,\mathcal{N}=0}(k_1,k_2,k_3,k_4)$. We first rewrite the answer
\be
\!\!\!\!\!\!\!\!\!\!\!\!\!\!\!\!\!\!\!\!
A^{1-{\rm loop}}_{1;\,\mathcal{N}=0}(k_1,k_2,k_3,k_4) = {2 i \spb1.2 \spb3.4\over \spa1.2 \spa3.4} I_4^{D = 4 - 2\e}[\mu^4] = -{2 \e (1-\e)i \spb1.2 \spb3.4\over \spa1.2 \spa3.4} I_4^{D = 8 - 2\e}
\ee
and then Feynman parametrize it:
\be
\!\!\!\!\!\!\!\!\!\!\!\!\!\!\!\!\!\!\!\!
A^{1-{\rm loop}}_{1;\,\mathcal{N}=0}(k_1,k_2,k_3,k_4) = -{2 i \spb1.2 \spb3.4\over \spa1.2 \spa3.4} \int_0^1 dx\int_0^{1-x}dy\int_0^{1-x-y}dz 
{\e (1-\e)\G(\e)\over \mathcal{D}(x,y,z)^\e}\,{\rm .}
\ee
Remarkably, the $\e$ expansion of the above starts at $\Ord(\e^0)$. Explicitly, we find 
\bea
\!\!\!\!\!\!\!\!\!\!\!\!\!\!\!\!\!\!\!\!
A^{1-{\rm loop}}_{1;\,\mathcal{N}=0}(k_1,k_2,k_3,k_4) &=& -{2 i \spb1.2 \spb3.4\over \spa1.2 \spa3.4} \int_0^1 dx\int_0^{1-x}dy\int_0^{1-x-y}dz + \Ord(\e)
\elale -{i \spb1.2 \spb3.4\over 3 \spa1.2 \spa3.4}+ \Ord(\e)\,{\rm .}
\eea
At first sight, this result might seem rather puzzling since, without the $\mu^4$ in the numerator, the integral $I_4^{D = 4 -2 \e}$ is UV finite and IR divergent. What has happened is that, in shifting to $D = 8 - 2\e$, we have induced a UV divergence (the integral now has the same number of powers of the loop momenta in the measure of integration as it has in the denominator) and the IR divergences effectively got regulated by the $\mu^2$ factors in the propagator denominators. The explicit $\e$ in the numerator coming from eq. (\ref{DSmu}) is canceling the induced UV pole, which is why the $\e$ expansion of $A^{1-{\rm loop}}_{1;\,\mathcal{N}=0}(k_1,k_2,k_3,k_4)$ starts at $\Ord(\e^0)$.

Although we have been focusing on scalars running in the loop we could equally well have performed the above analysis for a loop of fermions with one obvious additional complication: the need to sum over internal spin states in a Lorentz covariant way. Typical tree amplitudes with a pair of massive fermions will be built out of a string beginning with $\bar{u}^{\pm}(p)$ and ending with $u^\pm(p)$. In order to fuse together two such tree amplitudes across a unitarity cut, we simply use the spin sum identity 
\be
\sum_{s} u^s(p)\bar{u}^s(p) = \slashed{p} + \mu
\label{fermspsum}
\ee
heavily used in traditional perturbation theory~\cite{PeskinSchroeder}. In Subsection \ref{effgcomp}, we treat a gluon running in the loop as well. Due to the fact there is no straightforward massive counterpart (with two spin states) to the massless gluon, treating an internal gluon line requires a little more thinking. In the last few years, $D$-dimensional unitarity has been systematized by several different groups~\cite{Rocket1,Rocket2,BadgerGUD,Ossola}.

Also, we wish to remark that there is no reason for us to restrict ourselves to double cuts; as we shall see in the next section, we can profit enormously by using quintuple cuts in $D$ dimensions to determine individual pentagon coefficients one at a time. Quintuple cuts are conceptually quite similar to the quadruple cuts (actually contour integrals in $\mathbb{C}^4$ in this context) used in leading singularity computations, although they are a bit more difficult to work out. In the next section we will see that the leading singularity method supplemented by $D$ dimensional quintuple cuts allows one to efficiently calculate all-orders-in-$\e$ one-loop $\Nsym$ amplitudes.
\section{Efficient Computation and New Results For One-Loop $\Nsym$ Gluon Amplitudes Calculated To All Orders in $\e$}
\label{gluoncomp}
\subsection{Efficient Computation Via $D$ Dimensional Generalized Unitarity}
\label{effgcomp}
In order to harness the power of $D$-dimensional unitarity for the application at hand, we have to extend the results of Bern and Morgan to treat cut internal gluon lines. To be clear, many other authors have thought about extending the Bern-Morgan approach to integrand reconstruction (see {\it e.g.}~\cite{Rocket1,Rocket2,BadgerGUD,Ossola}). All of them either focus on getting numerical results or isolate terms that would be missed by four dimensional generalized unitarity. There are obviously many applications where it makes sense to follow one of these strategies. In this review, however, we have a different goal. We further develop the Bern-Morgan approach and show that it is a very efficient way to analytically reconstruct general one-loop integrands in $D$ dimensions. In fact, we expect that our approach will mesh well with the spinor integration reduction technique of~\cite{ABFKM,BFY}, which is applicable to general field theory amplitudes at one-loop. Although these references analyzed a variety of processes, they started with integrands obtained by other means in all cases except that of a complex scalar running in the loop. A general strategy for the analytical reconstruction of one-loop integrands in $D$ dimensions was not discussed. In what follows we fill in this gap.

As a simple example of our approach to $D$-dimensional unitarity, we derive the massless pentagon coefficient associated to $A^{1-{\rm loop}}_1(k_1^{1234},k_2^{1234},k_3,k_4,k_5)$. All we really need to do right now is extend Bern-Morgan to the case of purely gluonic external states with a massless vector running in the loop. Later on we will also treat the case where some of the external gluons are replaced by fermions. It seems likely that so far most researchers have found it expedient to side-step the question of how to properly treat a gluon running in the loop by exploiting supersymmetry decompositions as was done in Subsection \ref{GUD}. We argue that it is no more difficult to calculate directly. 

Of course, the pentagon coefficient of $A^{1-{\rm loop}}_1(k_1^{1234},k_2^{1234},k_3,k_4,k_5)$ also has contributions coming from scalar and fermion fields running in the loop. Using the massive scalar three-point vertices~\cite{BGKSmassive},
\bea
A^{tree}_{\mu^2}\left((-p)_s,k_1,
p'_{\bar{s}}\right) &&= -i \sqrt{2} p \cdot \pol^+(k_1)~~~~{\rm and}
\\ A^{tree}_{\mu^2}\left((-p)_s,k_1^{1234},
p'_{\bar{s}}\right) &&= -i \sqrt{2} p \cdot \pol^-(k_1){\rm ,}
\eea
(with $p'_{\bar s}$ determined by momentum conservation) and quintuple $D$ dimensional generalized unitarity cuts we can deduce the pentagon integral coefficient for the scalar loop contribution to the five-point MHV amplitude. In the above, the polarization vectors can be evaluated using 
eq.~(\ref{polvecs})
because we are implicitly using the four dimensional helicity scheme (see \ref{4DHS}) where the external polarization vectors are kept in four dimensions. The result of this calculation is
\bea
&&\!\!\!\!\!\!\!\!\!\!\!\!\!\!\!\!\!\!\!\!\!\!\!\!\!\!\!\!\!\!\!\!\!
A_{1;\,{\rm scalar}}^{1-{\rm loop}}\left(k_1^{1234},k_2^{1234},k_3,k_4,k_5\right)\Big|_{I_5} = A^{tree}_{\mu^2}\left((-p_*)_s,k_1^{1234},(p_*-k_1)_{\bar{s}}\right) 
\el \!\!\!\!\!\!\!\!\!\!\!\!\!\!\!\!\!\!\!\!\!\!\!\!\!\!\!\!\!\!\!\!\!
\times A^{tree}_{\mu^2}\left((-p_*+k_1)_s,k_2^{1234},(p_*-k_1-k_2)_{\bar{s}}\right) A^{tree}_{\mu^2}\left((-p_*+k_1+k_2)_s,k_3,(p_*+k_4+k_5)_{\bar{s}}\right)
\el \!\!\!\!\!\!\!\!\!\!\!\!\!\!\!\!\!\!\!\!\!\!\!\!\!\!\!\!\!\!\!\!\!
\times A^{tree}_{\mu^2}\left((-p_*-k_4-k_5)_s,k_4,(p_*+k_5)_{\bar{s}}\right)A^{tree}_{\mu^2}\left((-p_*-k_5)_s,k_5,(p_*)_{\bar{s}}\right) \,{\rm ,}
\label{s5ptpent}
\eea
where $p_*^\nu$ solves the on-shell conditions:
\cmb{-.6 in}{0 in}
\bea
p_*^2 - \mu^2 = 0 \qquad (p_*-k_1)^2 - \mu^2 &=& 0 \qquad (p_*-k_1-k_2)^2 - \mu^2 = 0 \nonumber \\
(p_*+k_4+k_5)^2 - \mu^2 = 0 & \qquad & (p_*+k_5)^2 - \mu^2 = 0 \,{\rm .}
\eea
\cme
It turns out that, in this case, the solution is unique and is given by~\cite{BFY} expanding the four dimensional, massive loop momentum with respect to a basis $K_1$, $K_2$, $K_3$, and $K_4$ of four-vectors:
\be
p^\nu = L_1 K_1^\nu + L_2 K_2^\nu + L_3 K_3^\nu + L_4 K_4^\nu
\label{5ptos1}
\ee
and then solving a system of linear equations for the $L_i$ coefficients. It makes sense to choose the $K$'s to be the four-vectors in the problem; in the present example we set
\be
K_1 = k_1 + k_2 \qquad K_2 = k_1 \qquad K_3 = -k_4-k_5 \qquad K_4 = -k_5{\rm .}
\ee
Explicitly, we have
\be
\left(\begin{array}{c}L_1\\L_2\\L_3\\L_4\end{array}\right) = {1\over2}\left(\begin{array}{cccc}K_1^2 & K_1\cdot K_2 & K_1\cdot K_3 & K_1 \cdot K_4 \\K_2\cdot K_1 & K_2^2 & K_2\cdot K_3 & K_2 \cdot K_4\\K_3\cdot K_1 & K_3\cdot K_2 & K_3^2 & K_3 \cdot K_4\\K_4\cdot K_1 & K_4\cdot K_2 & K_4\cdot K_3 & K_4^2\end{array}\right)^{-1} \left(\begin{array}{c}K_1^2\\K_2^2\\K_3^2\\K_4^2\end{array}\right)\,{\rm .}
\label{5ptos2}
\ee

Now that we are warmed up, we are ready to try the quintuple cut of the fermion loop contribution. The only reason that the fermion loop contribution is more complicated is that we have to sum over internal fermion spin states using eq. (\ref{fermspsum}); the net result of the sum over internal states for the scalar loop contribution is just an overall factor of two. Although Bern and Morgan did not literally give their fermions a mass $\mu$, our procedure is easily deduced from the discussion in their paper~\cite{BernMorgan}.

To reconstruct the one-loop integrand, we need tree amplitudes with two massive fermions and a gluon:
\bea
A^{tree}_{\mu^2}\left(p_{\bar{f}},k_1,(-p-k_1)_f\right) &=& -{i \over \sqrt{2}}\bar{u}(p)\slashed{\pol}^+(k_1) u(p+k_1)\\
A^{tree}_{\mu^2}\left(p_{\bar{f}},k_1^{1234},(-p-k_1)_f\right) &=&  -{i \over \sqrt{2}}\bar{u}(p)\slashed{\pol}^-(k_1) u(p+k_1)
\eea
where we don't worry about specifying the spins of the fermions because we will ultimately sum over them using (\ref{fermspsum}). For the quintuple cut of the fermion loop we find  
\begin{changemargin}{-0.8 in}{0 in}
\bea
&&A_{1;\,{\rm fermion}}^{1-{\rm loop}}\left(k_1^{1234},k_2^{1234},k_3,k_4,k_5\right)\Big|_{I_5} 
\\
&&~~~~~~
= - \left(-{i\over \sqrt{2}}\right)^5  \bar{u}(p_*)\slashed{\pol}^+(k_5) u(p_*+k_5) \bar{u}(p_*+k_5)\slashed{\pol}^+(k_4) u(p_*+k_4+k_5)
\nonumber\\ 
&&~~~~~~~~~~~~
\times \bar{u}(p_*+k_4+k_5)\slashed{\pol}^+(k_3) u(p_*-k_1-k_2) \bar{u}(p_*-k_1-k_2)\slashed{\pol}^-(k_2) u(p_*-k_1)
\nonumber\\
&&~~~~~~~~~~~~
\times \bar{u}(p_*-k_1)\slashed{\pol}^-(k_1) u(p_*) 
\\ 
&& ~~~~~~
=- \left({i\over \sqrt{2}}\right)^5 {\Tr}\Big[\slashed{\pol}^+(k_5)(\slashed{p}_*+\slashed{k}_5 + \mu)\slashed{\pol}^+(k_4) 
(\slashed{p}_*+\slashed{k}_4+\slashed{k}_5 + \mu)\\ &&
~~~~~~~~~~~~
\slashed{\pol}^+(k_3) (\slashed{p}_*-\slashed{k}_1-\slashed{k}_2 + \mu)\slashed{\pol}^-(k_2) (\slashed{p}_*-\slashed{k}_1 + \mu)\slashed{\pol}^-(k_1) (\slashed{p}_* + \mu) \Big] \,{\rm .}
\nonumber
\label{f5ptpent}
\eea
\end{changemargin}
In this context, the extra overall minus sign is a result~\cite{BernMorgan} of using three-point amplitudes with spinor strings of the form $\bar{u}(p)\slashed{\pol}^+(k_1) u(p+k_1)$, when really they should have spinor strings of the form $\bar{u}(p)\slashed{\pol}^+(k_1) u(-p-k_1)$. Now that we understand how to deal with a loop of fermions, it is natural to ask what the analogous prescription is for a loop of gluons. Clearly, to start we need to write down three-point gluon amplitudes
\cmb{-1.0 in}{0 in}
\bea
A^{tree}_{\mu^2}\left(-p_{g},k_1,
p'_g\right) &=&  i \sqrt{2}\left(\pol^+(k_1)\cdot p~ g_{\rho \sigma}+k_{1\,\rho}~\pol^+_\sigma(k_1)-k_{1\,\sigma}~ \pol^+_\rho(k_1)\right)\pol^{*\,\rho}(p)\pol^\sigma(p-k_1)\nn
\\
A^{tree}_{\mu^2}\left(-p_{g},k_1^{1234},
p'_g\right) &=&   i \sqrt{2}\left(\pol^-(k_1)\cdot p~ g_{\rho \sigma}+k_{1\,\rho}~\pol^-_\sigma(k_1)-k_{1\,\sigma}~ \pol^-_\rho(k_1)\right)\pol^{*\,\rho}(p)\pol^\sigma(p-k_1)\nn
\eea
\cme
without committing to a specific choice of polarization for the gluons with $p$-dependent external momenta and with $p'_g$ determined by momentum conservation. These degrees of freedom will eventually be summed over. Actually, the correct summation procedure is fairly obvious~\cite{PeskinSchroeder}. We can use the na\"{i}ve replacement 
\be
\sum_\lambda \pol^\lambda_\rho(k_1)\pol^{*\,\lambda}_\sigma(k_1) \rightarrow -g_{\rho \sigma} 
\ee
valid in Abelian gauge theory, provided that we correct for the fact that we are overcounting states by including the quintuple cut of a ghost loop. This is simple since the contribution from a ghost loop is nothing but the contribution from a complex scalar loop with an extra overall minus sign coming the fact that the ghost field obeys Fermi-Dirac statistics:
\be
\!\!\!\!\!\!\!\!\!\!\!\!
A_{1;\,{\rm ghost}}^{1-{\rm loop}}\left(k_1^{1234},k_2^{1234},k_3,k_4,k_5\right)\Big|_{I_5} = -A_{1;\,{\rm scalar}}^{1-{\rm loop}}\left(k_1^{1234},k_2^{1234},k_3,k_4,k_5\right)\Big|_{I_5}\,{\rm .}
\ee
Returning to the quintuple cut of the gluon loop, we have (using the notation $k_{12}=k_1+k_2$, etc)
\begin{changemargin}{-1.0 in}{0 in}
\bea
&&A_{1;\,{\rm gluon}}^{1-{\rm loop}}\left(k_1^{1234},k_2^{1234},k_3,k_4,k_5\right)\Big|_{I_5} 
\\
&&
= \left(i \sqrt{2}\right)^5  \pol^{*\,\rho_1}(p_*) \Big(\pol^-(k_1)\cdot p_*~ g_{\rho_1 \sigma_1}+k_{1\,\rho_1}~\pol^-_{\sigma_1}(k_1)
-k_{1\,\sigma_1}~ \pol^-_{\rho_1}(k_1)\Big)\pol^{\sigma_1}(p_*-k_1) 
\nonumber\\&&~~~~~~~
\pol^{*\,\rho_2}(p_*-k_1) \Big(\pol^-(k_2)\cdot (p_*-k_1)~ g_{\rho_2 \sigma_2}+k_{2\,\rho_2}~\pol^-_{\sigma_2}(k_2)
-k_{2\,\sigma_2}~ \pol^-_{\rho_2}(k_2)\Big)\pol^{\sigma_2}(p_*-k_{12})  
\nonumber\\&&~~~~~~~
\pol^{*\,\rho_3}(p_*-k_{12}) \Big(\pol^+(k_3)\cdot (p_*-k_{12})~ g_{\rho_3 \sigma_3}+k_{3\,\rho_3}~\pol^+_{\sigma_3}(k_3)
-k_{3\,\sigma_3}~ \pol^+_{\rho_3}(k_3)\Big)\pol^{\sigma_3}(p_*+k_{45})
\nonumber\\&&~~~~~~~
\pol^{*\,\rho_4}(p_*+k_{45}) \Big(\pol^+(k_4)\cdot (p_*+k_{45})~ g_{\rho_4 \sigma_4}+k_{4\,\rho_4}~\pol^+_{\sigma_4}(k_4)
-k_{4\,\sigma_4}~ \pol^+_{\rho_4}(k_4)\Big)\pol^{\sigma_4}(p_*+k_5)
\nonumber\\&&~~~~~~~
\pol^{*\,\rho_5}(p_*+k_5) \Big(\pol^+(k_5)\cdot (p_*+k_5)~ g_{\rho_5 \sigma_5}+k_{5\,\rho_5}~\pol^+_{\sigma_5}(k_5)-k_{5\,\sigma_5}~ \pol^+_{\rho_5}(k_5)\Big)\pol^{\sigma_5}(p_*)
\\&&
=\left(i \sqrt{2}\right)^5   \left(\pol^-(k_1)\cdot p_*~ g_{\rho_1 \sigma_1}+k_{1\,\rho_1}~\pol^-_{\sigma_1}(k_1)-k_{1\,\sigma_1}~ \pol^-_{\rho_1}(k_1)\right)\left(-g^{\sigma_1 \rho_2}\right) 
\el~~~~~~~~~~~~
\left(\pol^-(k_2)\cdot (p_*-k_1)~ g_{\rho_2 \sigma_2}+k_{2\,\rho_2}~\pol^-_{\sigma_2}(k_2)-k_{2\,\sigma_2}~ \pol^-_{\rho_2}(k_2)\right)\left(-g^{\sigma_2 \rho_3}\right) 
\el~~~~~~~~~~~~
\left(\pol^+(k_3)\cdot (p_*-k_{12})~ g_{\rho_3 \sigma_3}+k_{3\,\rho_3}~\pol^+_{\sigma_3}(k_3)-k_{3\,\sigma_3}~ \pol^+_{\rho_3}(k_3)\right)\left(-g^{\sigma_3 \rho_4}\right)
\el~~~~~~~~~~~~
\left(\pol^+(k_4)\cdot (p_*+k_{45})~ g_{\rho_4 \sigma_4}+k_{4\,\rho_4}~\pol^+_{\sigma_4}(k_4)-k_{4\,\sigma_4}~ \pol^+_{\rho_4}(k_4)\right)\left(-g^{\sigma_4 \rho_5}\right)
\el~~~~~~~~~~~~
\left(\pol^+(k_5)\cdot (p_*+k_5)~ g_{\rho_5 \sigma_5}+k_{5\,\rho_5}~\pol^+_{\sigma_5}(k_5)-k_{5\,\sigma_5}~ \pol^+_{\rho_5}(k_5)\right)\left(-g^{\sigma_5 \rho_1}\right)\,{\rm .}
\label{g5ptpent}
\eea
\end{changemargin}

Finally, we combine together all of the above results with the appropriate multiplicities:
\cmb{-0.7 in}{0 in}
\bea
&&A_{1}^{1-{\rm loop}}\left(k_1^{1234},k_2^{1234},k_3,k_4,k_5\right)\Big|_{I_5} = 3 A_{1;\,{\rm scalar}}^{1-{\rm loop}}\left(k_1^{1234},k_2^{1234},k_3,k_4,k_5\right)\Big|_{I_5} 
\el+4 A_{1;\,{\rm fermion}}^{1-{\rm loop}}\left(k_1^{1234},k_2^{1234},k_3,k_4,k_5\right)\Big|_{I_5}
+\Bigg(A_{1;\,{\rm gluon}}^{1-{\rm loop}}\left(k_1^{1234},k_2^{1234},k_3,k_4,k_5\right)\Big|_{I_5}
\el-A_{1;\,{\rm scalar}}^{1-{\rm loop}}\left(k_1^{1234},k_2^{1234},k_3,k_4,k_5\right)\Big|_{I_5}\Bigg) \,{\rm ,}
\label{5ptpent}
\eea
\cme
where we have dealt with the ghost loop contribution as discussed above. 

Naturally, we would like to compare our result to the existing literature. Unfortunately, (\ref{5ptpent}) is not yet written in the appropriate form. Obviously, eq. (\ref{5ptpent}) is expressed in terms of $4-2\e$ dimensional boxes and pentagons. This basis is called the geometric basis~\cite{SharpLS}. It has been known, probably since the work of~\cite{oneloopdimreg}, that there is a better basis for $\Nsym$ amplitudes composed of $4 - 2\e$ dimensional box integrals and $6 - 2\e$ dimensional pentagon integrals. Answers typically look much more compact when written in this basis because, in what we will refer to as the dual conformal basis in this paper\footnote{This notation will be motivated in Section \ref{supercomp}.}, the higher order in $\e$ terms are more cleanly separated from those that are present through $\Ord(\e^0)$. This is related to the fact that there will always be an explicit $\e$ out front of $I_5^{D=6-2 \e}$ and $I_5^{D=6-2 \e}$ is both UV and IR finite~\cite{oneloopdimreg}. The connection between the geometric basis and the dual conformal basis was worked out in~\cite{oneloopdimreg}:
\be
I_5^{D=4-2\e} = {1\over 2}\bigg[\sum_{j=1}^5 C_j I_{4}^{(j),\,D=4-2\e} + 2\e C_0 I_5^{D=6-2 \e} \bigg]\,{\rm .}
\label{PtoDCB}
\ee
In \ref{muint}, we define the $C_j$ and $C_0$, derive eq. (\ref{PtoDCB}), and discuss some related integral reduction identities.

At last we can straightforwardly check (numerically using {\it e.g.} S@M~\cite{S@M}) that, after projecting (\ref{5ptpent}) onto the dual conformal basis using eq. (\ref{PtoDCB}), the result of eq. (\ref{5ptpent}) agrees with that obtained in~\cite{oneloopselfdual}:
\be
\e ~\pol(1,2,3,4) A_{5;\langle 12 \rangle}^{\textrm{\scriptsize{MHV}}} I_5^{D=6-2 \e}
\ee
where we have made the useful definition
\be
\!\!\!\!\!\!\!\!\!\!\!\!\!\!\!\!\!\!\!\!
 \pol(i,j,m,n) \equiv 4 i \pol_{\mu \nu \rho \sigma} k_i^\mu k_j^\nu k_m^\rho k_n^\sigma = \spb{i}.j \spa{j}.m \spb{m}.n \spa{n}.i - \spa{i}.j \spb{j}.m \spa{m}.n \spb{n}.i \, {\rm .}
 \label{epstensor}
\ee
Evaluating the numerator algebra becomes slightly more involved for quintuple cuts of one-loop six-gluon amplitudes, but we will still be able to use the above procedure to great effect. 

We are finally in a position to outline the strategy that we will use to solve, say, $A_{1}^{1-{\rm loop}}(k_1^{1234},k_2^{1234},k_3,k_4,k_5,k_6)$ to all orders in $\e$.  This amplitude works well as an example because its full analytical form is known~\cite{oneloopselfdual}:
\cmb{-0.9 in}{0 in}
\bea
&&A_{1}^{1-{\rm loop}}(k_1^{1234},k_2^{1234},k_3,k_4,k_5,k_6) ={A_{6;~\langle 12 \rangle}^{\textrm{\scriptsize{MHV}}} \over 2} \bigg(- s_3 s_4 I_{4}^{(1,2),\,D=4-2\e}- {s_4 s_5} I_{4}^{(2,3),\,D=4-2\e} 
\el\qquad 
- {s_5 s_6} I_{4}^{(3,4),\,D=4-2\e} -{s_1 s_6 }I_{4}^{(4,5),\,D=4-2\e}  
- {s_1 s_2 }I_{4}^{(5,6),\,D=4-2\e}- {s_2 s_3 } I_{4}^{(1,6),\,D=4-2\e} 
\el\qquad 
+ (s_3 s_6 - t_2 t_3) I_{4}^{(1,4),\,D=4-2\e}
+ (s_1 s_4 - t_1 t_3) I_{4}^{(2,5),\,D=4-2\e} + (s_2 s_5 - t_1 t_2) I_{4}^{(3,6),\,D=4-2\e}
\el\qquad 
+ \e \sum_{i=1}^6 \pol(i+1, i+2, i+3, i+4) I_5^{(i),\,D=6-2\e}
+\e~ \textrm{tr}[\slashed{k_1}\slashed{k_2}\slashed{k_3}\slashed{k_4}\slashed{k_5}\slashed{k_6}] I_6^{D=6-2\e}\bigg) \,{\rm .}
\label{6ptMHV}
\eea
\cme
We will, of course, mostly be interested in evaluating the three six-gluon NMHV amplitudes not related by discrete symmetries,  but the strategy utilized for $A_{1}^{1-{\rm loop}}(k_1^{1234},k_2^{1234},k_3,k_4,k_5,k_6)$ carries over in a completely straightforward fashion to the other six-gluon amplitudes.

The general idea is that, while the leading singularity method does not fix everything to all orders in $\e$ starting at six points, the method is very powerful and {\it does} fix everything up to terms with trivial soft and collinear limits. To illustrate this point let us discuss to what extent the universal factorization properties of $A_{1}^{1-{\rm loop}}(k_1^{1234},k_2^{1234},k_3,k_4,k_5,k_6)$ under soft and collinear limits determine the analytic form of the amplitude, given that we already know $A_{1}^{1-{\rm loop}}(k_1^{1234},k_2^{1234},k_3,k_4,k_5)$ to all orders in $\e$. It turns out that there is only one function in $A_{1}^{1-{\rm loop}}(k_1^{1234},k_2^{1234},k_3,k_4,k_5,k_6)$ that is not constrained in this approach: One can check that 
$\textrm{tr}[\slashed{k_1}\slashed{k_2}\slashed{k_3}\slashed{k_4}\slashed{k_5}\slashed{k_6}]$
has no soft or collinear limits in any channel.\footnote{A soft or collinear limit for planar amplitudes is particularly simple because one only has to consider nearest-neighbor pairs of momenta. If unfamiliar, see~\cite{Dixon96rev} for an elementary discussion of planar soft and collinear limits.} Therefore any attempt to deduce the form of the one-loop six-gluon MHV amplitude from that of the one-loop five-gluon amplitude by demanding consistency of the soft and collinear limits will miss terms like that above. 

This ambiguity is reflected in the solution of the leading singularity equations for $A_{1}^{1-{\rm loop}}(k_1^{1234},k_2^{1234},k_3,k_4,k_5,k_6)$. Solving the system of $15\times 2 = 30$ equations in $15 + 6 = 21$ unknowns\footnote[1]{Here we would like to remind the reader that, if desired, the one-loop scalar hexagon integral may be expressed as a linear combination of six one-loop scalar pentagon integrals (see eq. (\ref{hexred})).} determines 20 of the unknown integral coefficients in terms of one of the pentagon coefficients, say that associated to $I_5^{(1),\,4-2\e}$. The point is that if we can evaluate one pentagon coefficient using $D$-dimensional unitarity, then the leading singularity equations, which require only four dimensional inputs, give us everything else. This is a much better strategy than trying to evaluate the quintuple cut of each pentagon independently because it allows one to solve for all the pentagon coefficients with a minimum of effort beyond that required to determine the coefficients of the boxes. 

Before going any further, we should clarify a potentially confusing point about the solution to the leading singularity equations for $A_{1}^{1-{\rm loop}}(k_1^{1234},k_2^{1234},k_3,k_4,k_5,k_6)$. Suppose we let $B_1$ be the coefficient of $I_5^{(1),\,4-2\e}$. Then a generic box coefficient, say that of $I_{4}^{(1,6),\,D=4-2\e}$, will have the form $\alpha_{16}+\beta_{16} B_1$. It may seem strange that the box coefficient associated to $I_{4}^{(1,6),\,D=4-2\e}$ depends on the pentagon coefficient $B_1$. This apparent paradox is resolved by projecting the geometric basis onto the dual conformal basis: the pentagons $I_{5}^{(1),\,D=4-2\e}$ and $I_{5}^{(6),\,D=4-2\e}$ each contribute to the coefficient of $I_{4}^{(1,6),\,D=4-2\e}$ in the dual conformal basis after the formula (\ref{PtoDCB}) is applied to them. Remarkably, these extra contributions conspire to cancel all of the $B_1$ dependence that was present in the coefficient of $I_{4}^{(1,6),\,D=4-2\e}$, considered as an element of the geometric basis.

In solving the leading singularity equations, we were free to choose any pentagon coefficient we wanted as the parameter undetermined by the system. The reason that we chose the coefficient of $I_5^{(1),\,4-2\e}$ is that it is particularly simple to determine this integral coefficient using quintuple cuts. This follows from the fact that $A^{tree}_{\mu^2}\left((p-k_1)_s,k_1^{1234},k_6,(-p-k_6)_{\bar{s}}\right)$, $A^{tree}_{\mu^2}\left((p-k_1)_{\bar{f}},k_1^{1234},k_6,(-p-k_6)_f\right)$, and $A^{tree}_{\mu^2}\left((p-k_1)_{g},k_1^{1234},k_6,(-p-k_6)_g\right)$ can each be represented by a single Feynman diagram:
\cmb{-.8 in}{0 in}
\bea
A^{tree}_{\mu^2}\left(p_s,k_1^{1234},k_6,p'_{\bar{s}}\right) &=& -{i \spab1.{p}.6^2 \over s_6 \spab1.p.1}
\label{mu4tree1}\\ 
A^{tree}_{\mu^2}\left(p_{\bar{f}},k_1^{1234},k_6,p'_f\right) &=& {i (p+k_6)\cdot \pol^+(k_6)\over \spab1.p.1}\bar{u}(p+k_6)\slashed{\pol}^-(k_1)u(p-k_1)
\label{mu4tree2}\\ 
A^{tree}_{\mu^2}\left(p_{g},k_1^{1234},k_6,p'_g\right) &=& -{2 i\pol^\rho(p-k_1) \pol^{*\,\sigma}(p+k_6) \over \spab1.p.1}\Big(\pol^-(k_1)\cdot p ~\pol^+(k_6)\cdot p ~g_{\rho \sigma} 
\nonumber\\&&\!\!\!\!\!\!\!\!\!\!\!\!\!\!\!\!\!\!\!\!\!\!\!\!\!\!\!\!\!\!\!\!
+ \pol^+(k_6)\cdot p ~k_{1\,\sigma} \pol^-_\rho(k_1)-\pol^+(k_6)\cdot p ~k_{1\,\rho}~ \pol^-_\sigma(k_1) + \pol^-(k_1)\cdot p ~k_{6\,\sigma}~ \pol^+_{\rho}(k_6) 
\nonumber\\&&\!\!\!\!\!\!\!\!\!\!\!\!\!\!\!\!\!\!\!\!\!\!\!\!\!\!\!\!\!\!\!\!
- \pol^-(k_1)\cdot p ~k_{6\,\rho}~ \pol^+_{\sigma}(k_6) - k_1\cdot k_6 \pol^-_\rho(k_1)\pol^+_\sigma(k_6)\Big)
\label{mu4tree3}\eea
\cme
Here $p_s$, $p_{\bar f}$ and $p_g$ are given by $(p-k_1)$ and $p'_{\bar s}$, $p'_{f}$ and $p'_g$ are equal to $(-p-k_6)$.
Using the same logic that was employed for the five-point pentagon coefficient calculated above and the results of eqs. (\ref{mu4tree1})-(\ref{mu4tree3}), it is straightforward to compute $B_1$. One subtlety is that the line $p^2-\mu^2$ is left uncut in the evaluation of this integral coefficient. As a result, the expression for $\mu^2$ is not given simply by $p_*^2$, the way that it was in the five-point example worked out in detail above. Instead, one has the relation $p_*^2-2 p_*\cdot k_1 = \mu^2$. Also, to use the framework of eqs. (\ref{5ptos1}) and (\ref{5ptos2}), we have to make the following adjustments: (\ref{5ptos1}) becomes
\be
p^\nu = L_1 K_1^\nu + L_2 K_2^\nu + L_3 K_3^\nu + L_4 K_4^\nu+k_1
\ee
and the $K_i$ four-vectors all need to be shifted by $-k_1$ ({\it i.e.} instead of $K_1 = k_1 + k_2 + k_3$, we have $K_1 = k_2 + k_3$). 

Before presenting the results of our all-orders-in-$\e$ six-gluon NMHV calculations, we make some remarks about how we expect the strategy outlined for six-gluon MHV amplitudes to generalize to higher-multiplicity amplitudes. First of all, we conjecture that the number of unconstrained by the one-loop soft/collinear bootstrap is controlled by kinematics as opposed to dynamics ({\it i.e.} independent of the $k$ in $\rm{N}^k$MHV). We interpret the fact that this is true at the six-point level (in the sense that the leading singularities for both MHV and NMHV amplitudes fix everything up to a single pentagon coefficient) as evidence for this proposal. If this conjecture is indeed correct, it follows that the number of terms in arbitrary $n$-point amplitudes left unconstrained by the soft/collinear bootstrap is equal to the number of unconstrained terms in the $n$-gluon MHV amplitudes. The number of such terms is 6 in the seven-gluon and 21 in the eight-gluon MHV amplitudes and we expect the answer to be the same for non-MHV helicity configurations as well.

Thus, we conclude this subsection by conjecturing that the number of terms left unconstrained by the soft/collinear bootstrap at the $n$-point level (pentagon coefficients undetermined by the leading singularity method) is
\be
\left(\begin{array}{c}n - 1 \\ 5\end{array}\right) = {(n-5)(n-4)(n-3)(n-2)(n-1)\over 120} \,{\rm ,}
\ee
equal to the number of pentagons at the $(n-1)$-point level. Loosely speaking, we can think of this result as the statement that, at the $n$-point level an independent object analogous to $\textrm{tr}[\slashed{k_1}\slashed{k_2}\slashed{k_3}\slashed{k_4}\slashed{k_5}\slashed{k_6}]$ can be constructed for each pentagon integral at the $(n-1)$-point level without spoiling any of the soft/collinear constraints relating $n$-point one-loop planar amplitudes to $(n-1)$-point one-loop planar amplitudes.
\subsection{The All-Orders in $\e$ Planar One-Loop $\Nsym$ NMHV Six-Gluon Amplitudes}
\label{gresults}
In this subsection, we give our formulae for the one-loop planar six-gluon NMHV pentagon coefficients in $\Nsym$ and discuss the structural similarities between our results and certain two-loop planar six-gluon integral coefficients entering into the NMHV amplitudes calculated in~\cite{KRV}. Our first task, of course, is to understand how many independent NMHV gluon amplitudes there are (delaying a discussion of the constraints coming from $\Nsym$ supersymmetry until Section \ref{supercomp}). Na\"{i}vely, there are a large number of possibilities, many of which are obviously related by parity\footnote{Recall that CP is a good symmetry of perturbative scattering amplitudes even in pure $\mathcal{N} = 0$ Yang-Mills.} or cyclic symmetry\footnote{Recall from \ref{planar} that, for example, the amplitudes $A_{1}^{1-{\rm loop}}(k_1^{1234},k_2^{1234},k_3^{1234},k_4,k_5,k_6)$ and $A_{1}^{1-{\rm loop}}(k_2^{1234},k_3^{1234},k_4,k_5,k_6,k_1^{1234})$ are equal by virtue of the color structure in the planar limit.}. In the end, it turns out that all possibilities are related  by discrete symmetries to $A_{1}^{1-{\rm loop}}\left(k_1^{1234},k_2^{1234},k_3^{1234},k_4,k_5,k_6\right)$, $A_{1}^{1-{\rm loop}}\left(k_1^{1234},k_2^{1234},k_3,k_4^{1234},k_5,k_6\right)$, or $A_{1}^{1-{\rm loop}}\left(k_1^{1234},k_2,k_3^{1234},k_4,k_5^{1234},k_6\right)$.

Now that we understand why it makes sense to focus on $A_{1}^{1-{\rm loop}}\left(k_1^{1234},k_2^{1234},k_3^{1234},k_4,k_5,k_6\right)$, $A_{1}^{1-{\rm loop}}\left(k_1^{1234},k_2^{1234},k_3,k_4^{1234},k_5,k_6\right)$, and $A_{1}^{1-{\rm loop}}\left(k_1^{1234},k_2,k_3^{1234},k_4,k_5^{1234},k_6\right)$, we present our results for these amplitudes. To begin, let us recall the results of the calculations performed in~\cite{BDDKNMHV}. The authors of that work determined the box coefficients for all NMHV gluon amplitudes in the dual conformal basis. The $6 - 2\e$ dimensional pentagon coefficients, however, were undetermined. It was found that
\cmb{-1.0 in}{0 in}
\bea
\nonumber 
A_{1}^{1-{\rm loop}}\left(k_1^{1234},k_2^{1234},k_3^{1234},k_4,k_5,k_6\right)
&&
=-{1\over2}B_1 \Bigg( s_4 s_5 I_4^{(2,3)}+s_1 s_2 I_4^{(5,6)}+s_6 t_1 I_4^{(3,5)}+s_3 t_1 I_4^{(2,6)}\Bigg) \\ \nonumber
&&
-{1\over2}B_2 \Bigg( s_5 s_6 I_4^{(3,4)}+s_2 s_3 I_4^{(6,1)}+s_1 t_2 I_4^{(4,6)}+s_4 t_2 I_4^{(1,3)}\Bigg) \\ \nonumber 
&&
-{1\over2}B_3 \Bigg( s_6 s_1 I_4^{(4,5)}+s_3 s_4 I_4^{(1,2)}+s_2 t_3 I_4^{(1,5)}+s_5 t_3 I_4^{(2,4)}\Bigg) \\ \nonumber 
&& 
+K_1 \e I_5^{(1),6-2 \e}+K_2 \e I_5^{(2),6-2 \e}+K_3 \e I_5^{(3),6-2 \e}
\\ && 
+K_4 \e I_5^{(4),6-2 \e}+K_5 \e I_5^{(5),6-2 \e}+K_6 \e I_5^{(6),6-2 \e} \, ,
\label{6ptgNMHV}
\eea
\bea
\nonumber A_{1}^{1-{\rm loop}}\left(k_1^{1234},k_2,k_3^{1234},k_4,k_5^{1234},k_6\right)&&=-{1\over2}G_1 \Bigg( s_4 s_5 I_4^{(2,3)}+s_1 s_2 I_4^{(5,6)}+s_6 t_1 I_4^{(3,5)}+s_3 t_1 I_4^{(2,6)}\Bigg) \\ \nonumber
&&-{1\over2}G_2 \Bigg( s_5 s_6 I_4^{(3,4)}+s_2 s_3 I_4^{(6,1)}+s_1 t_2 I_4^{(4,6)}+s_4 t_2 I_4^{(1,3)}\Bigg) \\ \nonumber 
&&-{1\over2}G_3 \Bigg( s_6 s_1 I_4^{(4,5)}+s_3 s_4 I_4^{(1,2)}+s_2 t_3 I_4^{(1,5)}+s_5 t_3 I_4^{(2,4)}\Bigg) \\ \nonumber 
&& +F_1 \e I_5^{(1),6-2 \e}+F_2 \e I_5^{(2),6-2 \e}+F_3 \e I_5^{(3),6-2 \e}
\\ 
&& +F_4 \e I_5^{(4),6-2 \e}+F_5 \e I_5^{(5),6-2 \e}+F_6 \e I_5^{(6),6-2 \e} \, , 
\eea
\cme
and
\cmb{-1.0 in}{0 in}
\bea
\nonumber A_{1}^{1-{\rm loop}}\left(k_1^{1234},k_2^{1234},k_3,k_4^{1234},k_5,k_6\right)&&=-{1\over2}D_1 \Bigg( s_4 s_5 I_4^{(2,3)}+s_1 s_2 I_4^{(5,6)}+s_6 t_1 I_4^{(3,5)}+s_3 t_1 I_4^{(2,6)}\Bigg) \\ \nonumber
&&-{1\over2}D_2 \Bigg( s_5 s_6 I_4^{(3,4)}+s_2 s_3 I_4^{(6,1)}+s_1 t_2 I_4^{(4,6)}+s_4 t_2 I_4^{(1,3)}\Bigg) \\ \nonumber 
&&-{1\over2}D_3 \Bigg( s_6 s_1 I_4^{(4,5)}+s_3 s_4 I_4^{(1,2)}+s_2 t_3 I_4^{(1,5)}+s_5 t_3 I_4^{(2,4)}\Bigg) \\ \nonumber 
&& +H_1 \e I_5^{(1),6-2 \e}+H_2 \e I_5^{(2),6-2 \e}+H_3 \e I_5^{(3),6-2 \e}
\\ 
&& +H_4 \e I_5^{(4),6-2 \e}+H_5 \e I_5^{(5),6-2 \e}+H_6 \e I_5^{(6),6-2 \e} \,.
\eea
\cme
All of the spin factors which entered into the box coefficients ($B_i$, $G_i$, and $D_i$) were determined. They are given by
\bea
B_1 &=& B_0 \label{B1}\\
B_2 &=& \bigg({\spab1.{2+3}.4\over t_2}\bigg)^4 B_0\Big|_{j\rightarrow j+1}+\bigg({\spa2.3 \spb5.6 \over t_2}\bigg)^4 B_0^{\langle~\rangle \leftrightarrow [~]}\Big|_{j\rightarrow j+1} \\
B_3 &=& \bigg({\spab3.{1+2}.6\over t_3}\bigg)^4 B_0\Big|_{j\rightarrow j-1}+\bigg({\spa1.2 \spb4.5 \over t_3}\bigg)^4 B_0^{\langle~\rangle \leftrightarrow [~]}\Big|_{j\rightarrow j-1} \, ,
\eea
\bea
G_1 &=& \bigg({\spab5.{4+6}.2\over t_1}\bigg)^4 B_0+\bigg({\spa1.3 \spb4.6 \over t_1}\bigg)^4 B_0^{\langle~\rangle \leftrightarrow [~]} \\
G_2 &=& \bigg({\spab3.{2+4}.6\over t_2}\bigg)^4 B_0^{\langle~\rangle \leftrightarrow [~]}\Big|_{j\rightarrow j+1}+\bigg({\spa5.1 \spb2.4 \over t_2}\bigg)^4 B_0\Big|_{j\rightarrow j+1} \\
G_3 &=& \bigg({\spab1.{2+6}.4\over t_3}\bigg)^4 B_0^{\langle~\rangle \leftrightarrow [~]}\Big|_{j\rightarrow j-1}+\bigg({\spa3.5 \spb6.2 \over t_3}\bigg)^4 B_0\Big|_{j\rightarrow j-1} \, ,
\eea
and
\bea
D_1 &=& \bigg({\spab4.{1+2}.3\over t_1}\bigg)^4 B_0+\bigg({\spa1.2 \spb5.6 \over t_1}\bigg)^4 B_0^{\langle~\rangle \leftrightarrow [~]} \\
D_2 &=& \bigg({\spab1.{2+4}.3\over t_2}\bigg)^4 B_0\Big|_{j\rightarrow j+1}+\bigg({\spa2.4 \spb5.6 \over t_2}\bigg)^4 B_0^{\langle~\rangle \leftrightarrow [~]}\Big|_{j\rightarrow j+1} \\
D_3 &=& \bigg({\spab4.{1+2}.6\over t_3}\bigg)^4 B_0\Big|_{j\rightarrow j-1}+\bigg({\spa1.2 \spb3.5 \over t_3}\bigg)^4 B_0^{\langle~\rangle \leftrightarrow [~]}\Big|_{j\rightarrow j-1} \, ,
\eea
where
\be
B_0 = i{\spa1.2 \spa2.3 \spb4.5 \spb5.6 \spab3.{1+2}.6 \spab1.{2+3}.4 t_1^3 \over s_1 s_2 s_4 s_5 (t_1 t_2-s_2 s_5)(t_1 t_3-s_1 s_4)} \, . \label{B0}
\ee
Using the strategy outlined it Subsection \ref{effgcomp}, we reproduce the above and, furthermore, find explicit expressions for the $K_i$, $G_i$, and $H_i$.

Although the raw answers obtained using the method described in the last subsection are already compact enough to fit on a single page, it is clearly desirable to find more compact formulae. In their work on the two-loop planar NMHV gluon amplitudes~\cite{KRV}, Kosower, Roiban, and Vergu derived explicit expressions for all possible $\mu$-integral hexabox coefficients (see Figure \ref{hexabox} for an illustration). Motivated by issues of IR consistency that we will elaborate on in Section \ref{supercomp}, we evaluated the answers they obtained numerically and were able to find a straightforward mapping between their results and ours. To explain this relationship, it is useful to consider a concrete example.

We consider the coefficient of the $s_1$-channel hexabox integral (see Figure \ref{hexabox}) that appears in the amplitude $A_{1}^{2-{\rm loop}}\left(k_1^{1234},k_2^{1234},k_3^{1234},k_4,k_5,k_6\right)$ calculated to  all orders in $\e$. 
\begin{figure}
\begin{center}
\resizebox{0.6\textwidth}{!}{\includegraphics{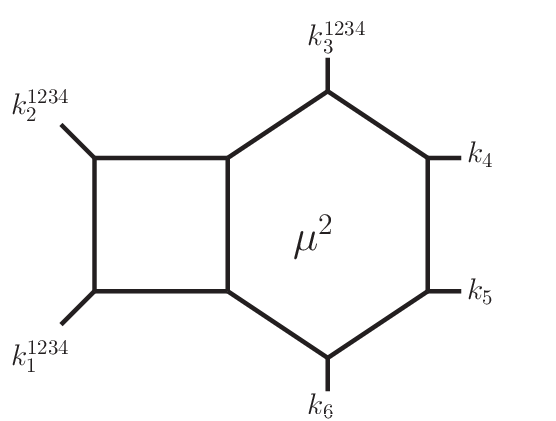}}
\end{center}
\caption{The $s_1$-channel hexabox $\mu$-integral topology of $A_{1}^{2-{\rm loop}}\left(k_1^{1234},k_2^{1234},k_3^{1234},k_4,k_5,k_6\right)$. The factor of $\mu^2$ should be thought of as belonging to the hexagon subdiagram.}
\label{hexabox}
\end{figure}
It turns out that this $\mu$-integral hexabox coefficient and $K_2$ (coefficient of the $s_1$-channel $6 - 2\e$ dimensional pentagon coefficient that appears in $A_{1}^{1-{\rm loop}}\left(k_1^{1234},k_2^{1234},k_3^{1234},k_4,k_5,k_6\right)$ are simply related:
\be
K_2 = {C_2 \over 2 s_1} \mathcal{K}_2 \, {\rm ,}
\label{hexpentrel}
\ee
where we have given the hexabox coefficient the convenient label $\mathcal{K}_2$ and $C_2$ is one of the variables that we used to define the reduction of the one-loop scalar hexagon integral to a sum of six scalar pentagons in \ref{muint}:
\be
I_6 = {1\over 2}\sum_{i = 1}^6 C_i I_5^{(i)}\,{\rm .}
\label{hexred}
\ee
This relation makes a certain amount of sense if we think about collapsing the box in the $\mu$-integral hexabox in Figure \ref{hexabox} to a point. This turns the hexabox $\mu$-integral into a pentagon $\mu$-integral. Evidently, $s_1$ appears because we are working with the $s_1$ channel hexabox and, perhaps, $C_2$ appears because we are relating an object with six external legs to one with five. In any case, the above relation will allow us to exploit extremely simple results found for the NMHV hexabox coefficients~\cite{KRV} to write beautiful formulas for the $K_i$, $F_i$, and $H_i$. We find
\cmb{-.9 in}{0 in}
\bea
K_1 &=& {i \over 2} C_1 {\Big(\spab2.{1+6}.5 \spab1.{2+3}.4 \spab3.{1+2}.6 +\spab5.{1+6}.2 \spa1.2 \spa2.3 \spb4.5 \spb5.6 \Big)^2\over s_6 s_3 \spab2.{6+1}.5 \spab5.{6+1}.2} 
\\
K_2 &=& {i \over 2} C_2 {\spab3.{1+2}.6^2 \spa1.2^2 \spb4.5^2 t_1^2 \over s_1 s_4 \spab3.{1+2}.6 \spab6.{1+2}.3} \\
K_3 &=& {i \over 2} C_3 {\spab1.{2+3}.4^2 \spa2.3^2 \spb5.6^2 t_1^2 \over s_2 s_5 \spab1.{2+3}.4 \spab4.{2+3}.1} \\
K_4 &=& {i \over 2} C_4 {\Big(\spab2.{1+6}.5 \spab1.{2+3}.4 \spab3.{1+2}.6 +\spab5.{1+6}.2 \spa1.2 \spa2.3 \spb4.5 \spb5.6 \Big)^2\over s_6 s_3 \spab2.{6+1}.5 \spab5.{6+1}.2} 
\\
K_5 &=& {i \over 2} C_5 {\spab3.{1+2}.6^2 \spa1.2^2 \spb4.5^2 t_1^2 \over s_1 s_4 \spab3.{1+2}.6 \spab6.{1+2}.3} \\
K_6 &=& {i \over 2} C_6 {\spab1.{2+3}.4^2 \spa2.3^2 \spb5.6^2 t_1^2 \over s_2 s_5 \spab1.{2+3}.4 \spab4.{2+3}.1} \, ,
\eea
\cme
\cmb{-.9 in}{0 in}
\bea
F_1 &=& {i \over 2} C_1 {\Big(\spab5.{6+1}.2 \spab3.{2+4}.6 \spab1.{3+5}.4 +\spab2.{6+1}.5 \spb6.2 \spb2.4 \spa1.5 \spa3.5 \Big)^2\over s_6 s_3 \spab5.{6+1}.2 \spab2.{6+1}.5} 	
\\
F_2 &=& {i \over 2} C_2 {\Big(\spab3.{1+2}.6 \spab1.{3+5}.4 \spab5.{4+6}.2 +\spab6.{1+2}.3 \spa1.3 \spa3.5 \spb2.6 \spb4.6 \Big)^2\over s_1 s_4 \spab3.{1+2}.6 \spab6.{1+2}.3} 	
\\
F_3 &=& {i \over 2} C_3 {\Big(\spab1.{2+3}.4 \spab5.{4+6}.2 \spab3.{5+1}.6 +\spab4.{2+3}.1 \spb2.4 \spb4.6 \spa3.1 \spa5.1 \Big)^2\over s_2 s_5 \spab1.{2+3}.4 \spab4.{2+3}.1} 	
\\
F_4 &=& {i \over 2} C_4 {\Big(\spab5.{6+1}.2 \spab3.{2+4}.5 \spab1.{3+5}.4 +\spab2.{6+1}.5 \spb2.6 \spb4.2 \spa5.1 \spa5.3 \Big)^2\over s_6 s_3 \spab5.{6+1}.2 \spab2.{6+1}.5} 	
\\
F_5 &=& {i \over 2} C_5 {\Big(\spab3.{1+2}.6 \spab1.{3+5}.4 \spab5.{4+6}.2 +\spab6.{1+2}.3 \spa1.3 \spa3.5 \spb2.6 \spb4.6 \Big)^2\over s_1 s_4 \spab3.{1+2}.6 \spab6.{1+2}.3}  	
\\
F_6 &=& {i \over 2} C_6 {\Big(\spab1.{2+3}.4 \spab5.{4+6}.2 \spab3.{5+1}.6 +\spab4.{2+3}.1 \spb2.4 \spb4.6 \spa3.1 \spa5.1 \Big)^2\over s_2 s_5 \spab1.{2+3}.4 \spab4.{2+3}.1} \, ,
\eea
\cme
and
\cmb{-.9 in}{0 in}
\bea
H_1 &=& {i \over 2} C_1 {\Big(\spab2.{6+1}.5 \spab1.{2+4}.3 \spab4.{1+2}.6 +\spab5.{6+1}.2 \spa1.2 \spa2.4 \spb3.5 \spb5.6 \Big)^2\over s_6 s_3 \spab2.{6+1}.5 \spab5.{6+1}.2} 
\\
H_2 &=& {i \over 2} C_2 {\spa1.2^2 \Big(\spab3.{1+2}.6 \spab4.{1+2}.3 \spb5.3 +\spab6.{1+2}.3 \spab4.{1+2}.6 \spb5.6\Big)^2\over s_1 s_4 \spab3.{1+2}.6 \spab6.{1+2}.3} 		
\\
H_3 &=& {i \over 2} C_3 {\spb5.6^2 \Big(\spab1.{2+3}.4 \spab4.{1+2}.3 \spa2.4 +\spab4.{2+3}.1 \spab1.{2+4}.3 \spa2.1\Big)^2\over s_2 s_5 \spab4.{2+3}.1 \spab1.{2+3}.4} 		
\\
H_4 &=& {i \over 2} C_4 {\Big(\spab2.{6+1}.5 \spab1.{2+4}.3 \spab4.{1+2}.6 +\spab5.{6+1}.2 \spa1.2 \spa2.4 \spb3.5 \spb5.6 \Big)^2\over s_6 s_3 \spab2.{6+1}.5 \spab5.{6+1}.2} \nonumber \\ \\
H_5 &=& {i \over 2} C_5 {\spa1.2^2 \Big(\spab3.{(1+2)}.6 \spab4.{1+2}.3 \spb5.3 +\spab6.{1+2}.3 \spab4.{1+2}.6 \spb5.6\Big)^2\over s_1 s_4 \spab3.{1+2}.6 \spab6.{1+2}.3} 		
\\
H_6 &=& {i \over 2} C_6 {\spb5.6^2 \Big(\spab1.{2+3}.4 \spab4.{1+2}.3 \spa2.4 +\spab4.{2+3}.1 \spab1.{2+4}.3 \spa2.1\Big)^2\over s_2 s_5 \spab4.{2+3}.1 \spab1.{2+3}.4} \, {\rm .}
\eea
\cme
We also checked these results against a Feynman diagram calculation.

These results have a couple of striking features of which we have only a partial understanding. The numerators of all the spin factors (divided by the appropriate $C_i$) are perfect squares. Furthermore, the pentagon coefficients possess a certain $i \rightarrow i+3$ symmetry:
\be
{K_1 \over C_1} = {K_4 \over C_4} \qquad
{K_2 \over C_2} = {K_5 \over C_5} \qquad
{K_3 \over C_3} = {K_6 \over C_6} \, {\rm .}
\label{mystrel}
\ee
with analogous formulas for the $F_i$ and $H_i$. As we will see in Section \ref{supercomp}, this $i \rightarrow i+3$ symmetry is related to the action of parity when the amplitude is written in a way that makes a hidden symmetry of the planar S-matrix manifest. In the next section we explore an interesting connection between all-orders-in-$\e$ one-loop $\Nsym$ amplitudes and the first two non-trivial orders in the $\alpha'$ expansion of tree-level superstring amplitudes. The explicit one-loop results presented so far in this review will provide us with useful explicit cross-checks on the relations we propose.
\section{New Relations Between One-Loop Amplitudes in $\Nsym$ Gauge Theory and Tree-Level Amplitudes in Open Superstring Theory}
\label{gsrel}
Before reviewing the scattering of massless modes in open superstring theory, we motivate what follows. Stieberger and Taylor~\cite{ST1} calculated the lowest-order, $\Ord(\alpha'^2)$, stringy corrections to $\Nsym$ tree-level gluon MHV amplitudes.\footnote{As mentioned in the introduction, tree-level amplitudes of massless particles in open superstring constructions compactified to four dimensions have a universal form~\cite{ST1}.} They found that their result\footnote{In Stieberger and Taylor's notation, say at the six-point level, $[[1]]_1 = s_1$, $\{\,[[1]]_1\,\} = s_1 + s_2 + s_3 + s_4 + s_5 + s_6$, $[[1]]_2 = t_1$, and $\{\,[[1]]_2\,\} = t_1 + t_2 + t_3$.},
\cmb{-1.6 in}{0 in}
\bea
A^{tree}_{str}\left(k_1^{1234},k_2^{1234},k_3,\cdots,k_n\right)\Bigl|_{\Ord(\alpha'^2)} &=& 
-{\pi^2\over 12} A_{n;\langle 12 \rangle}^{\textrm{\scriptsize{MHV}}}\Bigg(
 \sum_{k=1}^{\left[\frac{n}{2}-1\right]}\{\,[[1]]_k [[2]]_k\,\}-
\sum_{k=3}^{\left[\frac{n}{2}-1\right]}\{\,[[1]]_k [[2]]_{k-2}\,\}\nonumber\\
&& +~C^{(n)}+\sum_{j<k<\ell<m<n} \pol(j,k,\ell,m)\Bigg)\, ,\\
 &C^{(n)} =& \Bigg\{ \begin{array}{ll}
        -\{\,[[1]]_{\frac{n}{2}-2}  [[{n \over 2}+1]]_{\frac{n}{2}-2}\}  
& \mbox{$n>4$, even,}\nonumber \\
         -\big\{\,[[1]]_{\frac{n-5}{2}} [[\frac{n+1}{2}]]_{\frac{n-3}{2}}\big\} & \mbox{$n>5$, odd},\end{array}
         \label{stringMHVoal2}
\eea
\cme
was precisely equal to $-6 \zeta(2)$ times the analogous one-loop $\Nsym$ amplitude with a factor of $\mu^4$ inserted into the numerator of each basis integral, $A_{1}^{1-{\rm loop}}\left(k_1^{1234},k_2^{1234},k_3,\cdots,k_n\right)[\mu^4]\Bigl|_{\e\to0}$. This non-obvious connection was actually made by showing that both 
$${A^{tree}_{str}\left(k_1^{1234},k_2^{1234},k_3,\cdots,k_n\right)\Bigl|_{\Ord(\alpha'^2)}\over \spa1.2^4} ~~~~{\rm and}~~~~ {A_{1}^{1-{\rm loop}}\left(k_1^{1234},k_2^{1234},k_3,\cdots,k_n\right)[\mu^4]\Bigl|_{\e\to0} \over \spa1.2^4}$$
 are, apart from trivial constants, equal to the all-plus one-loop amplitude in pure Yang-Mills theory~\cite{oneloopselfdual}, $A_{1;\,\mathcal{N}=0}^{1-{\rm loop}}\left(k_1,k_2,\cdots,k_n\right)$. The only reason an equivalence between 
 $${A_{1}^{1-{\rm loop}}\left(k_1^{1234},k_2^{1234},k_3,\cdots,k_n\right)[\mu^4]\Bigl|_{\e\to0}\over\spa1.2^4} ~~~~{\rm and}~~~~ {A_{1;\,\mathcal{N}=0}^{1-{\rm loop}}\left(k_1,k_2,k_3,\cdots,k_n\right)}$$ is possible is that both have the same manifest invariance under cyclic shifts $i \rightarrow i+1$. It is hard to imagine that additional relationships between $\mathcal{N}=0$ and $\Nsym$ amplitudes could exist because, in general, there is no reason to expect $\mathcal{N}=0$ and $\Nsym$ amplitudes to have similar symmetry properties (for more general amplitudes there is no trick analogous to dividing the one-loop MHV amplitude by $\spa1.2^4$). Indeed, it is incredibly likely that this relation between pure Yang-Mills and $\Nsym$ is purely accidental. However, additional relations between superstring tree amplitudes and $\Nsym$ one-loop amplitudes are a more realistic possibility. It is this possibility that we discuss in this section. The new results presented are based on unpublished work done in collaboration with Lance J. Dixon~\cite{myfourth}.
\subsection{Organization of the Tree-Level Open Superstring S-matrix}
\label{BornInfeld}
For the simple case of a $U(1)$ gauge group, it has been known since the work of Fradkin and Tseytlin~\cite{FT} that the effective action governing the low-energy dynamics of open superstrings ending on a single Dirchlet 3-brane (though the connection between gauge symmetry and D-branes remained hidden until the work of Dai, Leigh, and Polchinski in~\cite{DLP} and Leigh in~\cite{L}) is nothing but a  supersymmetrization of the Born-Infeld action. This action, expressed in terms of the Maxwell field strength tensor,
\cmb{-1.0 in}{0 in}
\bea
\mathcal{L}_{BI} &=& {1 \over (2 \pi g \alpha')^2} \left(1 - \sqrt{{\rm Det}\big(~g_{\mu \nu} + (2 \pi g \alpha') F_{\mu \nu} ~\big)} \right) \nonumber \\
&=& {1 \over (2 \pi g \alpha')^2} \left(1 - \sqrt{1 + {(2 \pi g \alpha')^2\over2}F_{\mu \nu}F^{\mu \nu} - {(2 \pi g \alpha')^4\over16}~ (F_{\mu \nu} \tilde{F}^{\mu \nu})^2} \right) \nonumber \\
&=& -{1 \over 4} F_{\mu \nu} F_{\mu \nu} + 3~ \zeta(2) g^2 \alpha'^2 \left(F_{\mu \nu}F_{\nu \rho}F_{\rho \sigma}F_{\sigma \mu} - {1 \over 4}~\big(F_{\mu \nu}F_{\nu \mu}\big)^2\right) + \Ord\left(g^4 \alpha'^4 F^6\right)
\label{BI}
\eea
\cme
was proposed in~\cite{BI} as an alternative description of electrodynamics. In the context of string scattering, the constant $\alpha'$ is identified with the string tension. A natural generalization to the case of a $U(\Nc)$ gauge group is realized~\cite{T97} when the open superstrings under consideration end on a stack of $\Nc$ coincident $D_3$-branes. This situation is unfortunately much more complicated to describe with an effective action and there is no known analog of \eqn{BI}; the non-Abelian Born-Infeld action, as it is commonly called, is only known up to fourth order~\cite{Wulff,KKNSWrev} in $\alpha'$ (reference~\cite{KKNSWrev} is a review with many more references to the original literature). For us, only the first two non-trivial orders in this expansion play an important role. Due to the fact that there is no $\Nsym$ supersymmetrizable operator of mass dimension six\footnote{In this section we use lower-case $d$ for operator dimensions and upper-case $D$ for spacetime dimensions.} ($d = 6$) that one can write down in terms of non-Abelian field strengths and covariant derivatives, the first two non-trivial orders in the $\alpha'$ expansion are actually $\Ord(\alpha'^2)$ and $\Ord(\alpha'^3)$. In our conventions, the non-Abelian Born-Infeld action is given by
\cmb{-1 in}{0 in}
\bea
&&\mathcal{L}_{NABI} = -{1\over4} \Tr \left[ F_{\mu \nu} F_{\mu \nu} \right]
\nonumber \\ && + \zeta(2) g^2 \alpha'^2 \Tr \left[{1\over2}F_{\mu \nu} F_{\nu \rho}
                    F_{\rho \sigma} F_{\sigma \mu}
                +   F_{\mu \nu} F_{\nu \rho} 
                    F^{\sigma \mu} F_{\rho \sigma}  
        - {1\over8} F_{\mu \nu} F_{\rho \sigma}
                    F_{\nu \mu} F_{\sigma \rho}
   - {1\over4} F_{\mu \nu} F_{\nu \mu}
                    F_{\rho \sigma} F_{\sigma \rho } \right] 
 \nonumber\\ &&  - 8 ~\zeta(3) \alpha'^3 \Tr \left[\rule{0 cm}{.5 cm} {i g^3\over \sqrt{2}} \Big( F_{\mu \nu} F_{\nu \rho}
                    F_{\rho \sigma} F_{\tau \mu} F_{\sigma \tau} + F_{\mu \nu} F_{\sigma \tau}             F_{\nu \rho} F_{\tau \mu} F_{\rho \sigma} - {1\over2} F_{\mu \nu} F_{\nu \rho}  F_{\sigma \tau}     F_{\rho \mu} F_{\tau \sigma}\Big)\right.
\nonumber \\ && + g^2 \Big( {1\over2} (D_{\mu}F_{\nu \rho}) (D_{\mu} F_{\rho \sigma}) F_{\tau \nu} F_{\sigma \tau} 
 + {1\over2}(D_{\mu} F_{\nu \rho})  F_{\tau \nu} (D_{\mu} F_{\rho \sigma})  F_{\sigma \tau} - F_{\mu \nu} (D_{\mu} F_{\rho \sigma}) (D_{\tau} F_{\nu \rho})  F_{\sigma \tau}
 \nonumber\\ && \left. - {1\over8} (D_{\mu} F_{\nu \rho}) F_{\sigma \tau} (D_{\mu} F_{\rho \nu}) F_{\tau \sigma} + (D_{\tau} F_{\mu \nu}) F_{\rho \sigma} (D_{\mu} F_{\nu \rho})  F_{\sigma \tau} \Big)\right]\rule{0 cm}{1 cm} + \Ord\left(\alpha'^4 \Tr F^6\right)\,.
\label{nonABI}
\eea
\cme
The form reproduced above is very nearly that given in~\cite{BMM}, but their conventions are slightly different\footnote{It also appears that their overall normalization differs from ours by a factor of two.}. We use the conventions of~\cite{Dixon96rev}, in which
\be
F_{\mu \nu} = \partial_\mu A_\nu - \partial_\nu A_\mu - {i g \over \sqrt{2}} [A_\mu, A_\nu]
\ee
and
\be
D_\mu \Phi = \partial_\mu \Phi - {i g \over \sqrt{2}} [A_\mu, \Phi] \, .
\ee
Results essentially identical to those above appeared in~\cite{KS} (see also the later work of~\cite{Drummond03}) and the derivative terms at $\Ord(\alpha'^3)$ were obtained earlier in~\cite{Bilal}. We have normalized our $\Ord(\alpha'^2)$ and $\Ord(\alpha'^3)$ effective Lagrangians so that they reproduce the appropriate terms in the $\alpha'$ expansion of the string scattering results given in~\cite{ST2}, where a representative leading four-point color-ordered partial amplitude is\footnote{It has been known since the early days of superstring theory, that one can write a open superstring theory amplitude as
$$\a^{tree}_{str}\left(k_1^{h_1},~k_2^{h_2},~\cdots,~k_n^{h_n}\right) = 
g^{n-2} \sum_{\sigma \in S_n/\mathcal{Z}_n} {\rm Tr}[T^{a_{\sigma(1)}}T^{a_{\sigma(2)}}~\cdots T^{a_{\sigma(n)}}] A^{tree}_{str}\left(k_{\sigma(1)}^{h_{\sigma(1)}},~k_{\sigma(2)}^{h_{\sigma(2)}},~\cdots,~k_{\sigma(n)}^{h_{\sigma(n)}}\right)\,{\rm .}$$}
\cmb{-.5 in}{0 in}
\bea
A^{tree}_{str}(k_1^{1234},k_2^{1234},k_3,k_4) &=& A_{4;\langle 12 \rangle}^{\textrm{\scriptsize{MHV}}} {\Gamma(1+\alpha' s)\Gamma(1+\alpha' t)\over \Gamma(1 + \alpha'(s+t))} \nonumber \\
&=& A_{4;\langle 12 \rangle}^{\textrm{\scriptsize{MHV}}} \bigg( 1 - \zeta(2) s t ~\alpha'^2 + \zeta(3) s t (s+t) ~\alpha'^3 
\nonumber \\ && \qquad~~~~\;
- {\zeta(4) \over 4} s t (4 s^2+s t+4 t^2)~\alpha'^4 +\Ord(\alpha'^5)\bigg) \, .
\label{ST4pt}
\eea
\cme

\subsection{New Relations}
\label{results}
We now return to the observed correspondence between the results of~\cite{oneloopselfdual} and~\cite{ST1} discussed briefly at the beginning of this section. By comparing the two references it is easy to see that
\be
A^{tree}_{str}\left(k_1^{h_1},\cdots,k_n^{h_n}\right)\Bigl|_{\Ord(\alpha'^2)} = -6 \zeta(2) A_{1}^{1-{\rm loop}}\left(k_1^{h_1},\cdots,k_n^{h_n}\right)[\mu^4]\Bigl|_{\e\to0} \, {\rm ,}
\label{oldconj}
\ee
where the gluon helicity configuration is MHV and should of course match on both sides of eq. (\ref{oldconj}). 

Since our notation may not be completely obvious, we consider an illustrative example. Specifically, we check that eq. (\ref{oldconj}) holds for the five-gluon MHV amplitude $A_{1}^{1-{\rm loop}}\left(k_1^{1234},k_2^{1234},k_3,k_4,k_5\right)$. In terms of unevaluated scalar Feynman integrals~\cite{oneloopselfdual}, we have
\cmb{-1.0 in}{0 in}
\bea
&& A_{1}^{1-{\rm loop}}\left(k_1^{1234},k_2^{1234},k_3,k_4,k_5\right)  = 
{- A_{5;\langle 12 \rangle}^{\textrm{\scriptsize{MHV}}} \over 2} \Bigg( s_{2}s_{3} I_4^{(1),~D=4-2\e}+s_{3}s_{4} I_4^{(2),~D=4-2\e}
\\ &&  
+ s_{4}s_{5} I_4^{(3),~D=4-2\e}
+s_{5}s_{1} I_4^{(4),~D=4-2\e}+s_{1}s_{2} I_4^{(5),~D=4-2\e}-2 \e ~\pol(k_1,k_2,k_3,k_4) I_5^{D=6-2\e}\Bigg) \, {\rm .}
\nonumber
\eea
\cme
Applying the dimension shift operation of~\cite{oneloopselfdual} to the amplitude sends $\e \to \e-2$ and $I_n^D \to I_n^D [\mu^4]$:
\cmb{-1.0 in}{0 in}
\bea
&& A_{1}^{1-{\rm loop}}\left(k_1^{1234},k_2^{1234},k_3,k_4,k_5\right)[\mu^4] =  {- A_{5;\langle 12 \rangle}^{\textrm{\scriptsize{MHV}}} \over 2 } \Bigg( s_{2}s_{3} I_4^{(1),~D=4-2\e}[\mu^4] + s_3 s_{4} I_4^{(2),~D=4-2\e}[\mu^4]
\nonumber \\ 
&&   ~~~~~~~~~~~~~~~~~~~~
+  s_{4}s_{5} I_4^{(3),~D=4-2\e}[\mu^4]+s_{5}s_{1} I_4^{(4),~D=4-2\e}[\mu^4] + s_{1}s_{2} I_4^{(5),~D=4-2\e}[\mu^4]
\nonumber \\ 
&&  ~~~~~~~~~~~~~~~~~~~~
-2 (\e-2) ~\pol(k_1,k_2,k_3,k_4) I_5^{D=6-2\e}[\mu^4] \Bigg) \, .
\eea
\cme
Applying eq. (\ref{DSmu}) gives
\cmb{-1.0 in}{0 in}
\bea
&& A_{1}^{1-{\rm loop}}\left(k_1^{1234},k_2^{1234},k_3,k_4,k_5\right)[\mu^4] =  { A_{5;\langle 12 \rangle}^{\textrm{\scriptsize{MHV}}} \e (1-\e)\over 2} \Bigg( s_{2}s_{3} I_4^{(1),~D=8-2\e} + 
 s_{3}s_{4} I_4^{(2),~D=8-2\e} 
 \nonumber \\ 
 &&    ~~~~~~~~~~~~~~~~~~~~
 +s_{4}s_{5} I_4^{(3),~D=8-2\e}+s_{5}s_{1} I_4^{(4),~D=8-2\e} + 
 s_{1}s_{2} I_4^{(5),~D=8-2\e}
 \nonumber \\ 
 &&   ~~~~~~~~~~~~~~~~~~~~
  -2 (\e-2) ~\pol(k_1,k_2,k_3,k_4) I_5^{D=10-2\e} \Bigg) \, .
\eea
\cme
Finally, we take the limit as $\e\to0$. As explained in Subsection \ref{GUD}, the terms which survive are those proportional to the ultraviolet singularities of the dimensionally-shifted basis integrals.
\cmb{-1.0 in}{0 in}
\bea
 A_{1}^{1-{\rm loop}}\left(k_1^{1234},k_2^{1234},k_3,k_4,k_5\right)[\mu^4]\Bigl|_{\e\to0} &&=  { A_{5;\langle 12 \rangle}^{\textrm{\scriptsize{MHV}}} \over 2} \Bigg( s_{2}s_{3} \Big(~{1\over6}~\Big) + 
 s_{3}s_{4} \Big(~{1\over6}~\Big)  +  s_{4}s_{5} \Big(~{1\over6}~\Big)
 \nonumber \\ &&  +s_{5}s_{1} \Big(~{1\over6}~\Big)  + 
 s_{1}s_{2} \Big(~{1\over6}~\Big)  + 4 ~\pol(1,2,3,4) \Big(~{1\over 24}~\Big)  \Bigg)
\nonumber \\ && =  { A_{5;\langle 12 \rangle}^{\textrm{\scriptsize{MHV}}} \over 12} 
  \Big( \big\{ s_{2}s_{3} \big\}  +  \pol(k_1,k_2,k_3,k_4) \Big) \, ,
\eea
\cme
where, following \cite{ST1}, $\big\{ s_{2}s_{3} \big\}$ represents the sum of $s_{2}s_{3}$ and its four cyclic permutations.
Finally, plugging this expression into \eqn{oldconj} gives the following prediction for \\ $A^{tree}_{str}\left(k_1^{1234},k_2^{1234},k_3,k_4,k_5\right)\Bigl|_{\Ord(\alpha'^2)}$:
\be
\!\!\!\!\!\!\!\!\!\!\!\!\!\!\!\!\!\!\!\!
A^{tree}_{str}\left(k_1^{1234},k_2^{1234},k_3,k_4,k_5\right)\Bigl|_{\Ord(\alpha'^2)}=  -\zeta(2) {A_{5;\langle 12 \rangle}^{\textrm{\scriptsize{MHV}}} \over 2 } 
  \Big( \big\{ s_{2}s_{3} \big\}  +  \pol(k_1,k_2,k_3,k_4) \Big) \, .
\ee
By comparing to the all-$n$ result for the $\Ord(\alpha'^2)$ stringy corrections given at the beginning of this section, it is clear that the prediction of the conjecture for the $\Ord(\alpha'^2)$ piece of $A^{tree}_{str}\left(k_1^{1234},k_2^{1234},k_3,k_4,k_5\right)$ is correct. It is obvious from the above analysis that we would have been unsuccessful had we performed the dimension shift operation on the expression usually associated with the five-gluon one-loop MHV amplitude,
\bea
  {- A_{5;\langle 12 \rangle}^{\textrm{\scriptsize{MHV}}} \over 2} \Bigg( s_{2}s_{3} I_4^{(1),~D=4-2\e}+s_{3}s_{4} I_4^{(2),~D=4-2\e}
  + s_{4}s_{5} I_4^{(3),~D=4-2\e} + \\ 
  ~~~~~~~~~~
  +s_{5}s_{1} I_4^{(4),~D=4-2\e}+s_{1}s_{2} I_4^{(5),~D=4-2\e}\Bigg) \, , \nonumber
\label{ordepsans}
\eea
illustrating that eq. (\ref{oldconj}) is only applicable if one works to all orders in $\e$ on the field theory side. We wish to stress that, although we find the language of~\cite{oneloopselfdual} convenient, we could have used the coefficients of the UV poles of $\Nsym$ one-loop MHV amplitudes considered in $D = 8 - 2 \e$ to define the right-hand side of \eqn{oldconj} and nothing would have changed, apart from maybe an unimportant overall minus sign.

Now, suppose we want to generalize the Stieberger-Taylor relation. One obvious question is whether we can relax their requirement that the helicity configuration on both sides of (\ref{oldconj}) be MHV. Indeed, we will see that the relation actually holds for general helicity configurations. Fortunately, Stieberger and Taylor calculated all six-point NMHV open superstring amplitudes in~\cite{ST4} (unfortunately not in a form as elegant as eq. (\ref{stringMHVoal2})). As a first check, we verified that $A_{1}^{1-{\rm loop}}\left(k_1^{1234},k_2^{1234},k_3^{1234},k_4,k_5,k_6\right)$, $A_{1}^{1-{\rm loop}}\left(k_1^{1234},k_2^{1234},k_3,k_4^{1234},k_5,k_6\right)$, and $A_{1}^{1-{\rm loop}}\left(k_1^{1234},k_2,k_3^{1234},k_4,k_5^{1234},k_6\right)$ all satisfy (\ref{oldconj}). There exists, in fact, a more general way to argue that relation (\ref{oldconj}) should be helicity-blind. Furthermore, it is possible to show that one can use all-orders-in-$\e$ $\Nsym$ Yang-Mills amplitudes to derive the $\Ord(\alpha'^3)$ stringy corrections as well. It was pointed out in~\cite{DunbarTurner} that the $\Nsym$ theory considered in $D = 8 - 2\e$ has UV divergences and that the requirements that the counterterm Lagrangian respect $\Nsym$ supersymmetry and have $d = 8$ uniquely fix it to be the $\Nsym$ supersymmetrization of ${\rm Tr}[F^4]$ (2nd line of eq. (\ref{nonABI})), the {\it same} operator that appears at $\Ord(\alpha'^2)$ in the non-Abelian Born-Infeld action of~\cite{T97}. Now it is clear why we found that, up to a trivial constant, one-loop $\Nsym$ gluon amplitudes dimensionally shifted to $D = 8 - 2\e$ are equal to the $\Ord(\alpha'^2)$ stringy corrections to $\Nsym$ gluon tree amplitudes: The underlying effective Lagrangians are completely determined by dimensional analysis and $\Nsym$ supersymmetry. In other words, there is only one $\Nsym$ supersymmetrizable $d = 8$ operator built out of field-strength tensors and their covariant derivatives. 

This is not, however, the end of the story. That the non-Abelian Born-Infeld action is fixed to order $\Ord(\alpha'^2)$ by $\Nsym$ supersymmetry is perhaps more widely appreciated than the fact that it is fixed to order $\Ord(\alpha'^3)$ by $\Nsym$ supersymmetry. It is highly non-trivial to prove the above claim (see~\cite{KS,Collinucci}), but it is true; there is a unique $\Nsym$ supersymmetrizable linear combination of the available $d = 10$ operators (schematically, there are only two such operators, $D^2 F^4$ and $F^5$) built out of field strength tensors and their covariant derivatives. On the field theory side, Dunbar and Turner showed that $D = 10 - 2\e$ counterterm Lagrangians are built out of (an appropriate $\Nsym$ supersymmetrization of) the $d = 10$ operators $F^5$ and $D^2 F^4$. The results of~\cite{KS,Collinucci} clearly imply that this $\Nsym$ supersymmetric linear combination, being unique, coincides with the $\Ord(\alpha'^3)$ terms in the non-Abelian Born-Infeld action (eq. (\ref{nonABI})). As an additional check, we evaluated $A_{1}^{1-{\rm loop}}\left(k_1^{1234},k_2^{1234},k_3,k_4,k_5,k_6\right)[\mu^6]\Big|_{\e \to 0}$ and observed that, up to an overall factor of $60 \zeta(3)$, the results obtained precisely matched the appropriate stringy corrections (eq. (\ref{stringMHVoal2})). These observations indicate that an analogous relationship, 
\be
A^{tree}_{str}\left(k_1^{h_1},\cdots,k_n^{h_n}\right)\Bigl|_{\Ord(\alpha'^3)} = 60 \zeta(3) A_{1}^{1-{\rm loop}}\left(k_1^{h_1},\cdots,k_n^{h_n}\right)[\mu^6]\Bigl|_{\e\to0}
\label{newconj}
\ee
holds in this case (again for arbitrary helicity configurations).

To summarize, we have seen that quite a bit of non-trivial information about the low-energy dynamics of open superstrings is encoded in all-orders-in-$\e$ one-loop $\Nsym$ amplitudes. At this point, one might hope that the trend continues and the stringy corrections are all somehow encoded in the $\Nsym$ theory considered in some higher dimensional spacetime. Unfortunately, there is no analog of \eqn{oldconj} and \eqn{newconj} at $\Ord(\alpha'^4)$. It is not hard to see this explicitly at the level of four-point amplitudes. 

Based on our experience so far, one might expect the four-point MHV amplitude dimensionally shifted to $D=12-2\e$ to match the $\Ord(\alpha'^4)$ stringy correction given in eq. (\ref{ST4pt}) up to a multiplicative constant. However, a short calculation shows that
\be
A_{1}^{1-{\rm loop}}(k_1^{1234},k_2^{1234},k_3,k_4)[\mu^8]\Big|_{\e \to 0} = {s t (2 s^2+s t+2 t^2)\over 840} A_{4;\langle 12 \rangle}^{\textrm{\scriptsize{MHV}}}
\ee
which does {\it not} have the same $s$ and $t$ dependence as the $\Ord(\alpha'^4)$ stringy correction,
\be
A^{tree}_{str}(k_1^{1234},k_2^{1234},k_3,k_4)\Big|_{\Ord(\alpha'^4)} = - {\zeta(4) \over 4} s t (4 s^2+s t+4 t^2) A_{4;\langle 12 \rangle}^{\textrm{\scriptsize{MHV}}} \, .
\ee

Although it was originally hoped that $\Nsym$ supersymmetry would constrain the non-Abelian Born-Infeld action to all orders in $\alpha'$, it is now clear that this already fails to work at $\Ord(\alpha'^4)$~\cite{Collinucci}. Since it is illuminating, we repeat the argument of~\cite{Collinucci}. One can easily see that there must be more than one independent $\Nsym$ superinvariant at $\Ord(\alpha'^4)$ by comparing the $\Ord(\alpha'^4)$ terms in the Abelian Born-Infeld action to the $\Ord(\alpha'^4)$ terms in the non-Abelian Born-Infeld action responsible for the $\Ord(\alpha'^4)$ piece of the four-point tree open superstring amplitude. It is clear from eq. (\ref{BI}) that the Abelian Born-Infeld action doesn't contain any derivative terms. On the other hand, operators of the form $(D F)^4$ are the only dimension ten operators which can enter into and produce the observed $\Ord(\alpha'^4)$ four-point tree-level superstring amplitude~\cite{Bilal},
\bea
A^{tree}_{str}(k_1^{1234},k_2^{1234},k_3,k_4)\Bigl|_{\Ord(\alpha'^4)} = - {\zeta(4) \over 4} s t (4 s^2+s t+4 t^2) A_{4;\langle 12 \rangle}^{\textrm{\scriptsize{MHV}}} \, .
\label{restateordalp4}
\eea
Since the $\Ord(\alpha'^4)$ terms in the Abelian Born-Infeld action form an $\Nsym$ superinvariant by themselves (since they are present in the Abelian case where no derivative terms are allowed), the linear combination of operators of the form $(D F)^4$ responsible for the above result must be part of an distinct $\Nsym$ superinvariant.

Before leaving this section, we make one further remark about our results at $\Ord(\alpha'^2)$ that is relevant to $n$-gluon scattering. One might expect that the stringy corrections at this order in $\alpha'$ would obey a photon-decoupling relation exactly like the one in pure Yang-Mills at tree level, where replacing a single gluon by a photon produces a vanishing result. This turned out to be too simplistic. The ${\rm Tr}[F^4]$ operator that governs the dynamics at this order in $\alpha'$ can, in fact, couple one or even two photons to gluons. However, once you have replaced at least three external photons, the matrix elements do vanish, so long as at least one of the gluons touching the insertion of ${\rm Tr}[F^4]$ is off-shell. Figure \ref{ThreePhotonFigure} illustrates how this works. 

For example, replacing, for sake of argument, gluons $k_2^{1234}$, $k_3$, and $k_4$ by photons results in the identity
\cmb{-.6 in}{0 in}
\bea
&& 0 = A_{str}^{tree}(k_1^{1234},k_2^{1234},k_3,k_4,k_5)\Big|_{\Ord(\alpha'^2)}+A_{str}^{tree}(k_1^{1234},k_2^{1234},k_4,k_3,k_5)\Big|_{\Ord(\alpha'^2)}
\nonumber \\ 
&& ~+ A_{str}^{tree}(k_1^{1234},k_3,k_4,k_2^{1234},k_5)\Big|_{\Ord(\alpha'^2)}+A_{str}^{tree}(k_1^{1234},k_3,k_2^{1234},k_4,k_5)\Big|_{\Ord(\alpha'^2)}
\nonumber \\ 
&& ~+ A_{str}^{tree}(k_1^{1234},k_4,k_3,k_2^{1234},k_5)\Big|_{\Ord(\alpha'^2)}+A_{str}^{tree}(k_1^{1234},k_4,k_2^{1234},k_3,k_5)\Big|_{\Ord(\alpha'^2)}
\nonumber \\ 
&& ~+ A_{str}^{tree}(k_1^{1234},k_2^{1234},k_3,k_5,k_4)\Big|_{\Ord(\alpha'^2)}+A_{str}^{tree}(k_1^{1234},k_2^{1234},k_4,k_5,k_3)\Big|_{\Ord(\alpha'^2)}
\nonumber \\ 
&& ~+ A_{str}^{tree}(k_1^{1234},k_3,k_4,k_5,k_2^{1234})\Big|_{\Ord(\alpha'^2)}+A_{str}^{tree}(k_1^{1234},k_3,k_2^{1234},k_5,k_4)\Big|_{\Ord(\alpha'^2)}
\nonumber \\ 
&& ~+ A_{str}^{tree}(k_1^{1234},k_4,k_3,k_5,k_2^{1234})\Big|_{\Ord(\alpha'^2)}+A_{str}^{tree}(k_1^{1234},k_4,k_2^{1234},k_5,k_3)\Big|_{\Ord(\alpha'^2)}
\nonumber \\ 
&& ~+ A_{str}^{tree}(k_1^{1234},k_2^{1234},k_5,k_3,k_4)\Big|_{\Ord(\alpha'^2)}+A_{str}^{tree}(k_1^{1234},k_2^{1234},k_5,k_4,k_3)\Big|_{\Ord(\alpha'^2)}
\nonumber \\ 
&& ~+ A_{str}^{tree}(k_1^{1234},k_3,k_5,k_4,k_2^{1234})\Big|_{\Ord(\alpha'^2)}+A_{str}^{tree}(k_1^{1234},k_3,k_5,k_2^{1234},k_4)\Big|_{\Ord(\alpha'^2)}
\nonumber \\ 
&& ~+ A_{str}^{tree}(k_1^{1234},k_4,k_5,k_3,k_2^{1234})\Big|_{\Ord(\alpha'^2)}+A_{str}^{tree}(k_1^{1234},k_4,k_5,k_2^{1234},k_3)\Big|_{\Ord(\alpha'^2)}
\nonumber \\ 
&& ~+ A_{str}^{tree}(k_1^{1234},k_5,k_2^{1234},k_3,k_4)\Big|_{\Ord(\alpha'^2)}+A_{str}^{tree}(k_1^{1234},k_5,k_2^{1234},k_4,k_3)\Big|_{\Ord(\alpha'^2)}
\nonumber \\ 
&& ~+ A_{str}^{tree}(k_1^{1234},k_5,k_3,k_4,k_2^{1234})\Big|_{\Ord(\alpha'^2)}+A_{str}^{tree}(k_1^{1234},k_5,k_3,k_2^{1234},k_4)\Big|_{\Ord(\alpha'^2)}
\nonumber \\ 
&& ~+ A_{str}^{tree}(k_1^{1234},k_5,k_4,k_3,k_2^{1234})\Big|_{\Ord(\alpha'^2)}+A_{str}^{tree}(k_1^{1234},k_5,k_4,k_2^{1234},k_3)\Big|_{\Ord(\alpha'^2)}\, {\rm .}
\eea
\cme

An immediate out-growth of our three-photon decoupling relation for ${\rm Tr}[F^4]$ matrix elements is a plausible explanation of the observation~\cite{allplus} that, for the all-plus helicity configuration at one loop in pure Yang-Mills, replacing three gluons by photons always gives zero for the five- and higher-point amplitudes. Stieberger and Taylor showed that MHV ${\rm Tr}[F^4]$ matrix elements are closely related to the all-plus one-loop pure Yang-Mills amplitudes and, therefore, it is reasonable to expect the photon-decoupling identity discussed above for ${\rm Tr}[F^4]$ matrix elements to carry over to the all-plus one-loop pure Yang-Mills amplitudes as well.
\begin{figure}
\begin{center}
\resizebox{0.70\textwidth}{!}{\includegraphics{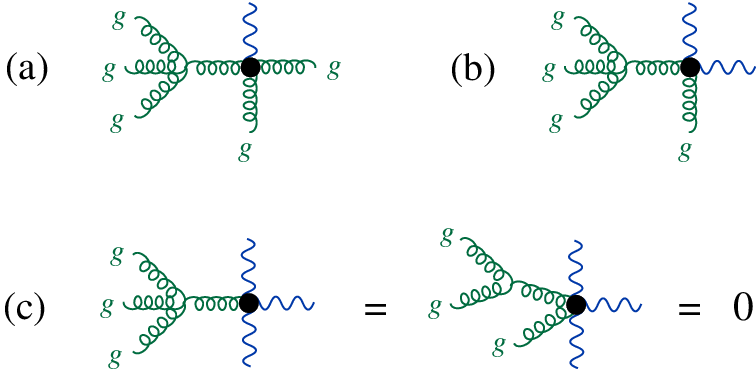}}
\end{center}
\caption{Matrix elements of ${\rm Tr}[F^4]$ for a number of gluons and 
(a) a single photon, or (b) two photons, can be non-vanishing, as explained
  above.  On the other hand, matrix elements with (c) three or more
  photons have to vanish (except for the case $n=4$ with four photons).}
\label{ThreePhotonFigure}
\end{figure}
\section{Dual Superconformal Symmetry and the Ratio of the Six-Point NMHV and MHV Superamplitudes at Two-Loops}
\label{supercomp}
In the discussion of previous sections, $\Nsym$ supersymmetry has played a somewhat peripheral role in that all results have been presented in component form and we have rather arbitrarily focused on the computation of $n$-gluon scattering amplitudes. In this section we begin by reviewing $\Nsym$ on-shell superspace, a powerful formalism that unifies all amplitudes with a given $k$-charge\footnote[1]{Recall that, so far, we have defined the $k$-charge of an amplitude operationally as how many complete copies of the set $\{1,2,3,4\}$ appear in the helicity labels of the amplitude's external lines ({\it e.g.} $\a(k_1,k_2^{123},k_3^4,k_4,k_5^{12},k_6^{34})$ has $k$-charge two).} into so-called $\Nsym$ {\it superamplitudes}. We then review the higher-loop structure of the MHV amplitudes in $\a_{n;2}$, the BDS ansatz, the light-like Wilson loop/MHV amplitude correspondence, and other important related results. After reminding the reader about these exciting recent developments, we will introduce a recently discovered hidden symmetry of the planar $\Nsym$ S-matrix, {\it dual superconformal symmetry}, and try to make use of it to supersymmetrize the pentagon coefficients of Section \ref{gluoncomp} derived for amplitudes with purely gluonic external states. We will see that it is possible to profitably exploit this recently discovered hidden symmetry (see references~\cite{JHennRev} and~\cite{JDrumRev} for a detailed review of some of the consequences of this symmetry) to derive elegant formulae for the higher-order contributions to $\a_{6;3}^{1-{\rm loop}}$. The final formula obtained is very simple and is the form of our results used in a recent study of the $\Nsym$ planar NMHV superamplitude at two loops~\cite{KRV}. As will be explained more below, one of the ideas tested in~\cite{KRV} is whether the NMHV superamplitude divided by the MHV superamplitude is dual superconformally invariant, as was proposed earlier in~\cite{DHKSdualconf} by Drummond, Henn, Korchemsky, and Sokatchev. The new results described in this section for $\a_{6;3}^{1-{\rm loop}}$ were obtained in collaboration with one of the authors of~\cite{KRV}, Cristian Vergu. This review is primarily concerned with weak coupling perturbation theory and, therefore, we will describe all developments in perturbation theory even though most of them were first seen non-perturbatively in the strong coupling regime of $\Nsym$ (via the AdS/CFT correspondence).
\subsection{On-Shell Superspace and All-Loop $\Nsym$ MHV Superamplitudes}
\label{gendisonshell}
The usual construction of the massless $\Nsym$ supermultiplet begins with the anticommutator of the supercharges 
\bea
\{Q^a_{~\alpha}, \bar{Q}_{b~\dot{\alpha}}\} &=& \D_b^{\,~a}~ P_{\alpha \dot{\alpha}}\elale
 \D_b^{\,~a}~ \lambda_{\alpha} \tilde{\lambda}_{\dot{\alpha}}\,,
\label{fundsusyalg}
\eea
in which one chooses $p^\mu = ({\bf p},0,0,{\bf p})$ to define a preferred reference frame. It then follows that some of the supercharges anticommute with themselves and everything else in this frame and others act as creation ($\bar{Q}_{b~\dot{1}}$) or annihilation ($Q^a_{~1}$) operators on the space of states. This approach is useful for some purposes ({\it e.g.} determining the particle content of the massless supermultiplet) but to describe scattering it is better to try and build a formalism where the supercharges act as creation and annihilation operators on the space of states in a manifestly Lorentz covariant way.\footnote{This alternative approach is not new~\cite{Nair,Sokatchev}, but its power was not properly appreciated until very recently~\cite{DHKSdualconf,EFK,SUSYBCFW}.} Our goal is readily accomplished if we introduce a set of four Grassmann variables, $\{\eta_\ell^1,\eta_\ell^2,\eta_\ell^3,\eta_\ell^4\}$, for each external four-vector, $p_\ell$, in the problem. Then one can easily see that (suppressing the $\ell$ label for now)
\be
Q^a_{~\alpha}= \lambda_{\alpha} \eta^a \qquad {\rm and} \qquad \bar{Q}_{b\,\dot{\alpha}} = \tilde{\lambda}_{\dot{\alpha}} {\partial \over \partial \eta^b}
\label{susycharges}
\ee
together satisfy (\ref{fundsusyalg}). Furthermore, the introduction of the $\eta^a$ allows one to build a super wavefunction (Grassmann coherent state) for each external line
\begin{eqnarray}
  \Phi(p,\eta) &=& G^{+}(p) + \G_a(p) \eta^a  + \frac{1}{2!2!}\e_{a b c d}S^{a c}(p)\eta^b \eta^d 
  + \frac{1}{3!} \e_{a b c d}\bar{\G}^{a}(p)\eta^b\eta^c\eta^d  \el
    + \frac{1}{4!}\e_{a b c d} G^{-}(p) \eta^a\eta^b\eta^c \eta^d\,.
  \label{supwvfun}
\end{eqnarray}
which makes it possible to consider $\Nsym$ scattering with half of the supersymmetries (the $Q^a_{~\alpha}$ supercharges which we have chosen to implement multiplicatively) made manifest. To convince the reader that (\ref{susycharges}) and (\ref{supwvfun}) make sense, we must construct the covariant analogs of $Q^i_{~1}$ and $\bar{Q}_{j~\dot{1}}$ in the traditional, non-covariant approach ({\it i.e.} we need to identify the relevant creation operators). In fact, given that $Q^a_{~\alpha} \lambda^\alpha = 0$ and $\bar{Q}_{b~\dot{\alpha}}\tilde{\lambda}^{\dot{\alpha}} = 0$ (the supercharges only have components parallel to $\lambda^\alpha$ and $\tilde{\lambda}^{\dot{\alpha}}$), we can read off the analogs of $Q^i_{~1}$ and $\bar{Q}_{j~\dot{1}}$: the annihilation and creation operators are simply the components of the supercharges along the directions of the spinors, $\hat{a}^c \equiv Q^c = \eta^c$ and $\hat{a}_d^{\dagger} \equiv \bar{Q}_{d} = \partial/\partial \eta^d$, and they satisfy the algebra
\be
\{Q^c, \bar{Q}_{d}\} = \D^{~\,c}_{d} \,{\rm .}
\ee
Now that we know what the creation operators are we can act on the super wavefunction (\ref{supwvfun}) in various combinations. All that we have to do to show that (\ref{supwvfun}) is complete and correct is find some combination of creation operators ($\eta$ derivatives) that isolate each term in the super wavefunction. Following~\cite{DHKSdualconf,origElvang}, we have 
\cmb{-1.0 in}{0 in}
\bea
&& \Phi(p,\eta)\Big|_{\eta^n = 0} = G^{+}(p) \qquad \bar{Q}_{a} \Phi(p,\eta)\Big|_{\eta^{n} = 0} = \G_{a}(p) \qquad  {1\over 2!}\bar{Q}_{a} \bar{Q}_{b} \e^{a b c d} \Phi(p,\eta)\Big|_{\eta^n = 0} = S^{c d}(p)
  \el {1\over 3!} \bar{Q}_{a} \bar{Q}_{b} \bar{Q}_{c} \e^{a b c d} \Phi(p,\eta)\Big|_{\eta^n = 0} = \bar{\G}^{d}(p) \qquad 
    {1\over 4!} \bar{Q}_{a} \bar{Q}_{b} \bar{Q}_{c} \bar{Q}_{d} \e^{a b c d} \Phi(p,\eta)\Big|_{\eta^n = 0} = G^{-}(p) \,{\rm .}
\eea
\cme

\vskip3truemm

Evidently, the on-shell superspace construction is well-defined and it therefore makes sense to speak about $\Nsym$ on-shell superamplitudes, $\a(\Phi_1,\cdots, \Phi_n)$, that take into account all elements of the planar\footnote{Clearly, at the moment, this is a choice we are making since supersymmetry commutes with color.} S-matrix with $n$ external states simultaneously. The $n$-point superamplitude is naturally expanded into $k$-charge sectors as\footnote{Due to supersymmetry, the $k = 0$ and $k = 1$ sectors (and by parity the $k = n - 1$ and $k = n$ sectors) are identically zero for non-degenerate kinematical configurations.}
\begin{equation}
    \a(\Phi_1,\cdots, \Phi_n) = \a_{n;2} + \a_{n;3} + \cdots + \a_{n;\, n-2} \,{\rm .}
    \label{kexpans}
\end{equation}
So far, we have defined $k$-charge at the level of component amplitudes. For example, $A(p_1^{1234},p_2^{1234},p_3,p_4)$ and $A(p_1^{1234},p_2^{123},p_3^{4},p_4)$ both have $k$-charge two because one needs two copies of $\{1,2,3,4\}$ to label their external states. At the level of the superamplitude, the $k$-charge of a given term on the right-hand side of (\ref{kexpans}) is determined by the number of Grassmann parameters that appear in it divided by four\footnote{The $SU(4)_R$ rotates the Grassmann parameters into each other and the superamplitude must be a singlet under R-symmetry transformations. This is impossible unless, for a given term, each $SU(4)$ index, $\{1,2,3,4\}$, appears the same number of times.}. We will refer to $\a_{n;k}$ (a $k$-charge sector of the superamplitude) as a superamplitude since there is usually no possibility of confusion. 

We now turn to the MHV tree-level superamplitude, $\a_{n;2}^{tree}$, which has the simplest superspace structure and can be completely determined by matching onto the Parke-Taylor formula of eq. (\ref{ParkeTayl}) (or any other component amplitude for that matter). Clearly, the simplest possible superspace structure is given by the eight-fold Grassmann delta function $\D^{(8)}\left(Q^{a\,\A}\right)$ by itself and this corresponds to the first term on the right-hand side of (\ref{kexpans}),
\be
\a_{n;2}^{tree} = {1\over 16}\prod_{a = 1}^4 \sum_{i,j = 1}^n \spa{i}.j \eta^a_i \eta^a_j~~ \hat{\a}_{n;2}^{tree} \,{\rm .}
\label{undetMHVsup}
\ee
Here we have used the well-known explicit formula for $\D^{(8)}\left(Q^{a\,\A}\right)$~\cite{EFK}. Suppose we are interested in computing $A^{tree}\left(p_1^{1234},p_2^{1234},p_3,\cdots,p_n\right)$ using $\a_{n;2}^{tree}$. To compute this amplitude one expands (\ref{undetMHVsup}) and extracts the coefficient of $\eta_1^1\eta_1^2\eta_1^3\eta_1^4\eta_2^1\eta_2^2\eta_2^3\eta_2^4$. We will denote this combination as $\eta_1^{1234} \eta_2^{1234}$. A short calculation shows that
\be
\hat{\a}_{n;2}^{tree} = {i \over \spa1.2 \spa2.3 \cdots \spa{n}.1}
\ee
and
\be
\a_{n;2}^{tree} = i{{1\over 16} \prod_{a = 1}^4 \sum_{i,j = 1}^n \spa{i}.j \eta^a_i \eta^a_j \over \spa1.2 \spa2.3 \cdots \spa{n}.1} \,{\rm .}
\label{MHVsup}
\ee
This formalism gives a unified description of all MHV tree amplitudes in $\Nsym$. In fact, for appropriate supersymmetry-preserving variants of dimensional regularization\footnote{We refer the interested reader to \ref{4DHS}, where we describe the four dimensional helicity scheme, the particular variant used in most multi-loop studies of $\Nsym$ scattering amplitudes.}, it turns out that the superspace structure in the MHV case is independent of the loop expansion~\cite{oneloopselfdual,dualSmat} and we can write
\cmb{-1.0 in}{0 in}
\bea
\a_{n;2} &=& i{{1\over 16} \prod_{a = 1}^4 \sum_{i,j = 1}^n \spa{i}.j \eta^a_i \eta^a_j \over \spa1.2 \spa2.3 \cdots \spa{n}.1}
\nonumber\\
&&\times\Bigg(1+\left({g^2 \Nc \mu^{2\e} e^{-\gamma_E \e} \over (4 \pi)^{2-\e}}\right) M_{1-{\rm loop}}+ \left({g^2 \Nc \mu^{2\e} e^{-\gamma_E \e}\over (4 \pi)^{2-\e}}\right)^2 M_{2-{\rm loop}}+\cdots\Bigg)
\label{MHVsupL}
\eea
\cme
as well, although the determination of $M_{L-{\rm loop}}$ may be quite non-trivial\footnote{It is important to point out here that there is a natural seperation of the $M_{L-{\rm loop}}$ functions into even and odd components.}. In the above we still suppress the tree-level gauge coupling and color structure, worrying only about relative factors between different loop orders.

Another special case of interest is the so-called anti-MHV three-point superamplitude. This superamplitude was given in~\cite{EFK}
\be
\bar{\a}_{3;2} = i {\prod_{a = 1}^4 \left(\spb2.3 \,\eta^a_1 + \spb3.1\, \eta^a_2 + \spb1.2\, \eta^a_3\right) \over \spb1.2 \spb2.3 \spb3.1}
\ee
and we reproduce it here for reference. As we shall see, the superspace structure of the six-point NMHV superamplitude is in some sense built out of pieces of $\bar{\a}_{3;2}$. 

Of course, ultimately, the superamplitude $\a_{6;3}$ will be of particular interest to us because it represents the desired supersymmetrization of the results derived in Section \ref{gluoncomp}. Before going in this direction, we conclude the subsection by discussing eq. (\ref{MHVsupL}) in more detail. In eq. (\ref{MHVsupL}),
\cmb{-1.0 in}{0 in}
\bea
\a_{n;2} &=& i{{1\over 16} \prod_{a = 1}^4 \sum_{i,j = 1}^n \spa{i}.j \eta^a_i \eta^a_j \over \spa1.2 \spa2.3 \cdots \spa{n}.1}
\nonumber\\
&&\times\Bigg(1+\left({g^2 \Nc \mu^{2\e} e^{-\gamma_E \e} \over (4 \pi)^{2-\e}}\right) M_{1-{\rm loop}}
+ \left({g^2 \Nc \mu^{2\e} e^{-\gamma_E \e}\over (4 \pi)^{2-\e}}\right)^2 M_{2-{\rm loop}}+\cdots\Bigg) \,{\rm ,}
\eea
\cme
after $\a_{n;2}^{tree}$ is factored out, the analytic structure at $L$ loops, $M_{L-{\rm loop}}$, can be determined by comparing to, say, $A^{L-{\rm loop}}_1(p_1^{1234},p_2^{1234},p_3,\cdots,p_n)$ modulo the Parke-Taylor amplitude. The determination of the $M_{L-{\rm loop}}$ functions was initiated by Bern, Dixon, Dunbar, and Kosower (BDDK) in~\cite{BDDKMHV} where they computed all one-loop MHV superamplitudes in $\Nsym$ (as usual, we will only be interested in the planar contributions). In all of the applications that follow, it will be useful to redefine the contribution from the $L$-th loop as follows:
\bea
\!\!\!\!\!\!\!\!\!\!\!\!\!\!\!\!\!\!\!\!\!\!\!\!\!\!\!\!\!\!\!\!\!\!\!
\left({g^2 \Nc \mu^{2\e} e^{-\gamma_E \e} \over (4 \pi)^{2-\e}}\right)^L M_{L-{\rm loop}} = \left({2 g^2 \Nc e^{-\gamma_E \e} \over (4 \pi)^{2-\e}}\right)^L\M^{(L)}(n,t_i^{[r]},\e) = a^L\M^{(L)}(n,t_i^{[r]},\e) \, {\rm ,}
\eea
where we have made the useful definitions
\be
a \equiv {\lambda e^{-\gamma_E \e} \over (4 \pi)^{2-\e}}~~~~{\rm and}~~~~t_i^{[r]} \equiv (p_i + \cdots + p_{i+r-1})^2 \,{\rm .}
\ee
Using this notation, eq. (\ref{MHVsupL}) reads 
\be
\a_{n;2} = i{{1\over 16} \prod_{a = 1}^4 \sum_{i,j = 1}^n \spa{i}.j \eta^a_i \eta^a_j \over \spa1.2 \spa2.3 \cdots \spa{n}.1}\Bigg(1+\sum_{L = 1}^\infty a^L\M^{(L)}(n,t_i^{[r]},\e)\Bigg) \,{\rm .}
\ee

Let us now describe the results of BDDK in~\cite{BDDKMHV} where the structure of $\M^{(1)}(n,t_i^{[r]},\e)$ for all $n$ was determined through $\Ord(\e^0)$. It was found that:
\cmb{-0.8 in}{0 in}
\bea
\M^{(1)}(n,t_i^{[r]},\e) &=&
 C_\Gamma \sum_{i=1}^{n} \Bigg( -{ 1 \over \e^2 } \bigg(
{ \mu^2  \over -t_i^{[2]} } \bigg)^{\e}
-\sum_{r=2}^{\left[{n\over 2}\right] -1}
\sum_{i=1}^n
  \ln \bigg({ -t_i^{[r]}\over -t_i^{[r+1]} }\bigg)
  \ln \bigg({ -t_{i+1}^{[r]}\over -t_i^{[r+1]} }\bigg)\nonumber\\
  &&\qquad\quad~
+D_n\left(t_i^{[r]}\right) + L_n\left(t_i^{[r]}\right) +{ n \pi^2 \over 6 }\Bigg)
+ \Ord(\e)\, {\rm ,}
\label{UnivFunc}
\eea
\cme
where $C_{\G}$ is given by
$$C_\G = {\G(1+\e)\G(1-\e)^2 \over 2 \G(1-2\e)}\,{\rm .}$$
The form of $D_n\left(t_i^{[r]}\right)$ and $L_n\left(t_i^{[r]}\right)$ depends upon whether $n$ is odd or even.
For $n=2m+1$,
$$
D_{2m+1}= -\sum_{r=2}^{m-1} \Bigg( \sum_{i=1}^{n}
\li2 \bigg[ 1- { t_{i}^{[r]} t_{i-1}^{[r+2]}
\over t_{i}^{[r+1]} t_{i-1}^{[r+1]} } \biggr]  \Bigg)\, {\rm ,}
$$
$$
L_{2m+1}= -{ 1\over 2} \sum_{i=1}^n
  \ln \bigg({ -t_{i}^{[m]}\over -t_{i+m+1}^{[m]}  } \bigg)
  \ln \bigg({ -t_{i+1}^{[m]}\over -t_{i+m}^{[m]} } \bigg)\, {\rm ,}
$$
whereas for $n=2m$,
$$
D_{2m}= -\sum_{r=2}^{m-2} \Bigg( \sum_{i=1}^{n}
\li2 \bigg[ 1- { t_{i}^{[r]} t_{i-1}^{[r+2]}
\over t_{i}^{[r+1]} t_{i-1}^{[r+1]} }  \bigg]  \Bigg)
-\sum_{i=1}^{n/2} \li2 \bigg[ 1- { t_{i}^{[m-1]}t_{i-1}^{[m+1]}
\over t_{i}^{[m]}t_{i-1}^{[m]}} \bigg]\, {\rm ,}
$$
$$
L_{2m}=-{1\over 4} \sum_{i=1}^n
  \ln \bigg({ -t_{i}^{[m]}\over -t_{i+m+1}^{[m]}  } \bigg)
  \ln \bigg({ -t_{i+1}^{[m]}\over -t_{i+m}^{[m]} } \bigg)\, {\rm .}
$$
The above only holds for $n \geq 5$. For $n = 4$ we have
\cmb{-0.5 in}{0 in}
\bea
\M^{(1)}(4,t_i^{[r]},\e) = C_\G \bigg\{
-{2 \over \e^2} \Big[ \left( {-s\over\mu^2}\right)^{-\e}+ \left({-t\over\mu^2}\right)^{-\e} \Big]
+ \ln^2\left( {-s \over - t} \right) + \pi^2 \bigg\} \,{\rm .}
\label{4ptanal}
\eea
\cme

Subsequently, the functions $\M^{(2)}(4,t_i^{[r]},\e)$ and $\M^{(2)}(5,t_i^{[r]},\e)$ were determined through terms of $\Ord(\e^0)$ in~\cite{ABDK} and~\cite{TwoLoopFive} respectively. Remarkably, the following relationships were found:
\cmb{-0.5 in}{0 in}
\bea
\M^{(2)}(4,t_i^{[r]},\e)\Big|_{\Ord(\e^0)} - {1\over 2} \M^{(1)}(4,t_i^{[r]},\e)^2\Big|_{\Ord(\e^0)} &=& \alpha~ \M^{(1)}(4,t_i^{[r]},2\e)\Big|_{\Ord(\e^0)} + \beta \label{4pt2LBDS} \\
\M^{(2)}(5,t_i^{[r]},\e)\Big|_{\Ord(\e^0)} - {1\over 2} \M^{(1)}(5,t_i^{[r]},\e)^2\Big|_{\Ord(\e^0)} &=& \alpha~ \M^{(1)}(5,t_i^{[r]},2\e)\Big|_{\Ord(\e^0)} + \beta  \,{\rm ,}
\label{5pt2LBDS}
\eea
\cme
where both sides of the above are only considered through $\Ord(\e^0)$. $\alpha$ and $\beta$ are transcendentality two and four numbers respectively.\footnote{For example, $\zeta(2)$ is transcendentality two and $\zeta(4)$ is transcendentality four. One also speaks of functions carrying transcendentality ({\it e.g.} $\li2[1-{-s \over -t}]$ has transcendentality two).} Generically, $L$-loop planar amplitudes in $\Nsym$ are built out of transcendentality $2 L$ numbers and functions~\cite{KotikovLipatov}. In the above, $\alpha$ and $\beta$ have the transcendentality that they do because both sides of eqs. (\ref{4pt2LBDS}) and (\ref{5pt2LBDS}) are expected to have uniform transcendentality four. 

Given these striking results, Bern, Dixon, and Smirnov proposed~\cite{BDS} the following ansatz for the analytical structure of all planar MHV superamplitudes,
\cmb{-0.9 in}{0 in}
\bea
\ln\left(1+\sum_{L = 1}^\infty a^L \M^{(L)}(n,t_i^{[r]},\e)\right) = \sum_{L = 1}^\infty a^L\left(f^{(L)} \M^{(1)}(n,t_i^{[r]},L \e) 
+ C^{(L)}+E^{(L)}(n,\e)\right) \,{\rm ,}\nonumber\\
\label{BDSansatz}
\eea
\cme
which they checked for $n = 4$ through three loops. In eq. (\ref{BDSansatz}) above, $f^{(L)}$ and $C^{(L)}$ are numbers of the appropriate transcendentality ($2(L-1)$ and $2 L$ respectively) and $E^{(L)}(n,\e)$ contains higher-order in $\e$ contributions that are unimportant because, usually, both sides of (\ref{BDSansatz}) are expanded to some order in $a$ and then higher-order in $\e$ terms are dropped to put all of the $n$ dependence on the right-hand side into the function  $\M^{(1)}(n,t_i^{[r]},L \e)$. Actually, a structure like this is expected in gauge theory on general grounds for the $\e$ pole terms; the IR divergences of planar non-Abelian gauge theory amplitudes are well-understood and known to exponentiate~\cite{Catani,StermanYeomans}. What is really novel about eq. (\ref{BDSansatz}) is that it holds also for the finite terms.

In fact, the so-called BDS ansatz (eq. (\ref{BDSansatz})) is known to be valid to all loop orders if $n = 4 ~{\rm or}~5$~\cite{DHKSward}. We will explain this in Subsection \ref{DSI} after introducing dual superconformal symmetry. For higher multiplicity, however, life is not so simple. It was proven in~\cite{AMlargen,BLSV1,BDKRSVV} that the BDS ansatz is incomplete at two loops and six points. For this case, which will be the one of primary interest to us, eq. (\ref{BDSansatz}) must be modified:
\cmb{-0.9 in}{0 in}
\bea
\M^{(2)}(6,t_i^{[r]},\e)\Big|_{\Ord(\e^0)} - {1\over 2}\M^{(1)}(6,t_i^{[r]},\e)^2\Big|_{\Ord(\e^0)}= \alpha ~\M^{(1)}(6,t_i^{[r]},2\e)\Big|_{\Ord(\e^0)} + \beta + R_{6}^{(2)}\left(t_i^{[r]}\right)\,{\rm ,}\nonumber\\
\label{6ptstruct}
\eea
\cme
The new term on the right-hand side is called the two-loop, six-point remainder function. It is IR finite and highly constrained. For example, in order to be consistent with the known results for four and five point scattering at two loops, $R_6^{(2)}$ must have vanishing soft and collinear limits in all channels. Also, we know from the discussion above that $R_6^{(2)}$ should be a function of uniform transcendentality four. Furthermore, as we will see in the next section, the remainder function is not an arbitrary function of the $t_i^{[r]}$. In fact, for generic kinematics it is a function of only three independent variables.
\subsection{Light-Like Wilson-Loop/MHV Amplitude Correspondence}
\label{WL/MHV2}
Inspired by the AdS/CFT correspondence, a very surprising connection was suggested~\cite{origAldayMald} between two {\it a priori} completely unrelated observables. One of the observables, 
\be
{\a_{n;2} \over \a^{tree}_{n;2}} = 1 + \sum_{L = 1}^\infty a^L \M^{(L)}(n,t_i^{[r]},\e)
\label{analstruc}
\ee
was discussed at length in Subsection \ref{gendisonshell}. The other, the expectation value of an $n$-gon (denoted $C_n$) light-like Wilson loop
\be
W[C_n] \equiv {1\over \Nc} \langle0|{\rm Tr}\bigg[P\bigg\{ {\rm exp}\left(i g \oint_{C_n} dx^\nu A^a_\nu(x) t^a\right)\bigg\}\bigg]|0\rangle\,{\rm ,}
\label{WLdef}
\ee
has not been introduced so far, so we will analyze its definition in some detail. Of course, we will also have to understand, at least in principle, how to calculate the set of objects introduced above perturbatively if our goal is to establish a connection between eqs. (\ref{analstruc}) and (\ref{WLdef}). 

The Wilson loop appearing in eq. (\ref{WLdef}) is of a special type: the contour $C_n$ defining it has cusps connected by light-like segments. The expectation value of unions of light-like Wilson lines enter into the calculation of certain universal soft functions in QCD. These soft functions are important because they control the resummation of large logarithms that often appear at the edges of phase-space when one tries to na\"{i}vely compute next-to-leading (or higher) corrections to cross-sections for processes in QCD. As we shall see, $n$-cusp light-like Wilson loop expectation values also play an important in $\Nsym$, but in a rather different way.

It is now time to return to eq. (\ref{WLdef}) and define the quantities entering into the expression on the right-hand side. In the above, the gauge connection, $A_\nu^a$ is contracted with the $SU(\Nc)$ fundamental representation gauge group generators, $t^a$. The quantity $A_\nu^a t^a$ is integrated around the closed contour $C_n$ depicted in Figure \ref{hexWL} (for $n = 6$). Each cusp of $C_n$ is labeled $x_i^\nu$ and the lines between adjacent cusps have lengths $(x_i - x_{i+1})^2 = 0$. The distances between non-adjacent cusps are in general non-zero. If we introduce the notation $x_{i j}^2 = (x_i - x_j)^2$, we have nine distinct non-zero distances for $n = 6$: $\{x_{1 3}^2,\,x_{2 4}^2,\,x_{35}^2,\,x_{46}^2,\,x_{51}^2,\,x_{62}^2,\,x_{14}^2,\,x_{25}^2,\,x_{36}^2\}$. 
\begin{figure}
\begin{center}
\resizebox{.6\textwidth}{!}{\includegraphics{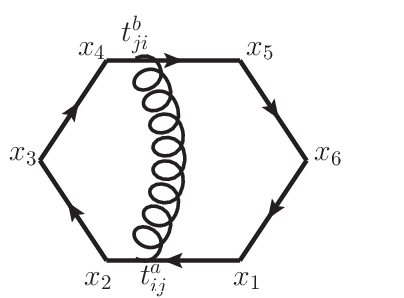}}
\end{center}
\caption{The Feynman diagram for one contribution to $W[C_6]$.}
\label{hexWL}
\end{figure}

Now that we understand how to interpret $W[C_n]$, following~\cite{DKS4pt}, we calculate it to order $\Ord(g^2)$ (lowest non-trivial order). Expanding the path-ordered exponential
gives 
\cmb{-0.9 in}{0 in}
\bea
P\bigg\{ {\rm exp}\left(i g \oint_{C_n} dx^\nu A^a_\nu(x) t^a\right)\bigg\} &=& 1 + i g \oint_{C_n} dx^\nu A^a_\nu(x) t^a \nonumber\\
&+&{1\over 2!}(i g)^2 \oint_{C_n}\oint_{C_n}  dx^\rho dy^\sigma A^a_\rho(x) A^b_\sigma(y) t^a_{i j}t^b_{j k} + \cdots ~{\rm .}
\eea 
\cme
Truncating the above at $\Ord(g^2)$ and taking its vacuum expectation value gives
\be
1 + {1\over 2!}(i g)^2 \oint_{C_n}\oint_{C_n}  dx^\rho dy^\sigma \langle0|A^a_\rho(x) A^b_\sigma(y)|0\rangle t^a_{i j}t^b_{j k}
\ee
since $\langle0|A^a_\nu(x) t^a|0\rangle = 0$ by virtue of Lorentz invariance. Finally, we take the trace over generator matrices, tack on the overall factor of $1/\Nc$, and obtain $W[C_n]$ through $\Ord(g^2)$:
\be
W[C_n]\Big|_{\Ord(g^2)} = 1 - {g^2\over 2! \Nc} \oint_{C_n}\oint_{C_n}  dx^\rho dy^\sigma \langle0|A^a_\rho(x) A^b_\sigma(y)|0\rangle t^a_{i j}t^b_{j i} \,{\rm .}
\label{WLsetup}
\ee
Since $\langle0|A^a_\rho(x) A^b_\sigma(y)|0\rangle$ is just the well-known two-point correlation function for the Yang-Mills field in position space,
\be
\langle0|A^a_\rho(x) A^b_\sigma(y)|0\rangle = {-g_{\rho \sigma} \D^{a b} \mu^{2\e} \pi^\e e^{\gamma_E \e}\over 4 \pi^2(-(x-y)^2)^{1-\e}}
\ee
it is clear that Wilson loop expectation values are conveniently described by Feynman diagrams. For example, if we parametrize our $n$-gon loop,  for $1 \leq i \leq n$, as
\be
\{x^\nu(\tau_i) = x_i^\nu - \tau_i x_{i\,i+1}^\nu,\, y^\nu(\tau_i) = x_i^\nu - \tau_i x_{i\,i+1}^\nu | 0\leq \tau \leq 1\} \,{\rm ,}
\ee
the $\Ord(g^2)$ contribution to $W[C_6]$ shown in Figure \ref{hexWL} can be calculated by integrating over the positions on lines $x_1 - x_2$ and $x_4 - x_5$ where the gluon stretched between them can be absorbed/emitted\footnote[1]{Due to the fact that we have two integrals over the entire closed contour, we pick up a factor of $2!$ (from interchanging the roles of $x^\rho$ and $y^\sigma$) that cancels against the factor of $2!$ in the denominator of eq. (\ref{WLsetup}).}:
\cmb{-.9 in}{0 in}
\bea
&&-{g^2 \over \Nc} \int_{x_1^\rho}^{x_2^\rho} dx^\rho \int_{x_4^\sigma}^{x_5^\sigma} dy^\sigma {-g_{\rho \sigma} \D^{a b} \mu^{2\e} \pi^\e e^{\gamma_E \e}\over 4 \pi^2(-(x-y)^2)^{1-\e}} t^a_{i j}t^b_{j i}
\nonumber\\ &&\qquad\quad 
=-{g^2 \over \Nc} \int_{0}^{1} (-d\tau_1 x_{12}^\rho) \int_{0}^{1} (-d\tau_4 x_{45}^\sigma) {-g_{\rho \sigma} \mu^{2\e} \pi^\e e^{\gamma_E \e} \over 4 \pi^2(-(x_1-x_4-\tau_1 x_{12} + \tau_4 x_{45})^2)^{1-\e}} t^a_{i j}t^a_{j i}
\nonumber\\ && \qquad\quad
={\mu^{2\e}g^2 \pi^\e e^{\gamma_E \e}\over 4 \pi^2 \Nc}\int_{0}^{1} d\tau_1 \int_{0}^{1} d\tau_4 {x_{12}\cdot x_{45} \over (-(x_1-x_4-\tau_1 x_{12} + \tau_4 x_{45})^2)^{1-\e}} C_F \Nc
\nonumber\\ &&\qquad\quad
={\mu^{2\e}g^2 \pi^\e e^{\gamma_E \e} C_F \over 4 \pi^2}\int_{0}^{1} d\tau_1 \int_{0}^{1} d\tau_4 {x_{12}\cdot x_{45} \over (-(x_1-x_4-\tau_1 x_{12} + \tau_4 x_{45})^2)^{1-\e}}
\eea
\cme
On general grounds, we expect such a contribution to be a real number for $(x_1-x_4-\tau_1 x_{12} + \tau_4 x_{45})^2 < 0$ and $\e$ sufficiently small, real, and positive. In this review we will never have to leave the region where these conditions are satisfied.
\begin{figure}
\begin{center}
\resizebox{.9\textwidth}{!}{\includegraphics{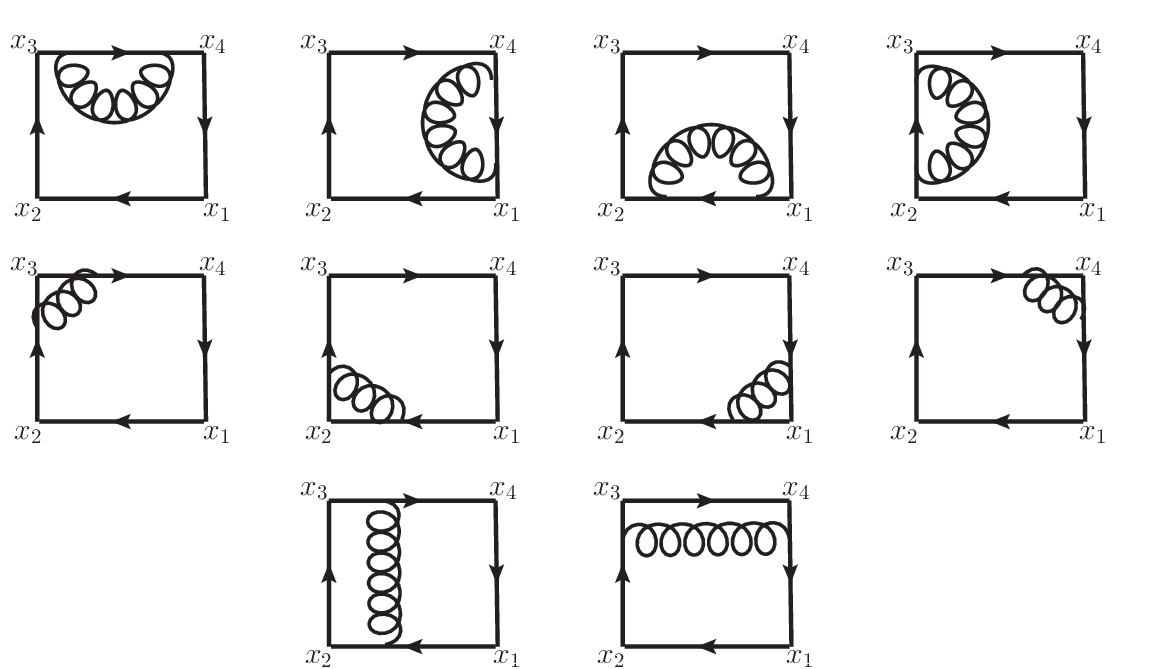}}
\end{center}
\caption{The complete set of Feynman diagrams required to calculate $W[C_4]$ to $\Ord(g^2)$.}
\label{WL4f}
\end{figure}
Of course, for $n = 6$, we will have to add a very large number of topologically distinct contributions in order to obtain a gauge invariant result. It will be simpler and get the point across just as effectively if we follow~\cite{DKS4pt} and calculate $W[C_4]$ in detail to $\Ord{(g^2)}$. The complete set of diagrams for the $\Ord{(g^2)}$ correction to $W[C_4]$ are shown in Figure \ref{WL4f}. In this case the only non-zero invariants are $x_{13}^2$ and $x_{24}^2$. Clearly, the first line of diagrams in Figure \ref{WL4f} vanish once the light-like character of the Wilson loop is taken into account. The second line of diagrams are divergent due to presence of the cusps. These divergences come from the regions of parameter-space where positions of absorption/emission approach a cusp. Such divergences are short distance and therefore ultraviolet in nature. Finally, we will see that the last line of diagrams are finite in four dimensions. If we denote the diagram in class $(\ell)$ that has a gluon stretched between lines $x_i - x_{i+1}$ and $x_j - x_{j+1}$ as $\mathcal{W}_{ij}^{(\ell)}$, 
we have 
\bea
\mathcal{W}_{i i}^{(1)} &=& 0 ~{\rm for~all}~1 \leq i \leq 4 \\
\mathcal{W}_{1 2}^{(2)} &=& \mathcal{W}_{3 4}^{(2)} = -{g^2 C_F e^{\gamma_E \e}(-x_{13}^2 \pi \mu^2)^\e \over 8 \pi^2 \e^2}\\
\mathcal{W}_{2 3}^{(2)} &=& \mathcal{W}_{1 4}^{(2)} = -{g^2 C_F e^{\gamma_E \e}(-x_{24}^2 \pi \mu^2)^\e \over 8 \pi^2 \e^2}\\
\mathcal{W}_{1 3}^{(3)} &=& \mathcal{W}_{2 4}^{(3)} = {g^2 C_F e^{\gamma_E \e}\left(\ln^2\left({x_{13}^2 \over x_{24}^2}\right)+\zeta(2)\right) \over 16 \pi^2} \,{\rm .}
\eea
Taking into account the fact that, in the large $\Nc$ limit,
$$C_F = {\Nc^2-1\over 2\Nc} \longrightarrow {\Nc \over 2} \, {\rm ,}$$
we make the replacement
\be
{g^2 C_F e^{\gamma_E \e} \pi^\e \over  8 \pi^2} \longrightarrow a
\ee
and find that the $\Ord(a)$ analytic structure of $W[C_4]$ is given by~\cite{DKS4pt}
\be
\!\!\!\!\!\!\!\!\!\!\!\!\!\!\!\!\!\!\!\!\!\!\!\!
W[C_4]\Big|_{\Ord(a)} = -{1 \over \e^2}\bigg(\left(-x_{13}^2 \mu^2\right)^\e+\left(-x_{24}^2 \mu^2\right)^\e\bigg)+{1\over 2}\left(\ln^2\left({x_{13}^2 \over x_{24}^2}\right)+\pi^2\right) + \Ord(\e)\,{\rm .}
\label{4ptWLfin}
\ee
This is a remarkable result. Recall eq. (\ref{4ptanal}), where we wrote down the one-loop analytic structure of the four-point MHV superamplitude:
\be
\!\!\!\!\!\!\!\!\!\!\!\!\!\!\!\!
\M^{(1)}(4,t_i^{[r]},\e') = C_\G \bigg\{
-{2 \over \e'^2} \Big[ \left( {-s\over\mu'^2}\right)^{-\e'}+ \left({-t\over\mu'^2}\right)^{-\e'} \Big]
+ \ln^2\left( {-s \over - t} \right) + \pi^2 \bigg\} \,{\rm ,}
\label{4ptanal2}
\ee
where $C_\G$ is given by
$$C_\G = {\G(1+\e')\G(1-\e')^2 \over 2 \G(1-2\e')}\,{\rm .}$$
Up to some redefinition of $a$, $\e$, and $\mu$, the above expression for $W[C_4]$ at lowest non-trivial order matches the above formula for $\M^{(1)}(4,t_i^{[r]},\e')$ exactly if we make the identifications
\be
s \leftrightarrow x_{13}^2 ~~~~{\rm and}~~~~ t \leftrightarrow x_{24}^2\,{\rm .} 
\ee
This surprising connection captures the essence of the light-like Wilson loop/MHV amplitude correspondence in planar $\Nsym$. Even more remarkably, the work of~\cite{DKS4pt} generalizes. There is now a large body of evidence for the following relation
\be
\ln \Bigg({\a_{n;2}\over \a_{n;2}^{tree}}\Bigg)\Bigg|_{{\rm finite};\,\Ord(a^L)}  = \ln \bigg(W[C_n]\bigg)\Bigg|_{{\rm finite};\,\Ord(a^L)} + D_n^{(L)}\, {\rm ,}
\label{equiv}
\ee
valid for all multiplicity and for all-loop orders. In the above, $D_n^{(L)}$ is transcendentality $2 L$ number. As one might guess from eqs. (\ref{4ptWLfin}) and (\ref{4ptanal2}) there is also a relation between the IR poles on the amplitude side and UV poles on the Wilson loop side. As hinted at above, one must make some non-trivial redefinitions of parameters in order to make this precise. See~\cite{DixonStermanMagnea} for a discussion of the IR poles. 

The key observation is that there is a superconformal symmetry (see \ref{sconf}) acting on the Wilson loop in a natural way because it is defined in a configuration space (where the Lagrangian density that possesses this symmetry is constructed). Ultraviolet divergences in the Wilson loop due to the presence of cusps breaks the subgroup of conformal transformations in a controlled fashion. The action of the conformal symmetry is anomalous and one can derive non-perturbatively valid anomalous conformal Ward identities that fix the finite part of $W[C_n]$ that comes from the breaking of the conformal symmetry up to an additive constant at all loop orders~\cite{DHKSward}. What remains must be a function of the conformal cross-ratios. For example, at the six-point level, there are three such cross-ratios
\be
u_1 = {x_{1 3}^2 x_{4 6}^2\over x_{1 4}^2 x_{3 6}^2} \qquad u_2 = {x_{2 4}^2 x_{5 1}^2\over x_{2 5}^2 x_{1 4}^2} \qquad u_3 = {x_{3 5}^2 x_{6 2}^2\over x_{3 6}^2 x_{2 5}^2}\,{\rm ,}
\label{crsrts}
\ee
each of which is invariant under conformal transformations. Now recall that the special conformal transformations can be obtained by conjugating the spatial translations by the conformal inversion operator, $I$ (see \ref{sconf}). Furthermore, it is straightforward to see that $(x_{i j})^{\alpha \dot{\alpha}} = (x_i^\mu - x_j^\mu)(\sigma_\mu)^{\alpha \dot{\alpha}}$ transforms under inversion as:
\be
I[x_{i j}] = x_i^{-1} - x_j^{-1} = -x_j^{-1} (x_i - x_j)x_i^{-1} = -x_j^{-1}x_{i j}x_i^{-1}\,{\rm .}
\ee
Due to the fact that $u_2$ and $u_3$ are obtained by cyclicly permuting $u_1$, we can rest assured that they are conformally invariant if $u_1$ is. We see that
\be
I[u_1] = {I[x_{13}^2] I[x_{46}^2] \over I[x_{1 4}^2] I[x_{3 6}^2]} = {{x_{13}^2 \over x_1^2 x_3^2}{x_{46}^2\over x_4^2 x_6^2} \over {x_{1 4}^2 \over x_1^2 x_4^2} {x_{3 6}^2 \over x_3^2 x_6^2}} = {x_{1 3}^2 x_{4 6}^2\over x_{1 4}^2 x_{3 6}^2}
\ee
and $u_1$ is indeed invariant under inversion. This actually implies the invariance of $u_1$ under the full conformal group, since it is obviously invariant under Poincar\'{e} transformations and dilatations.

We are now in a position to make some remarks about the analytic structure of the $n$-point MHV superamplitudes in $\Nsym$. As we will discuss more in the next subsection, the fact the Wilson loop/MHV amplitude correspondence of eq. (\ref{equiv}) exists implies the existence of a novel hidden symmetry of the planar $\Nsym$ S-matrix through the identification $x_{i}^\mu - x_{i+1}^\mu = p_i^\mu$. This symmetry is ``hidden'' because it cannot have its origin in the Lagrangian (it acts naturally in momentum space). This hidden symmetry is called dual superconformal invariance for reasons that should now be clear. In fact, the dual conformal subgroup already tells us quite a bit of useful information about the analytic structure of the MHV superamplitude. For instance, the reason that the BDS ansatz gives the exact finite part for $n = 4~{\rm or} ~5$ external states is obvious once one understands that the ansatz is just the contribution of the conformal anomaly and that the conformal anomaly is exact for four or five points; due to the light-like nature of the Wilson loops under consideration, no conformally invariant cross-ratios can even be written down for four or five particles in $\Nsym$. In fact, if dual conformal symmetry was not broken by IR divergences, we would expect the full non-perturbative answer to be just a constant times the appropriate tree amplitude.

We can also make precise the arguments of the two-loop six-point remainder function mentioned in Subsection \ref{gendisonshell}. Recall eq. (\ref{6ptstruct}) for the analytic structure of $\ln(1+\sum_{L = 1} a^L \M^{(L)}(6,t_i^{[r]},\e))$ expanded up to second order in perturbation theory:
\cmb{-1.0 in}{0 in}
\bea
\M^{(2)}(6,t_i^{[r]},\e)\Big|_{\rm finite} - {1\over 2}\M^{(1)}(6,t_i^{[r]},\e)^2\Big|_{\rm finite}= \alpha ~\M^{(1)}(6,t_i^{[r]},2\e)\Big|_{\rm finite} + \beta + R_{6}^{(2)}\left(t_i^{[r]}\right)\,{\rm .}\nonumber\\
\eea
\cme
Given everything that we have discussed so far, it is clear that the six-point two-loop remainder function
$R_{6}^{(2)}\left(t_i^{[r]}\right)$ must actually be a function of three dual conformally invariant cross-ratios. If we use the dictionary
\cmb{-.2 in}{0 in}
\bea
&&x_{13}^2 \leftrightarrow s_{1} \qquad x_{24}^2 \leftrightarrow s_{2} \qquad x_{35}^2 \leftrightarrow s_{3} \qquad x_{46}^2 \leftrightarrow s_{4}\nonumber \\
x_{15}^2 & \leftrightarrow & s_{5} \qquad x_{26}^2 \leftrightarrow s_{6} \qquad x_{14}^2 \leftrightarrow t_{1}\qquad
x_{25}^2 \leftrightarrow t_{2} \qquad
x_{36}^2 \leftrightarrow t_{3}
\eea
\cme
we see that, from the point of view of dual conformal symmetry, eq. (\ref{crsrts}) becomes
\be
u_1 = {s_1 s_4\over t_1 t_3} \qquad u_2 = {s_2 s_5\over t_2 t_1} \qquad u_3 = {s_3 s_6\over t_3 t_2}
\ee
and we have
\be
R_{6}^{(2)}\left(t_i^{[r]}\right) = R_{6}^{(2)}(u_1,u_2,u_3) \, {\rm .}
\ee
So far, we have really only used the dual conformal subgroup of the dual superconformal symmetry. In the next section we will describe the full dual symmetry group~\cite{DHKSdualconf}.
\subsection{Dual Superconformal Invariance and the Pentagon Coefficients of the Planar $\Nsym$ One-Loop Six-Point NMHV Superamplitude}
\label{DSI}
To realize the dual superconformal generators on their dual superspace~\cite{DHKSdualconf} we introduce variables $\theta_{i\,\alpha}^{a}$ to solve the $\delta^{(8)}$($Q^a_{~\alpha}$) supercharge conservation constraint in much the same way that the $x_{i\,\alpha\dot{\alpha}}$ of the last subsection solve the $\delta^{(4)}$($P_{\alpha \dot{\alpha}}$) momentum conservation constraint. In other words,
\be
\theta_{i\,\alpha}^{a}-\theta_{i+1\,\alpha}^{a} = \lambda_{i\,\alpha}\eta_i^a
\ee
is the supersymmetric complement of the relation
\be
x_{i\,\alpha\dot{\alpha}} - x_{i+1\,\alpha\dot{\alpha}} = \lambda_{i\,\alpha}\tilde{\lambda}_{i\,\dot{\alpha}} \,{\rm .}
\ee
Intuitively, since (dual) superconformal symmetry naturally acts in (momentum) position space and position and momentum are not mutually compatible observables, we expect the algebra of the ordinary superconformal group (see \ref{sconf})  and the dual superconformal group to be somehow entangled. This intuition is correct; the sketch below shows that there is indeed some overlap between the non-trivial generators of the superconformal (left-hand side) and the dual superconformal (right-hand side) groups:
\bea
~~~~~P_{\alpha \dot{\alpha}}~~~~~&&~~~~~\mc{K}_{\alpha \dot{\alpha}} \nonumber \\
Q^a_{~\alpha}~~~~~~~~~~\bar{Q}_{b\,\dot{\alpha}} &=& \bar{\mc{S}}_{b\,\dot{\alpha}}~~~~~~~~~~\mc{S}^a_{~\alpha} \nonumber \\
S_{a\,\alpha}~~~~~~~~~~\bar{S}^b_{~\dot{\alpha}} &=& \bar{\mc{Q}}^b_{~\dot{\alpha}}~~~~~~~~~~\mc{Q}_{a\,\alpha} \nonumber \\
~~~~~K_{\alpha \dot{\alpha}}~~~~~&&~~~~~\mathcal{P}_{\alpha \dot{\alpha}}
\label{dsalgstruc}
\eea
In the above, the generators $Q^a_{~\alpha}$ and $P_{\alpha \dot{\alpha}}$ on the superconformal side and $\mc{Q}_{a\,\alpha}$ and $\mathcal{P}_{\alpha \dot{\alpha}}$ on the dual superconformal side are actually realized in a pretty trivial fashion and were just included to make the table look more symmetrical:
\bea
Q^a_{~\alpha} &=& \sum_{i = 1}^n \lambda_{i\,\alpha}\eta^a_i\qquad {\rm and} \qquad P_{\alpha \dot{\alpha}} = \sum_{i = 1}^n \lambda_{i\,\alpha}\tilde{\lambda}_{i\,\dot{\alpha}}\nonumber \\
\mc{Q}_{a\,\alpha} &=& \sum_{i = 1}^n {\partial\over \partial \T_{i}^{a\,\alpha}}\qquad {\rm and} \qquad \mathcal{P}_{\alpha \dot{\alpha}} = \sum_{i = 1}^n {\partial\over \partial x_{i}^{~\,\alpha \dot{\alpha}}}\,{\rm .}
\eea
The generators $S_{a\,\alpha}$ and $K_{\alpha \dot{\alpha}}$ on the superconformal side and $\mc{S}^a_{~\alpha}$ and $\mc{K}_{\alpha \dot{\alpha}}$ on the dual superconformal side are a lot more complicated:
\cmb{-1.0 in}{0 in}
\bea
S_{a\,\alpha} &=& \sum_{i = 1}^n {\partial\over \partial \lambda_{i}^{\,~\alpha}}{\partial\over \partial \eta^a_i}\qquad {\rm and} \qquad K_{\alpha \dot{\alpha}} = \sum_{i = 1}^n {\partial\over \partial \lambda_{i}^{\,~\alpha}}{\partial\over \partial \tilde{\lambda}_{i}^{~\,\dot{\alpha}}}\nonumber \\
\mc{S}^a_{~\alpha} &=& \sum_{i = 1}^n \left(-\theta^b_{i\,\alpha}\theta^{a\,\beta}_{i}{\partial\over \partial \T_{i}^{b\,\B}}+x_{i\,\alpha}^{~~\,\dot{\beta}}\theta^{a\,\beta}_{i}{\partial\over \partial x_{i}^{~\,\B \dot{\B}}}+\lambda_{i\,\alpha}\theta^{a\,\gamma}_{i}{\partial\over \partial \lambda_{i}^{\,~\g}}+x_{i+1\,\alpha}^{\,~~~~~\dot{\beta}}\eta^a_i{\partial\over \partial \tilde{\lambda}_{i}^{~\,\dot{\B}}}-\theta^b_{i+1\,\alpha}\eta^a_i{\partial\over \partial \eta^b_i}\right) 
\nonumber \\
\mc{K}_{\alpha \dot{\alpha}} &=& \sum_{i = 1}^n \left(x_{i\,\alpha}^{~~\,\dot{\beta}}x_{i\,\dot{\alpha}}^{~~\,\beta}{\partial\over \partial x_{i}^{~\,\B \dot{\B}}}+x_{i\,\dot{\alpha}}^{~~\,\beta}\theta_{i\,\alpha}^b{\partial\over \partial \T_{i}^{b\,\B}}+x_{i\,\dot{\alpha}}^{\,~~\beta}\lambda_{i\,\alpha}{\partial\over \partial \lambda_{i}^{\,~\B}}+x_{i+1\,\alpha}^{~~~~~\,\dot{\beta}}\tilde{\lambda}_{i\,\dot{\alpha}}{\partial\over \partial \tilde{\lambda}_{i}^{~\,\dot{\B}}}+\tilde{\lambda}_{i\,\dot{\alpha}}\theta^b_{i+1\,\alpha}{\partial\over \partial \eta_{i}^b}\right)\,{\rm .}\nn
\eea
\cme
Finally, the generators $\bar{Q}_{b\,\dot{\alpha}}$ and $\bar{S}^b_{~\dot{\alpha}}$ on the superconformal side and $\bar{\mc{S}}_{b\,\dot{\alpha}}$ and $\bar{\mc{Q}}^b_{~\dot{\alpha}}$ on the dual superconformal side:
\bea
&&\!\!\!\!\!\!\!\!\!\!\!\!\!\!\!\!\!\!\!\!\!\!\!
\bar{Q}_{b\,\dot{\alpha}} = \sum_{i = 1}^n \tilde{\lambda}_{i\,\dot{\alpha}}{\partial\over \partial \eta^b_{i}}\qquad {\rm and} \qquad \bar{S}^b_{~\dot{\alpha}} = \sum_{i = 1}^n \eta_{i}^b{\partial\over \partial \tilde{\lambda}_{i}^{~\,\dot{\alpha}}} \label{QbarSbar} \\
&&\!\!\!\!\!\!\!\!\!\!\!\!\!\!\!\!\!\!\!\!\!\!\!
\bar{\mc{S}}_{b\,\dot{\alpha}} = \sum_{i = 1}^n \left(x_{i\,\dot{\alpha}}^{\,~~\beta}{\partial\over \partial \T_{i}^{b\,\B}}+\tilde{\lambda}_{i\,\dot{\alpha}}{\partial\over \partial \eta^b_{i}}\right)\qquad {\rm and} \qquad \bar{\mc{Q}}^b_{~\dot{\alpha}} = \sum_{i = 1}^n \left(\theta^{b\,\alpha}_{i}{\partial\over \partial x_{i}^{~\,\alpha \dot{\alpha}}}+\eta_{i}^{b}{\partial\over \partial \tilde{\lambda}_{i}^{~\,\dot{\alpha}}}\right)\,{\rm .}
\nonumber
\eea
actually match up if we restrict to the on-shell superspace introduced in Section \ref{supercomp} (by ignoring all $\theta_{i\,\alpha}^{a}$ terms).

We refer the reader to~\cite{JDrumRev} and~\cite{BeiRev} in this volume for a more detailed discussion of the superconformal and dual superconformal algebras. To check the (anti)commutation relations one needs generators not listed above; they are included in \ref{sconf}. As one can see from the form of eq. (\ref{dsalgstruc}), the superconformal and dual superconformal algebras do not commute. It turns out that their closure is a Yangian~\cite{DHPYangian1,BHMPYangian2}. We refer the reader to reference~\cite{BeiRev} for a discussion of this structure in this volume. It is worth pointing out, however, that since the dual superconformal generators are first order differential operators, one may expect them to be better behaved at the quantum level than the usual superconformal generators.

In~\cite{DHKSdualconf}, Drummond, Henn, Korchemsky, and Sokatchev constructed a set of six dual superconformally invariant functions,
\be
R_{1 4 6} = {\D^{(4)}\left(\spb4.5 \eta_6^a+\spb5.6\eta_4^a+
\spb6.4\eta_5^a\right) \prod_{i = 1}^6 \spa{i}.{i+1}\over t_1 \spa1.2 \spa2.3 \spab1.{5+6}.4 \spab3.{4+5}.6\spb4.5\spb5.6}
\ee
and its cyclic permutations, which are then used to write all of the one-loop box coefficients of $\a^{1-{\rm loop}}_{6;3}$ in a way that meshes well with dual superconformal symmetry. More precisely, they found that they could express all the leading singularities in the computation of the NMHV superamplitude in a manifestly dual superconformally invariant way using $R_{1 4 6}$ and its cyclic permutations. In $\Nsym$ there is a choice of basis (the dual conformal basis introduced in Section \ref{gluoncomp}) where the dual superconformal properties of the theory at loop level are as manifest as possible. At the one-loop level, this basis consists of $D = 4 - 2\e$ box integrals and $D = 6 - 2\e$ pentagon integrals. 

The simplicity of the results for boxes suggests that we should try to play the same game for the (now known) pentagon coefficients of the NMHV superamplitude. This is actually not as straightforward as it sounds, due to the fact that the $R_{pqr}$ above do not form a linearly independent set~\cite{DHKSdualconf,DHsuperBCFW}. In fact, for each pentagon topology, it is possible to fit an ansatz of the form
\cmb{-0.9 in}{0 in}
\bea
\a^{tree}_{6;3}= C_i \a^{tree}_{6;2} \left(z_1^{(i)} R_{413} + z_2^{(i)} R_{524} + z_3^{(i)} R_{635}+ z_4^{(i)} R_{146}+z_5^{(i)} R_{251}+z_6^{(i)} R_{362}\right)
\eea
\cme
using just five component amplitudes\footnote{Though perhaps counterintuitive, it is actually possible to do this using purely gluonic amplitudes.}. Fortunately, there is an obvious, preferred, maximally symmetric solution: $z_i^{(i)} = z_{i+3}^{(i)}$. For example, for the pentagon coefficient of $I_5^{(5)}$, we set $z_5^{(5)} = z_2^{(5)}$. This choice then forces 
\be
\left(z_1^{(5)}\right)^{\langle~\rangle \leftrightarrow [~]} = z_4^{(5)} \qquad \left(z_3^{(5)}\right)^{\langle~\rangle \leftrightarrow [~]} = z_6^{(5)}
\ee
as well. The other topologies behave in a completely analogous fashion. To simplify the result, it is convenient to work numerically with complex spinors. It is then possible to recognize the origin of the imaginary parts of the $z_i$ as coming from the natural odd six-point invariant
$$\spb1.2\spa2.3\spb3.4\spa4.5\spb5.6\spa6.1-\spa1.2\spb2.3\spa3.4\spb4.5\spa5.6\spb6.1\,{\rm .}$$
We can now determine the rest of the structure by experimenting with real-valued candidate expressions that respect all the constraints of the problem and have the right BCFW shifts in all channels. In the end, we find
\cmb{-1.5 in}{0 in}
\bea
 &&\a^{1-{\rm loop}}_{6;3} =
 \el \cdots + \frac{i}{6} \e ~\a^{tree}_{6;2}\sum_{i = 1}^6 C_i\bigg(
   {1\over 2} \left(2 s_{i+1} s_{i-2} - t_{i} t_{i+1}\right) t_{i-1} \left(R_{i+2\,i-1\,i+1} + R_{i-1\,i+2\,i-2}\right) \nonumber \\
  && - \Big([i \,i+1] \langle i+1\, i+2\rangle [i+2 \,i+3] \langle i+3\, i+4\rangle [i+4 \,i+5] \langle i+5\,i\rangle \nonumber \\
  &&- \langle i\, i+1\rangle [i+1 \,i+2] \langle i+2\, i+3\rangle [i+3\, i+4] \langle i+4\, i+5\rangle [i+5\, i]\Big) \left(R_{i+2 \, i-1\,i+1} - R_{i-1\,i+2\,i-2}\right) \nonumber \\
  && + {1\over 2} \left(2 s_{i-1} s_{i+2} - t_{i-1} t_{i+1}\right) t_{i} \left(R_{i+3\,i\,i+2} + R_{i\,i+3\,i-1}\right) + {1\over 2} \left(2 s_{i+3} s_{i} - t_{i} t_{i-1}\right) t_{i+1} \left(R_{i+1\,i-2\,i} + R_{i-2\,i+1\,i-3}\right) \nonumber \\
  && -\Big([i \,i+1] \langle i+1\, i+2\rangle [i+2 \,i+3] \langle i+3\, i+4\rangle [i+4 \,i+5] \langle i+5\,i\rangle \nonumber \\
  &&- \langle i\, i+1\rangle [i+1 \,i+2] \langle i+2\, i+3\rangle [i+3\, i+4] \langle i+4\, i+5\rangle [i+5\, i]\Big) \left(R_{i+1\,i-2\,i} - R_{i-2\,i+1\,i-3}\right) \bigg) I^{(i),\,D = 6 - 2\e}_5 \,{\rm .} \nn
\label{sd}
\eea
\cme
Remarkably, when written in this form, the pentagon contributions to the one-loop six-point NMHV superamplitude are related by cyclic symmetry. We can use this fact to explain relation (\ref{mystrel}), reproduced below for convenience:
\be
{K_1 \over C_1} = {K_4 \over C_4} \qquad
{K_2 \over C_2} = {K_5 \over C_5} \qquad
{K_3 \over C_3} = {K_6 \over C_6} \, {\rm .}
\ee
Examining eq. (\ref{sd}), it is trivial to see that the only piece of a given pentagon that does not return to itself under $i \rightarrow i+3$, is the $C_i$ coefficient out front. Evidently, relation (\ref{mystrel}) is a property of the full superamplitude because the symmetric choice
$$z_2^{(i)} = z_5^{(i)}$$
in our ansatz was necessary to manifest the cyclic symmetry of the superamplitude; writing the superamplitude in the form given by eq. (\ref{sd}) shows that there are not enough independent R-invariant structures, to support a full $i \rightarrow i+6$ symmetry for the coefficients divided by their $C_i$. That there are only three independent R-invariant structures can be understood as a consequence of parity invariance in the superamplitude; parity acts on R-invariants by shifting their indices from $i$ to $i+3$. 

Now that we have in hand a pretty formula for the pentagon coefficients of $\a^{1-{\rm loop}}_{6;3}$ built out of dual superconformal invariants, it would be nice if there was some application of our result. It is to this that we turn in the next subsection.
\subsection{Ratio of the Six-Point NMHV and MHV Superamplitudes at Two-Loops}
\label{rationfunc}
In~\cite{DHKSdualconf}, Drummond, Henn, Korchemsky, and Sokatchev made an interesting all-loop prediction based on a remarkable one-loop calculation in their paper. They calculated the parity even part of the NMHV ratio function, $R_{{\rm NMHV}} \equiv \a_{6;3}/\a_{6;2}\,$, to $\Ord(a)$ and found a dual superconformally invariant function. This is a non-trivial result because both $\a_{6;3}$ and $\a_{6;2}$ have IR divergences. The universal, helicity-blind structure of the IR divergences guarantees that the NMHV ratio function is finite to all loop orders. However, at loop level the dual superconformal symmetry is anomalous. One way to circumvent this problem might be to write
\be
\a_{6;3} = \a_{6;2} \Big(R_{\rm NMHV} + \Ord(\e)\Big)
\label{ratiofunc}
\ee
to all loop orders and hope that all of the messiness associated with dual superconformal anomalies resides in the $\a_{6;2}$ prefactor.\footnote{Recently, Beisert, Henn, McLoughlin, and Plefka developed a technique to address these anomalies directly by deforming the dual superconformal generators~\cite{BHMPYangian2}. For more details we refer the reader to~\cite{BeiRev} in this volume.} It is not {\it a priori} clear that the ratio function, $R_{\rm NMHV}$, should have any special properties. For example, as discussed in~\cite{DHKSdualconf}, it is not obvious that the dual superconformal generator $\bar{\mc{Q}}^a_{~\dot{\alpha}}$ annihilates the ratio function, because this generator (eq. (\ref{QbarSbar})) is  sensitive to the dependence of $R_{\rm NMHV}$ on the dual variables, $x_{i\,\alpha \dot{\alpha}}$, and the dependence of the finite parts of $\a_{6;3}$ and $\a_{6;2}$ on the dual variables is fairly complicated (even at $\Ord(a)$). Therefore it is interesting to check by explicit calculation that $R_{\rm NMHV}$ is given by a dual superconformally invariant function. We have already seen that pulling a factor of $\a_{6;2}^{tree}$ out of $\a_{6;3}^{1-{\rm loop}}$ is a natural operation and simplifies the formula for the one-loop NMHV superamplitude. The question is whether $\a_{6;3}^{1-{\rm loop}}$ simplifies when one factors out the entire one-loop MHV superamplitude.

DHKS carried out this analysis and they found that $R_{\rm NMHV}^{1-{\rm loop}}$ could be expressed in terms of R-invariants and linear combinations of two mass hard, two mass easy, and one mass boxes (see eqs. (\ref{6ptgMHV}) and (\ref{6ptgNMHV})). When evaluated through $\Ord(\e^0)$, these box integrals give rise to logarithms and dilogarithms. After simplifying all logarithms and dilogarithms, non-trivial cancellations occur and DHKS found the simple dual superconformally invariant result:\footnote{In eq. (\ref{1Lratfunc}) the $\pi^2/3$ factors are inessential and depend on precisely how one defines the analytic structure of the MHV amplitude. We follow the conventions of DHKS in~\cite{DHKSgenunit}.}
\bea
\!\!\!\!\!\!\!\!\!\!\!\!\!\!\!\!\!\!\!\!
R_{\rm NMHV}^{1-{\rm loop}} &=& {1 \over 4} \sum_{i=1}^6 R_{i\,i+3\,i+5} \bigg(-\ln\left(u_i\right)\ln\left(u_{i+1}\right)+\ln\left(u_{i+1}\right)\ln\left(u_{i+2}\right)+\ln\left(u_{i+2}\right)\ln\left(u_i\right) \nonumber \\
&+&\li2\left(1-u_i\right)+\li2\left(1-u_{i+1}\right)+\li2\left(1-u_{i+2}\right)-{\pi^2\over 3}\bigg) \,{\rm .}
\label{1Lratfunc}
\eea
It is important to note that, in eq. (\ref{1Lratfunc}) above, the index $i$ is understood to be mod 3 for the $u_i$ and mod 6 for the $R_{i\,i+3\,i+5}$. Given the validity of eq. (\ref{ratiofunc}) at one loop and six points, it is reasonable to suspect that something similar will happen at higher loops as well. However, a na\"{i}ve extrapolation from one to higher loops is often dangerous. For example, the BDS ansatz is exact at the one-loop $n$-point level, but is incomplete at two loops and six points, as discussed in Subsection \ref{gendisonshell}. NMHV configurations first appear at the six-point level and, consequently, the first really non-trivial check of (\ref{ratiofunc}) is at two loops and six points. To this end, Kosower, Roiban, and Vergu recently computed the two-loop six-point NMHV superamplitude and they verified (\ref{ratiofunc}) for the parity even part of the ratio function~\cite{KRV}. Before they could check (\ref{ratiofunc}) at $\Ord(a^2)$, they had to resolve a technical problem related to $\e$ poles induced by $\mu$-integrals at the two-loop level. 

In order to understand the problem we need to recall the discussion of Subsection \ref{gresults} where we introduced $\mu$-integral hexabox integrals. We did not properly define this integral in Subsection \ref{gresults} because it was not necessary at the time. The $\mu$-integral hexabox integral depicted in Figure \ref{hexabox} is given by
\cmb{-1.6 in}{0 in}
\bea
&&I_{(4,6)}^{(2);\,D = 4 - 2\e}[\mu^2] = \int {dp^{4-2\e}\over (2\pi)^{4-2\e}}  
{1\over p^2 (p-k_2)^2(p-k_1-k_2)^2}\int {dq^4\over (2 \pi)^4}\int {d^{-2\e} \mu \over (2\pi)^{-2\e} }{\mu^2 \over (q+p)^2+2 \vec{\mu}\cdot p-\mu^2}\times
\el
\times{\mu^2 \over (q^2-\mu^2)((q-k_3)^2-\mu^2)((q-k_3-k_4)^2-\mu^2)((q-k_3-k_4-k_5)^2-\mu^2)((q+k_1+k_2)^2-\mu^2)}\,{\rm .} \nn
\eea
\cme
In this case it turns out that, to leading order, the above integral factorizes~\cite{BDKRSVV} and we can write 
\cmb{0 in}{0 in}
\bea
&I_{(4,6)}^{(2);\,D = 4 - 2\e}[\mu^2]& = I_3^{(2);\,D = 4 - 2\e} I_{6}^{D = 4 - 2\e}[\mu^2] = \left(-{1 \over \e^2}(-s_{1})^{-1-\e}\right)\left(-\e I_6^{D = 6 - 2\e}\right) 
\elale \left(-{1 \over \e^2}(-s_{1})^{-1-\e}\right)\left(-{\e \over 2} \sum_{i = 1}^6 C_i I_5^{(i);\,D = 6 - 2\e}\right) \,{\rm ,}
\label{hexaboxrel}
\eea
\cme
where the last equality follows from eq. (\ref{hexred}). One can check numerically that (\ref{hexaboxrel}) is valid through $\Ord(\e^0)$; the hexabox $\mu$-integral can be evaluated through $\Ord(\e^0)$, apart from trivial factors, is a $1/\e$ pole times a certain linear combination of the finite one-loop functions $I_5^{(i);\,D = 6}$. In our discussion of planar gluon NMHV amplitudes in Section \ref{gluoncomp}, we noted a close connection between the one-loop pentagon coefficients we calculated and appropriate $\mu$-integral hexabox coefficients. For the sake of concreteness, we go back to the particular example discussed in Subsection \ref{gresults}, where we wrote down the relationship between the coefficients of $\e \,I_5^{(2);\,D = 6 - 2\e}$ and $I_{(4,6)}^{(2);\,D = 4 - 2\e}[\mu^2]$:
\be
K_2 = {C_2 \over 2 s_1} \mathcal{K}_2 \, {\rm .}
\ee
If we use the above relation to express the $\mathcal{K}_2$ in terms of $K_2$, we find that the contribution from this NMHV $\mu$-integral hexabox to the ratio function at $\Ord(a^2)$ looks like 
\be
\mc{K}_2 I_{(4,6)}^{(2);\,D = 4 - 2\e}[\mu^2] = {(-s_1)^{-\e} K_2 \over \e \,C_2} \sum_{i=1}^6 C_i I_5^{(i);\,D = 6} + \Ord(\e^0)
\label{hexaboxepole}
\ee
To see how this is all related to our one-loop NMHV pentagon coefficients, let us take a step back and remember what we're trying to calculate. Since we want $R_{\rm NMHV}$ to two loops\footnote{In eq. (\ref{ratexp}), $\hat{\a}_{6;3}^{L-{\rm loop}}$ denotes the superamplitude with a factor of $\a_{6;2}^{tree}$ stripped off.} 
\cmb{-1.0 in}{0 in}
\bea
&&R_{\rm NMHV} = {\hat{\a}_{6;3}^{tree} + a \,\hat{\a}_{6;3}^{1-{\rm loop}} + a^2 \hat{\a}_{6;3}^{2-{\rm loop}} +\cdots\over 1 + a\, \M^{(1)}(n,t_i^{[r]},\e)+a^2 \M^{(2)}(n,t_i^{[r]},\e)+\cdots} 
\nonumber\\ &&\quad\quad
=\hat{\a}_{6;3}^{tree} + a \left(\hat{\a}_{6;3}^{1-{\rm loop}}-\hat{\a}_{6;3}^{tree} \M^{(1)}(n,t_i^{[r]},\e)\right) 
\nonumber\\ &&\quad\quad
+ a^2 \left(\hat{\a}_{6;3}^{2-{\rm loop}} - \hat{\a}_{6;3}^{1-{\rm loop}} \M^{(1)}(n,t_i^{[r]},\e) 
+ \hat{\a}_{6;3}^{tree} \left(\M^{(1)}(n,t_i^{[r]},\e)^2 - 
\M^{(2)}(n,t_i^{[r]},\e)\right)\right) 
\nonumber\\ &&\quad\quad
+ \Ord(a^3)\,{\rm ,}
\label{ratexp}
\eea
\cme
we see that there are other places for us to look for $1/\e$ poles at $\Ord(a^2)$ besides the actual two loop contributions. It is possible for one-loop contributions of $\Ord(\e)$ to hit the universal soft singular terms (see eq. (\ref{1LIR})) in another one-loop contribution and interfere to produce $1/\e$ singularities. For instance, there will be a contribution of the form
\be
\!\!\!\!\!\!\!\!\!\!\!\!\!\!\!\!\!\!\!\!\!\!\!\!\!\!\!\!\!\!\!\!\!\!\!\!
-\left(\e\, K_2 I_5^{(2);\,D = 6 - 2 \e}\right)\left(-{1 \over \e^2}\sum^6_{i=1}\left(-s_{i\,i+1}\right)^{-\e}\right) = {K_2 \sum^6_{i=1}\left(-s_{i\,i+1}\right)^{-\e} \over \e} I_5^{(2);\,D = 6 - 2 \e} + \Ord(\e^0)\nn
\label{pentagonepole}
\ee
coming from the cross-term $-\hat{\a}_{6;3}^{1-{\rm loop}} \M^{(1)}(n,t_i^{[r]},\e)$. This shows that, to obtain all IR divergent contributions to the parity even part of $R_{\rm NMHV}$, the even terms in the one-loop NMHV pentagon coefficients of eq. (\ref{sd}) must be included. Indeed, the authors of~\cite{KRV} have checked at the level of superamplitudes using our results that the even part of $R_{\rm NMHV}$ is dual superconformally invariant. It is now possible to explain why the hexabox coefficients derived by Kosower, Roiban, and Vergu in~\cite{KRV} are so similar to our one-loop pentagon coefficients. A close connection between them is necessary for all of the exotic IR divergent contributions (those that have their origin in $\mu$-integrals) to cancel out in the calculation of the ratio function.

Actually, with a modest amount of additional effort, we can simplify our NMHV pentagon coefficients further and explicitly make contact with the form used by Kosower, Roiban, and Vergu in carrying out their analysis of the two-loop NMHV ratio function. Kosower, Roiban, and Vergu make use of a particular rearrangement of eq. (\ref{sd}). This rearrangement is very nice because, with it, the usual MHV level notions\footnote{At the MHV level, the ``even components'' are simply those terms in the amplitude with no explicit factors of $\pol(i,j,k,\ell)$ and the ``odd components'' are those terms with such factors. We remind the reader that $\pol(i,j,k,\ell)$ was defined in eq. (\ref{epstensor}).} of ``even components'' and ``odd components'' actually make sense in the context of the one-loop NMHV amplitude as well. 

Recall the form of (\ref{sd}) and collect all terms in the above proportional to each R-invariant structure\footnote{There are six such structures: $R_{3 6 2} + R_{6 3 5}$, $R_{4 1 3} + R_{1 4 6}$, $R_{5 2 4} + R_{2 5 1}$, $R_{3 6 2} - R_{6 3 5}$, $R_{4 1 3} - R_{1 4 6}$, and $R_{5 2 4} - R_{2 5 1}$.}:
\cmb{-1.0 in}{0 in}
\bea
 &&\a^{1-{\rm loop}}_{6;3} = \cdots + \frac{i}{6} \e ~\a^{tree}_{6;2}\left\{{\frac{1}{2}}\sum_{i = 1}^6 C_i I^{(i),\,D = 6 - 2\e}_5\right\}\sum_{i = 1}^3 \left(2 s_{i+1} s_{i-2} - t_{i} t_{i+1}\right) t_{i-1} \left(R_{i+2\,i-1\,i+1} + R_{i-1\,i+2\,i-2}\right) \nonumber \\
  && + \frac{i}{6} \e ~\a^{tree}_{6;2}\sum_{i = 1}^3 (-1)^{i} \left( C_i I^{(i),\,D = 6 - 2\e}_5 - C_{i+1} I^{(i+1),\,D = 6 - 2\e}_5 + C_{i-3} I^{(i-3),\,D = 6 - 2\e}_5 - C_{i-2} I^{(i-2),\,D = 6 - 2\e}_5\right)\times
  \el\times \Big(\spb1.2 \spa2.3 \spb3.4 \spa4.5 \spb5.6 \spa6.1 - \spa1.2 \spb2.3 \spa3.4 \spb4.5 \spa5.6 \spb6.1\Big) \left(R_{i+2 \, i-1\,i+1} - R_{i-1\,i+2\,i-2}\right) \,{\rm .}
\label{sd2}
\eea
\cme
Using eq. (\ref{hexred}), reproduced below for the convenience of reader,
\be
I^{D = 6 - 2\e}_6 = {1 \over 2}\sum_{i = 1}^6 C_i I^{(i),\,D = 6 - 2\e}_5
\ee
the first line of eq. (\ref{sd2}) can be put into a form that bears a close resemblance to the even components of the higher order pieces of the one-loop MHV superamplitude; it is proportional to the one-loop scalar hexagon integral (see eq. (\ref{6ptMHV})), $I^{D = 6 - 2\e}_6$. 

In fact, a similar simplification is possible for the terms proportional to $R_{i+2 \, i-1\,i+1} - R_{i-1\,i+2\,i-2}$ as well, although it is not at all obvious. We have numerically checked that
\be
C_i = {2(-1)^i \pol(i+1,i+2,i+3,i+4)\over \spb1.2 \spa2.3 \spb3.4 \spa4.5 \spb5.6 \spa6.1 - \spa1.2 \spb2.3 \spa3.4 \spb4.5 \spa5.6 \spb6.1}\,.
\ee
Using this relation we see that the terms in eq. (\ref{sd2}) not directly proportional to $I^{D = 6 - 2\e}_6$ bear a striking resemblance to the odd components of the higher order pieces of the one-loop MHV superamplitude (again, see eq. (\ref{6ptMHV})). Putting everything together, we find
\cmb{-1.0 in}{0 in}
\bea
&&\a^{1-{\rm loop}}_{6;3} = \cdots + \frac{i}{6} \e ~\a^{tree}_{6;2} I^{D = 6 - 2\e}_6 \sum_{i = 1}^3 \left(2 s_{i+1} s_{i-2} - t_{i} t_{i+1}\right) t_{i-1} \left(R_{i+2\,i-1\,i+1} + R_{i-1\,i+2\,i-2}\right) \nonumber \\
  && + \frac{i}{3} \e ~\a^{tree}_{6;2}\sum_{i = 1}^3 \bigg( \pol(i+1,i+2,i+3,i+4) I^{(i),\,D = 6 - 2\e}_5 + \pol(i+2,i+3,i+4,i+5) I^{(i+1),\,D = 6 - 2\e}_5 \nonumber \\
  &&- \pol(i-2,i-1,i,i+1) I^{(i-3),\,D = 6 - 2\e}_5 - \pol(i-1,i,i+1,i+2) I^{(i-2),\,D = 6 - 2\e}_5\bigg)\left(R_{i+2 \, i-1\,i+1} - R_{i-1\,i+2\,i-2}\right)\nn
\label{sdfinal}
\eea
\cme
for the higher-order components of the planar one-loop NMHV superamplitude. Eq. (\ref{sdfinal}) is particularly important because it was the form utilized by Kosower, Roiban, and Vergu for their analysis in~\cite{KRV}. It is now clear that, indeed, the notions of even and odd that were used in the context of the planar one-loop MHV superamplitude make sense at the NMHV level as well.
\section{Summary}
\label{sum}
In this review we discussed several recent developments in the theory of the $\Nsym$ S-matrix. After reviewing some of the most important computational techniques in Section \ref{revcomp}, we discussed a simple refinement of the $D$ dimensional unitarity technique of Bern and Morgan in Subsection \ref{effgcomp}. One notable feature of our approach is that all integrands are reconstructed in $D$ dimensions directly from tree amplitudes without any need for supersymmetric decompositions. We also discuss how our approach to $D$ dimensional unitarity meshes well with the leading singularity method in the context of all-orders-in-$\e$ one-loop $\Nsym$ calculations. In Subsection \ref{gresults} we presented simple formulae for the higher-order in $\e$ pentagon coefficients of the planar one-loop six-gluon NMHV amplitudes $A^{1-{\rm loop}}_{1}(k_1^{1234},k_2^{1234},k_3^{1234},k_4,k_5,k_6)$, $A^{1-{\rm loop}}_{1}(k_1^{1234},k_2^{1234},k_3,k_4^{1234},k_5,k_6)$, and $A^{1-{\rm loop}}_{1}(k_1^{1234},k_2,k_3^{1234},k_4,k_5^{1234},k_6)$. Na\"{i}vely, these results may seem rather useless because, if one only cares about the massless $\Nsym$ S-matrix, one never needs the pentagon coefficients.

However, we argue in Section \ref{gsrel} that, actually, the higher-order in $\e$ pentagon coefficients are useful because they contain non-trivial information about tree-level scattering of massless modes in open superstring theory. After reviewing the non-Abelian Born-Infeld action in Subsection \ref{BornInfeld}, we argued in Subsection \ref{results} that matrix elements of the non-Abelian Born-Infeld action at $\Ord(\alpha'^2)$ and $\Ord(\alpha'^3)$ can be predicted from all-orders-in-$\e$ $\Nsym$ amplitudes dimensionally shifted to either $D = 8-2\e$ or $D = 10-2\e$. As an amusing by-product of our analysis, we were able to use another close connection between the one-loop all-plus amplitudes in pure Yang-Mills and our stringy corrections at $\Ord(\alpha'^2)$ to understand the vanishing of the all-plus amplitudes when three or more gluons are replaced by photons for $n > 4$.  

At this point, in Section \ref{supercomp}, we explained how to supersymmetrize the results of Subsection \ref{gresults}. To this end, we introduced the $\Nsym$ on-shell superspace in Subsection \ref{gendisonshell} and discussed some important examples of $\Nsym$ superamplitudes. We then show in Subsection \ref{DSI} (after reviewing a few more of the developments that led to the discovery of dual superconformal invariance towards the end of Subsection \ref{gendisonshell} and in Subsection \ref{WL/MHV2}) that the superamplitude takes on a beautifully simple form if expressed in terms of the R-invariants of Drummond, Henn, Korchemsky, and Sokatchev. Remarkably, in this form, it is manifest that the pentagon coefficients are related by cyclic symmetry. Finally, in Subsection \ref{rationfunc}, we explain the relevance of our results to the study of the dual superconformal properties of (the parity even part of) the two-loop NMHV ratio function in dimensional regularization.  Our higher-order-in-$\e$ pentagon coefficients can interfere with $1/\e^2$ poles in the one-loop MHV superamplitude to produce contributions of order $1/\e$. Thus, the results written down in Subsection \ref{DSI} in terms of dual superconformal R-invariants are necessary to produce a finite result for the two-loop NMHV ratio function in on-shell superspace if one is working in dimensional regularization. To this end, we further improve the results presented in Subsection \ref{DSI} by seperating them into even and odd components. This decomposition is clearly natural because the results of Subsection \ref{DSI} simplify still more. In particular, the even components of the higher order pieces of the one-loop NMHV superamplitude can be rearranged and put into a form where they are actually proportional to the one-loop hexagon integral in $D = 6 - 2 \e$. This feature of the even components was exploited recently in a study of the two-loop NMHV ratio function by Kosower, Roiban, and Vergu~\cite{KRV}.

There has been a tremendous amount of recent progress on the planar $\Nsym$ S-matrix\footnote{Most, if not all, of the works cited here were significantly influenced by the seminal work of Witten~\cite{witten}.}~\cite{newpap1,newpap2,newpap3,newpap4,newpap5,newpap6,newpap7,newpap8,newpap9,newpap10,newpap11,newpap12,newpap13,newpap14,newpap15,newpap16,newpap17,newpap18,newpap19,newpap20,newpap21,newpap22,sugg1,sugg2,newpap23,newpap24,newpap25,newpap26,newpap27,newpap28,newpap29,DelDuca1,DelDuca2,DelDuca3,newpap30,newpap32,newpap33,newpap34,newpap35,newpap36}, some of which are reviewed elsewhere in this volume. A recurring theme in recent papers on the subject is the idea that one should be able to learn everything there is to know about the planar $\Nsym$ S-matrix using only four dimensional information. In Subsection \ref{effgcomp} we showed that, generically, four dimensional generalized unitarity cuts are not sufficient to determine one-loop planar $\Nsym$ scattering amplitudes to all orders in the dimensional regularization parameter. Clearly, the analysis of Subsection \ref{results} suggests that the one-loop pentagon coefficents missed by the leading singularity method are important and should be determined independently ({\it e.g.} by using maximal generalized unitarity in $D$ dimensions). However, in the spirit of the recent developments, we should first check whether, perhaps, our predictions for the $\Ord(\alpha'^2)$ and $\Ord(\alpha'^3)$ stringy corrections to $\Nsym$ amplitudes don't really rely on {\it all} pentagon coefficients but only some linear combination thereof.

Recall from the discussion of Section \ref{gluoncomp} that, at the one-loop $n$-point level, amplitudes computed via the leading singularity method are not uniquely determined to all orders in $\e$ but have
\be
{(n-5)(n-4)(n-3)(n-2)(n-1)\over 120}
\ee
pentagon coefficients that must be determined by some other method. It is conceivable that, after performing the dimension shift operation and summing over all contributions, all the undetermined coefficients actually drop out. In fact, there is evidence that this happens for $\Nsym$ amplitudes dimensionally shifted to $D = 8 - 2\e$; we checked that we could derive the appropriate tree-level stringy corrections at $\Ord(\alpha'^2)$ for both MHV and NMHV $n = 6$  amplitudes, $n = 7$ MHV amplitudes, and $n = 8$ MHV amplitudes using the leading singularity method (without $D$ dimensional unitarity). However, for the $\Ord(\alpha'^3)$ stringy corrections, this no longer works. For example, one can check that the one-loop $\Nsym$ six-point MHV amplitude cannot be used to compute $A^{tree}_{str}\left(k_1^{1234},k_2^{1234},k_3,k_4,k_5,k_6\right)$ at $\Ord(\alpha'^3)$ unless {\it all} pentagon coefficients in the amplitude are determined. Our conclusion is that there are still some questions that can be answered by calculating $\Nsym$ amplitudes to all orders in $\e$ that cannot (at least not obviously) be answered by calculating amplitudes in a framework that requires only four dimensional inputs. 

In this review we have seen that our approach to all-orders one-loop $\Nsym$ scattering amplitudes opens up several interesting avenues of exploration.  Besides the connection that we found between stringy corrections and dimensionally shifted one-loop amplitudes, it seems plausible, for example, that a variant of our approach will be the right way to think about computing scattering amplitudes at a generic point in the $\Nsym$ moduli space. It will also be quite interesting to see whether our approach to $D$ dimensional integrand construction is useful for theories with less or no supersymmetry. We expect that the approach advocated here will continue to be relevant and useful for future higher-loop studies of $\Nsym$ and more general amplitudes in dimensional regularization. For a more detailed exposition of the work presented here we refer the interested reader to~\cite{myfifth}. Besides containing a great deal more background material, the work upon which this review is based also contains an additional section where we realize in an elegant and explicit way the form of the six-point NMHV superamplitude written down by Elvang et. al. in~\cite{EFK}. Although this formula is somewhat outside the main line of this review, we feel that it may be of interest to some readers.
\ack
I am very grateful to Radu Roiban, Marcus Spradlin, and Anastasia Volovich for inviting me to write this contribution to the ``Scattering Amplitudes in Gauge Theories''series. I am also very grateful to Matt Strassler and Lance Dixon for their guidance, support, and collaboration over the last few years. I would like to thank the theory group at SLAC for its hospitality the last few summers and Carola Berger, Zvi Bern, Darren Forde, Tanju Gleisberg, and Daniel Ma\^{i}tre for enlightening conversations and correspondence. I am also very grateful to Cristian Vergu for enlightening correspondence and collaboration. I am especially grateful to Fernando Alday, Nima Arkani-Hamed, Jacob Bourjaily, Freddy Cachazo, James Drummond, Henriette Elvang, Johannes Henn, Juan Maldacena, David McGady, and Jaroslav Trnka for helping me to better understand exciting recent developments in planar $\Nsym$ gauge theory and related topics. A special acknowledgment goes to Johannes Henn, Karl Landsteiner, Andy O'Bannon, Matthew Strassler, and Cristian Vergu for providing me with some very useful comments and advice on earlier drafts of this work. In this work, figures were drawn primarily using the Jaxodraw~\cite{JaxD} package.


\appendix 

\section{$~~~~~~~~~~~~~~$Dimensional Regularization}

\label{dimregs}
In this appendix, we begin in \ref{4DHS} where we remind the reader that simply declaring that dimensional regularization will be used to regulate divergences is not meaningful because there are several different variants of dimensional regularization. We describe the salient features of one scheme, called the four dimensional helicity scheme, which is particularly useful for regulating the divergences in maximally supersymmetric gauge theories. In \ref{muint} we give an explicit derivation of eq. (\ref{PtoDCB}), which played an important role in the body of this work. Finally, in \ref{IR1Ldiv}, we briefly talk about the general structure of IR divergences in planar one-loop $\Nsym$ scattering amplitudes.
\subsection{$~~~~~~~~~~~~~~$The Four Dimensional Helicity Scheme}
\label{4DHS}
The four dimensional helicity scheme is a variant of dimensional regularization introduced to simplify the renormalization of supersymmetric gauge theories. In this Subsection, we discuss its salient features and contrast it to the 't Hooft-Veltman scheme. The four dimensional helicity scheme is the variant of dimensional regularization that we implicitly work in throughout the main text. The criterion that one uses to decide if a regulator is appropriate for a given quantum field theory is whether the proposed regulator preserves all the symmetries of the model. Despite the many successes of the 't Hooft-Veltman scheme in the Standard Model, it is not an appropriate regulator for supersymmetric models because examples exist (see {\it e.g.} \cite{DRTJones}) where its use explicitly violates certain supersymmetric Ward identities.

In 2002 the four dimensional helicity scheme~\cite{4dimHS} was proposed\footnote[1]{Very recently, calculations were presented in~\cite{Kilgore} which imply that the four dimensional helicity scheme is not a generally applicable renormalization scheme in the way that, say, dimensional reduction is. We are not in a position to evaluate the claims made by the author of~\cite{Kilgore} at the present time. Regardless, nothing in this review is affected by the discussion in~\cite{Kilgore} because UV divergences are absent in maximally supersymmetric gauge theory.} as a variant of dimensional regularization fully consistent with supersymmetry to all orders in perturbation theory. As the name suggests, all external momenta and wavefunctions are kept in four dimensions; only the loop momenta are continued to $D$ dimensions. The rules for objects built out of $\e_{\mu \nu \rho \sigma}$ are the same in the 't Hooft-Veltman scheme. The main insight of~\cite{4dimHS} was that one must introduce an additional scale, called the spin dimension, which is taken to be the dimension in which the wavefunctions of all virtual particles circulating in loops live. If supersymmetry is to be preserved, the spin dimension, $D_s$, must be treated as follows.
\begin{enumerate}[i.]
\item{Perform all index contractions as if $D_s > D > 4$.}
\item{After the amplitude is a function only of the loop momenta, external momenta, external wavefunctions, $D$, and $D_s$, set $D_s = 4$.}
\end{enumerate}
It is useful to note that if $D_s$ is set to $D$ we recover the 't Hooft-Veltman scheme.
\subsection{$~~~~~~~~~~~~~~$A Useful Integral Reduction Identity Involving Dimensionally-Shifted Integrals at the One-Loop Level}
\label{muint}
In this Subsection we derive eq. (\ref{PtoDCB}) explicitly to supplement the streamlined discussion of Subsection \ref{GUD}. This exercise should also help the reader understand why the coefficients of the pentagon integrals in the dimensionally shifted basis defined in \ref{GUD} contain an explicit factor of $\e$, whereas the box integrals in it do not.
We begin with the dimensionally regulated one-loop integrals of eq. (\ref{IntDef}): 
\be
I_n^{D=4-2 \e}
= i (-1)^{n+1} (4\pi)^{2-\e} \int 
    {d^{4-2\e} \ell \over (2\pi)^{4-2\e} }
  { 1 \over \ell^2 \ldots
    (\ell-\sum_{i=1}^{n-1} K_i )^2 } \,{\rm .}
\label{IntegralDef}
\ee
It has been known for a very long time how to write down an expression for (\ref{IntegralDef}) as a Feynman parameter integral~\cite{PeskinSchroeder}:
\be
I_n^{D=4-2 \e}
= \G(n-2+\e) \int_0^1 d^n x_i {\D\left(1-\sum_{i=1}^n x_i\right) \over \left(\left(\sum_{i=1}^n x_i p_{i-1}\right)^2-\sum_{i=1}^n x_i p_{i-1}^2\right)^{n-2+\e}} \, {\rm ,}
\ee
where $p_i = \sum_{j=1}^{i} K_j$. There is, however, a particularly nice, symmetric way of rewriting this expression~\cite{oneloopdimreg}. The above formula collapses to\footnote{One can verify this relation directly after eliminating one of the variables through the relation $\sum_{i=1}^n x_i = 1$ on both sides of the equation.}
\be
I_n^{D=4-2 \e}
= \G(n-2+\e) \int_0^1 d^n x_i {\D\left(1-\sum_{i=1}^n x_i\right) \over \left(\sum_{i,j=1}^n p_{i-1}\cdot p_{j-1}x_i x_j\right)^{n-2+\e}} \, {\rm .}
\ee 
This representation of $I_n^{D=4-2 \e}$ will enter into our derivation of eq. (\ref{PtoDCB}). The idea is to evaluate the same integral in two different ways. 

Following~\cite{oneloopdimreg}, we define
\be
I_n^{D=4-2 \e}[\ell^2] \equiv i (-1)^{n+1} (4\pi)^{2-\e} \int 
    {d^{4-2\e} \ell \over (2\pi)^{4-2\e} }
  { \ell^2 \over \ell^2 \ldots
    (\ell-\sum_{i=1}^{n-1} K_i )^2 } \, {\rm .}
\ee
Of course, this integral can be trivially reduced by canceling the numerator against the first propagator denominator. 
\be
I_n^{D=4-2 \e}[\ell^2] = -I_{n-1}^{D=4-2 \e} \, {\rm .}
\label{easyway}
\ee
However, we are also free to evaluate it as a Feynman parameter integral. Going through the usual Feynman parametrization procedure we find something of the form
\bea
&&I_n^{D=4-2 \e}[\ell^2] = i (-1)^{n+1} (4\pi)^{2-\e} \G(n) \int_0^1 d^n x_i \D\left(1-\sum_{i=1}^n x_i\right) \times
\el
\times \int 
    {d^{4-2\e} q \over (2\pi)^{4-2\e} }
  { q^2 + \sum_{i,j=1}^n p_{i-1}\cdot p_{j-1}x_i x_j +\sum_{i=1}^n x_i p_{i-1}^2\over \left(q^2 - \sum_{i,j=1}^n p_{i-1}\cdot p_{j-1}x_i x_j\right)^{n-2+\e}} \, {\rm .}
\eea
These integrals are easily carried out by using the standard formulae~\cite{PeskinSchroeder}
\bea
 &&i (-1)^{n+1} (4\pi)^{2-\e} \int  {d^{4-2\e} q \over (2\pi)^{4-2\e} } {1\over (q^2-\Delta)^n} = {\G(n-2+\e)\over \G(n)\Delta^{n-2+\e}}
 \el
{\rm and}~ i (-1)^{n+1} (4\pi)^{2-\e} \int  {d^{4-2\e} q \over (2\pi)^{4-2\e} } {q^2\over (q^2-\Delta)^n} = -{(2-\e)\G(n-3+\e)\over \G(n)\Delta^{n-3+\e}} \, {\rm .}
\eea
$I_n^{D=4-2 \e}[\ell^2]$ becomes 
\begin{changemargin}{-.6 in}{0 in}
\bea
&&I_n^{D=4-2 \e}[\ell^2] = \int_0^1 d^n x_i \D\left(1-\sum_{i=1}^n x_i\right) \Bigg(-{(2-\e)\G(n-3+\e)\over \left(\sum_{i,j=1}^n p_{i-1}\cdot p_{j-1}x_i x_j\right)^{n-3+\e}} 
\el
+ {\G(n-2+\e) \sum_{i,j=1}^n p_{i-1}\cdot p_{j-1}x_i x_j \over \left(\sum_{i,j=1}^n p_{i-1}\cdot p_{j-1}x_i x_j\right)^{n-2+\e}} + {\G(n-2+\e) \sum_{i=1}^n x_i p_{i-1}^2 \over \left(\sum_{i,j=1}^n p_{i-1}\cdot p_{j-1}x_i x_j\right)^{n-2+\e}}\Bigg)\,{\rm .}
\eea
\end{changemargin}
This simplifies nicely:
\begin{changemargin}{-.6 in}{0 in}
\bea
&&I_n^{D=4-2 \e}[\ell^2] = \int_0^1 d^n x_i \D\left(1-\sum_{i=1}^n x_i\right)\Bigg(-{(2-\e)\G(n-2+(\e-1))\over \left(\sum_{i,j=1}^n p_{i-1}\cdot p_{j-1}x_i x_j\right)^{n-2+(\e-1)}}
\el + {(n-3+\e)\G(n-2+(\e-1)) \over \left(\sum_{i,j=1}^n p_{i-1}\cdot p_{j-1}x_i x_j\right)^{n-2+(\e-1)}} + {\G(n-2+\e) \sum_{i=1}^n x_i p_{i-1}^2 \over \left(\sum_{i,j=1}^n p_{i-1}\cdot p_{j-1}x_i x_j\right)^{n-2+\e}}\Bigg)
\elale
\int_0^1 d^n x_i \D\left(1-\sum_{i=1}^n x_i\right) \left({(n-5+2\e)\G(n-2+(\e-1)) \over \left(\sum_{i,j=1}^n p_{i-1}\cdot p_{j-1}x_i x_j\right)^{n-2+(\e-1)}} \right.\nn
&&\qquad\qquad\qquad\qquad\qquad\qquad\qquad\qquad
\left.+ {\G(n-2+\e) \sum_{i=1}^n x_i p_{i-1}^2 \over \left(\sum_{i,j=1}^n p_{i-1}\cdot p_{j-1}x_i x_j\right)^{n-2+\e}}\right)
\elale
(n-5+2\e)I_n^{D=6-2 \e} + \sum_{i=1}^n p_{i-1}^2 I_n^{D=4-2 \e}[x_i]\,{\rm .}
\label{hardway}
\eea
\end{changemargin}
Equating the last line of (\ref{hardway}) with the right-hand side of(\ref{easyway}),
\be
-I_{n-1}^{D=4-2 \e} = (n-5+2\e)I_n^{D=6-2 \e} + \sum_{i=1}^n p_{i-1}^2 I_n^{D=4-2 \e}[x_i]
\ee
we finally obtain a non-trivial relation between scalar integrals. 

In fact, all of the above analysis goes through unchanged if $I_{n}^{D=4-2 \e}[\ell^2]$ is replaced by $I_{n}^{D=4-2 \e}[(\ell-p_{i-1})^2]$, allowing us to derive a total of $n$ relations that can be written in a unified way as
\be
-I_{n-1}^{(i);\,D=4-2 \e} = (n-5+2\e)I_n^{D=6-2 \e} + 2 \sum_{j=1}^n S_{i j} I_n^{D=4-2 \e}[x_i]\,{\rm ,}
\ee
where we have introduced the daughter-integral notation (which first appeared in Subsection \ref{GUD}) and the matrix $S_{i j}$ defined as
\bea
&& S_{i j} = -{1 \over 2}~ (p_i + ... + p_{j-1})^2,~~~~~ i \neq j
\el 
S_{i j} = 0,~~~~~~~~~~~~~~~~~~~~~~~~~~~~~~~i = j \, ,
\eea
where both $i$ and $j$ are to be taken mod $n$. Solving for $I_n^{D=4-2\e}[x_i]$, we obtain
\be
I_n^{D=4-2\e}[x_i] = {1\over 2}\bigg[\sum_{j=1}^n S_{i j}^{-1} I_{n-1}^{(j),~D=4-2\e} + (n-5+2\e)C_i I_n^{D=6-2 \e} \bigg]\,{\rm ,}
\label{interrel}
\ee
where $C_i = \sum_{j=1}^n S_{i j}^{-1}$. Finally, we can exploit the identity $\sum_{i=1}^n x_i = 1$ and sum over the index $i$ in the above. This yields
\be
I_n^{D=4-2\e} = {1\over 2}\bigg[\sum_{j=1}^n C_j I_{n-1}^{(j),~D=4-2\e} + (n-5+2\e)C_0 I_n^{D=6-2 \e} \bigg]\,{\rm ,}
\label{finalrel}
\ee
where $C_0 = \sum_{i=1}^n C_i$. This is the final form of our desired relation.

This formula for $n = 5$, 
\be
I_5^{D=4-2\e} = {1\over 2}\bigg[\sum_{j=1}^5 C_j I_{4}^{(j),~D=4-2\e} + 2\e C_0 I_5^{D=6-2 \e} \bigg]\,{\rm ,}
\label{5ptrel}
\ee
turns out to be very useful in the analysis of one-loop $\Nsym$ amplitudes in dimensional regularization because the five-point scalar integral is related to a linear combination of four-point scalar integrals plus a five-point integral in $D=6-2\e$ dimensions that has an explicit factor of $\e$ out front. Furthermore, it turns out that the $D=6-2\e$ scalar integral has no poles in $\e$. This then implies that eq. (\ref{5ptrel}) corresponds to a special case that relates the five-point integral to four-point integrals, up to $\Ord(\e)$ contributions that can be neglected if one is only interested in computing one-loop amplitudes to $\Ord(\e^0)$. For us, this relation provides a convenient way to separate higher order in $\e$ contributions from those that contribute only through $\Ord(\e^0)$.

Eq. (\ref{finalrel}) for $n = 6$ also appears throughout the main text (recall eq. (\ref{hexred})). To see the utility of (\ref{finalrel}) for this value of $n$, one needs to know something more about the matrix $S_{i j}$ and its rank. In deriving eq. (\ref{interrel}), we implicitly assume that $S_{i j}$ is invertible. Actually this is only a valid assumption for $n \leq 6$ and $n = 6$ is the borderline case. It is well-known that, beginning at the six-point level, additional non-linear constraints on scattering processes exist coming from the fact that it is no longer possible to find an $n-1$ dimensional linearly independent subset of the $n$ external momenta~\cite{oneloopdimreg}. 

To be more concrete, let us specialize to $n = 6$ and count the degrees of freedom for external momenta in $D = 4$. The sum of all momenta is zero by construction, so clearly at most five of the external momenta are linearly independent. However, it must be the case that any one of these five momenta can be expressed as a linear combination of the other four, simply because the vector space that we're working in is four dimensional. More precisely, we have the six relations
\be
{\rm Det}(k_i \cdot k_j)_{r} = 0 {\rm ,}
\label{Gram}
\ee
where the $r$ subscript is to be interpreted as an instruction to delete the $r$-th column of the matrix $(k_i \cdot k_j)$. It turns out that the changing the value of $r$ doesn't change the left-hand side of (\ref{Gram}) and, therefore, all six  equations give the same constraint\footnote{{\it A priori} (\ref{Gram}) could have been identically satisfied. It turns out that this is not the case and, as a result, there is a non-trivial constraint on the kinematics.} on the kinematics. The object $C_0$ is proportional to ${\rm Det}(k_i \cdot k_j)_{r}$ and therefore can be set equal to zero. This results in
\be
I_6^{D=4-2\e} = {1\over 2}\sum_{i=1}^6 C_i I_5^{(i),~D=4-2\e}
\ee
a special case of (\ref{finalrel}) for $n = 6$.
\subsection{$~~~~~~~~~~~~~~$IR Structure of One-Loop Planar Amplitudes in $\Nsym$}
\label{IR1Ldiv}
In this subsection, we review the results of references~\cite{Kunszt} and~\cite{GieleGlover} where all possible IR divergences at one loop in massless gauge theories were classified. Actually, the one-loop IR divergences in $\Nsym$ are a little bit simpler than in the general case. In general, one expects two distinct epsilon pole structures at one loop: poles that have their origin in purely soft or soft-collinear virtual particles and poles that have their origin in purely collinear virtual particles. The purely collinear singularities are governed by terms that have the schematic form~\cite{GieleGlover}
\be
{1 \over \e} \left({\mu^2 \over -\tilde{s}}\right)^\e A^{tree}\left(k_1^{h_1},~\cdots,~k_{n}^{h_n}\right)\, ,
\ee
where $\tilde{s}$ is some kinematic scale. However, there are clearly no divergences of this form in the integral basis of reference~\cite{allNMHV} (valid for planar $\Nsym$ through $\Ord(\e^0)$). Rather, one sees divergences of the schematic form
\be
{1 \over \e^2} \left({\mu^2 \over -\tilde{s}}\right)^\e A^{tree}\left(k_1^{h_1},~\cdots,~k_{n}^{h_n}\right)
\ee
in those integral functions, which correspond to soft-collinear and soft singularities. We conclude that the virtual IR divergences in planar $\Nsym$ one-loop scattering processes have their origin in soft or soft-collinear virtual particles connecting pairs of adjacent external lines\footnote{If we did not restrict ourselves to planar contributions, then the virtual particles could connect non-adjacent external lines as well.}. Quantitatively, if we take the index $i$ in what follows to be mod $n$, we have
\cmb{-.6 in}{0 in}
\bea
&&A^{1-{\rm loop}}_1 \left(k_1^{h_1},~\cdots,~k_{n}^{h_n}\right)\Bigg|_{\rm singular} = 
\el -{1 \over \e^2}{g^2 \Nc \mu^{2\e} e^{-\gamma_E \e}\over (4\pi)^{2-\e}} {\G(1+\e)\G^2(1-\e) \over \G(1-2\e)}\sum^n_{i=1}\left({\mu^2 \over -s_{i\,i+1}}\right)^\e A^{tree}\left(k_1^{h_1},~\cdots,~k_{n}^{h_n}\right)
\label{1LIR}
\eea
\cme
for color-ordered partial amplitudes in the Euclidean kinematical region (defined in Subsection \ref{GUD}).
\section{$~~~~~~~~~~~~~~$$\Nsym$ Superconformal Symmetry}
\label{sconf}
In this work we study the Yang-Mills theory based on the four-dimensional $\Nsym$ supersymmetric extension of the Poincar\'{e} group. This extension, called the $\Nsym$ superconformal group, is an example of a {\it Lie supergroup}, a generalization of a Lie group that possesses a $\mathbb{Z}_2$ graded Lie algebra. The $\Nsym$ superconformal group is normally discussed in the context of the Lagrange density of the $\Nsym$ gauge theory. For our purposes, we are more interested in fleshing out the discussion of Subsection \ref{DSI}. In Subsection \ref{DSI} particular realizations of both the ordinary and the closely related dual $\Nsym$ superconformal symmetries were discussed in the context of an on-shell chiral superspace construction. It turns out that, in classifying the symmetries of superamplitudes on this chiral on-shell superspace, one actually needs to consider the action of the generators of the {\it centrally extended} $\Nsym$ superconformal group. We begin by briefly describing each (ordinary) symmetry operation. 

Of course, it is not so easy to give the reader a feeling for the fermionic symmetries. Consequently, we argue by analogy to the appropriate even (under the $\mathbb{Z}_2$ grading) cases when discussing the symmetries associated with odd generators. In the end, we are mostly interested in representations of the appropriate Lie superalgebras on the on-shell superspace. Therefore, in the second part of this appendix, we write down the $\Nsym$ superconformal and dual superconformal algebras and give explicit representations thereof.

First, we remind the reader that the Poincar\'{e} group by itself is nothing but the isometry group of Minkowski space. As such it contains
\bea
{\rm spacetime ~translations:}&&~~~~~~~~~~x'^\mu = x^\mu + r^\mu ~~~~~{\rm and}\\
{\rm spacetime ~rotations:}&&~~~~~~~~~~x'^\mu = M^\mu_\nu x^\nu \,{\rm .}
\eea
Since there are four coordinates to translate in, three pairing of coordinate axes ($\{x,y\}$, $\{x,z\}$, and $\{y,z\}$) to define spatial rotations in, and three spatial directions to boost in, the dimension of the Poincar\'{e} group is ten. In this appendix, we will follow the conventions used in the main text and label generators using spinor notation. Spatial translations are generated by the momentum operator, $\mathbb{P}_{\alpha \dot{\alpha}}$, and spacetime rotations are generated by $\mathbb{M}_{\alpha \beta}$ and $\bar{\mathbb{M}}_{\dot{\alpha} \dot{\beta}}$. 

Now, suppose that one adds four fermionic directions to $\mathbb{R}^{1,3}$ labeled by $a$ for $a \in \{1,2,3,4\}$. One certainly expects any well-behaved theory to be invariant under the full isometry group of the space on which the theory sits. We clearly have to allow for translations along the new fermionic directions
\be
{\rm spacespace ~translations:}~~~~~~~~~~\theta'^a_\mu = \theta^a_\mu + \eta^a_\mu
\ee
in addition to the spacetime translations discussed above. Superspace translations are generated by the so-called supercharges, $\mathbb{Q}^a_{~\,\alpha}$ and $\bar{\mathbb{Q}}_{a\,\dot{\alpha}}$. There are sixteen of these fermionic generators in all because there are four fermionic coordinate axes and the supercharges carry spacetime indices as well. 

The $SU(4)_R$ symmetry discussed extensively in the main text fits neatly into this picture: the R-symmetry acts by rotating the supercharges into one other. As is well-known, $SU(4)_R$ has fifteen generators, $\mathbb{R}^a_{~\,b}$. On on-shell superspace, $\mathbb{R}^a_{~\,b}$ is realized as
\be
R^a_{~\,b} = \sum_{i = 1}^n \left(\eta_i^a{\partial \over \partial \eta_{i}^b}-{1 \over 4}\D_b^{\,~a}\eta^c_i{\partial \over \partial \eta_{i}^c}\right)\, ;
\ee
the trace part by itself is not a symmetry of the theory. However, in attempting to implement $\Nsym$ supersymmetry on on-shell scattering amplitudes, one discovers that it {\it is} possible to build an additional symmetry generator by appropriately modifying the generator of the trace part that we were initially tempted to discard. This new generator, $\mathbb{Z}$, is called the central charge of the Lie superalgebra due to the fact that it commutes with all of the other generators and is related to the helicity quantum number of on-shell superamplitudes. On on-shell superspace, the generator of the central charge,
\be
Z = \sum_{i = 1}^n \left(1 + {1\over 2}\lambda_i^{~\,\A}{\partial \over \partial \lambda_i^{~\,\A}}-{1\over 2}\tilde{\lambda}_i^{~\,\dot{\alpha}}{\partial \over \partial \tilde{\lambda}_i^{~\,\dot{\alpha}}}-{1\over 2}\eta_i^a{\partial \over \partial \eta_i^a}\right)\,,
\ee
is identified with one minus the helicity operator summed over states~\cite{DHPYangian1}:
\be
Z = \sum_{i = 1}^n \left(1-h_i\right)\,{\rm .}
\ee
By construction, each superfield $\Phi(p,\eta)$ has helicity $+ 1$ (see eq. (\ref{supwvfun})). Therefore, $Z$ annihilates all on-shell superamplitudes and it follows that there is indeed an additional bosonic symmetry as claimed. Clearly, the construction of $Z$ is tied up with the chirality of the on-shell superspace since we arbitrarily chose to work with superfields of helicity $+ 1$, rather than $-1$.

Now, there is a natural extension of the Poincar\'{e} group that provides five additional bosonic generators. What we are alluding to is the well-known conformal group which, in addition to the ten dimensional Poincar\'{e} group, consists of 
\bea
{\rm dilatations:}&&~~~~~~~x'^\mu = \alpha\, x^\mu~~~~~{\rm and}\\
{\rm special~conformal~transformations:}&&~~~~~~~x'^\mu = {x^\mu - r^\mu x^2 \over 1- 2 r \cdot x + r^2 x^2} \,{\rm .}
\eea
The dilatation operation, generated by $\mathbb{D}$, is just a rescaling of the coordinates and, at the level of operators, it measures the classical scaling dimension. The special conformal transformations, generated by $\mathbb{K}_{\alpha \dot{\alpha}}$, are a bit more difficult to understand, as their action on Minkowski space looks rather complicated. A nice way to proceed is as follows. If we introduce the discrete operation of conformal inversion
\be
{\rm inversion:}~~~~~~~x'^\mu = {x^\mu \over x^2} \equiv I[x^\mu] \,{\rm ,}
\ee
it turns out~\cite{DHKSdualconf} that one can think of the special conformal symmetries as being generated by an inversion, a translation, and another inversion applied in succession:
\be
\mathbb{K}_{\alpha \dot{\alpha}} = I\, \mathbb{P}_{\alpha \dot{\alpha}} \,I \,{\rm .}
\label{invK}
\ee  

Na\"{i}vely one might think that we have now successfully identified all generators. However, it turns out that we are still missing the analogs of special conformal transformations along the fermionic directions~\cite{Minwalla}. Indeed, we can identify sixteen new fermionic generators, the generators of the special supersymmetry transformations, along the lines of eq. (\ref{invK}):
\be
\mathbb{S}_a^{\alpha} = I\, \bar{\mathbb{Q}}_{a\,\dot{\alpha}}\,I~~~~{\rm and}~~~~\bar{\mathbb{S}}^{a\,\dot{\alpha}} = I\, \mathbb{Q}^a_{~\,\alpha}\,I \,{\rm .}
\ee
Now that we have actually succeeded in identifying all symmetry generators, we can give explicit forms for them and write down the (anti)commutation relations that they ought to satisfy.
\subsection{$~~~~~~~~~~~~~~$The Ordinary and Dual $\Nsym$ Superconformal Algebras and Differential Operator Representations Thereof}
\label{scalg}
We first present, in spinor notation, the non-trivial (anti)commutation relations of the ordinary $\Nsym$ superconformal algebra:
\cmb{-.5 in}{0 in}
\bea
&&\left[ \mathbb{D}, \mathbb{P}_{\alpha \dot{\alpha}} \right] = \mathbb{P}_{\alpha \dot{\alpha}} \qquad ~~~~~\left[ \mathbb{D}, \mathbb{K}_{\alpha \dot{\alpha}} \right] = -\mathbb{K}_{\alpha \dot{\alpha}} 
\el
\left[ \mathbb{D}, \mathbb{Q}^a_{~\,\alpha} \right] = {1\over 2} \mathbb{Q}^a_{~\,\alpha} \qquad ~~\left[ \mathbb{D}, \bar{\mathbb{Q}}_{a\,\dot{\alpha}} \right] = {1\over 2}\bar{\mathbb{Q}}_{a\,\dot{\alpha}}
\el
\left[ \mathbb{D}, \mathbb{S}_{a\,\alpha} \right] = -{1\over 2} \mathbb{S}_{a\,\alpha} \qquad ~\left[ \mathbb{D}, \bar{\mathbb{S}}^a_{~\,\dot{\alpha}} \right] = -{1\over 2}\bar{\mathbb{S}}^a_{~\,\dot{\alpha}}
\el
\left[ \mathbb{K}_{\alpha \dot{\alpha}}, \mathbb{Q}^a_{~\,\beta} \right] = \e_{\A \B}\bar{\mathbb{S}}^a_{~\,\dot{\alpha}} \qquad ~\left[ \mathbb{K}_{\alpha \dot{\alpha}}, \bar{\mathbb{Q}}_{a\,\dot{\beta}} \right] = \e_{\dot{\A} \dot{\B}} \mathbb{S}_{a\,\alpha}
\el
\left[ \mathbb{K}_{\alpha \dot{\alpha}}, \mathbb{P}_{\beta \dot{\beta}} \right] = \e_{\A \B}\e_{\dot{\A}\dot{\B}} \mb{D}+{1
\over 2}\e_{\dot{\A}\dot{\B}}\mb{M}_{\A \B}+{1
\over 2}\e_{\A \B}\bar{\mb{M}}_{\dot{\A}\dot{\B}}
\el
\left[ \mathbb{P}_{\A \dot{\A}}, \mathbb{S}_{a\,\B} \right] = \e_{\A \B} \bar{\mathbb{Q}}_{a\,\dot{\alpha}} \qquad ~~\left[ \mathbb{P}_{\A \dot{\A}}, \bar{\mathbb{S}}^a_{~\,\dot{\B}} \right] = \e_{\dot{\A} \dot{\B}}\mathbb{Q}^a_{~\,\alpha}
\el
\left[ \mathbb{M}_{\A \B}, \mathbb{M}_{\g \D} \right] = \e_{\A \g} \mathbb{M}_{\B \D} + \e_{\B \g} \mathbb{M}_{\A \D}+\e_{\A \D} \mathbb{M}_{\B \g}+\e_{\B \D} \mathbb{M}_{\A \g}
\el
\left[ \bar{\mathbb{M}}_{\dot{\A} \dot{\B}}, \bar{\mathbb{M}}_{\dot{\g} \dot{\D}} \right] = \e_{\dot{\A} \dot{\g}} \bar{\mathbb{M}}_{\dot{\B} \dot{\D}} + \e_{\dot{\B} \dot{\g}} \bar{\mathbb{M}}_{\dot{\A} \dot{\D}}+\e_{\dot{\A} \dot{\D}} \bar{\mathbb{M}}_{\dot{\B} \dot{\g}}+\e_{\dot{\B} \dot{\D}} \bar{\mathbb{M}}_{\dot{\A} \dot{\g}}
\el
\left[ \mathbb{M}_{\A \B}, \mathbb{S}_{a\,\g} \right] = \e_{\B \g} \mathbb{S}_{a\,\A}+\e_{\A \g} \mathbb{S}_{a\,\B} \qquad ~~\left[ \bar{\mathbb{M}}_{\dot{\A} \dot{\B}}, \bar{\mathbb{S}}^a_{~\,\dot{\g}} \right] = \e_{\dot{\A} \dot{\g}}\bar{\mathbb{S}}^a_{~\,\dot{\B}}+\e_{\dot{\B} \dot{\g}}\bar{\mathbb{S}}^a_{~\,\dot{\A}}
\el
\left[ \mathbb{M}_{\A \B}, \mathbb{Q}^a_{~\,\g} \right] = \e_{\B \g} \mathbb{Q}^a_{~\,\A}+\e_{\A \g} \mathbb{Q}^a_{~\,\B} \qquad \left[ \bar{\mathbb{M}}_{\dot{\A} \dot{\B}}, \bar{\mathbb{Q}}_{a\,\dot{\g}} \right] = \e_{\dot{\A} \dot{\g}}\bar{\mathbb{Q}}_{a\,\dot{\B}}+\e_{\dot{\B} \dot{\g}}\bar{\mathbb{Q}}_{a\,\dot{\A}}
\el
\left[ \mathbb{M}_{\A \B},\mathbb{K}_{\g \dot{\g}}\right] = \e_{\B \g} \mathbb{K}_{\A \dot{\g}}+\e_{\A \g} \mathbb{K}_{\B \dot{\g}} \qquad ~\left[ \bar{\mathbb{M}}_{\dot{\A} \dot{\B}}, \mathbb{K}_{\g \dot{\g}} \right] = \e_{\dot{\A} \dot{\g}}\mathbb{K}_{\g \dot{\B}}+\e_{\dot{\B} \dot{\g}}\mathbb{K}_{\g \dot{\A}}
\el
\left[ \mathbb{M}_{\A \B},\mathbb{P}_{\g \dot{\g}}\right] = \e_{\B \g} \mathbb{P}_{\A \dot{\g}}+\e_{\A \g} \mathbb{P}_{\B \dot{\g}} \qquad ~~~\left[ \bar{\mathbb{M}}_{\dot{\A} \dot{\B}}, \mathbb{P}_{\g \dot{\g}} \right] = \e_{\dot{\A} \dot{\g}}\mathbb{P}_{\g \dot{\B}}+\e_{\dot{\B} \dot{\g}}\mathbb{P}_{\g \dot{\A}}
\el
\left[ \mb{R}^a_{~\,b},\mb{R}^c_{~\,d} \right] = \D_b^{~\,c}\mb{R}^a_{~\,d}-\D_d^{~\,a}\mb{R}^c_{~\,b}
\el
\left[ \mb{R}^a_{~\,b},\mathbb{Q}^c_{~\,\A} \right] = \D_b^{~\,c}\mathbb{Q}^a_{~\,\A} - {1\over 4}\D_b^{~\,a}\mathbb{Q}^c_{~\,\A} \qquad \left[ \mb{R}^a_{~\,b},\bar{\mathbb{Q}}_{c\,\dot{\A}} \right] = -\left(\D_c^{~\,a}\bar{\mathbb{Q}}_{b\,\dot{\A}} - {1\over 4}\D_b^{~\,a}\bar{\mathbb{Q}}_{c\,\dot{\A}}\right)
\el
\left[ \mb{R}^a_{~\,b},\mathbb{S}_{c\,\A} \right] = -\left(\D_c^{~\,a}\mathbb{S}_{b\,\A} - {1\over 4}\D_b^{~\,a}\mathbb{S}_{c\,\A}\right) \qquad \left[ \mb{R}^a_{~\,b},\bar{\mathbb{S}}^c_{~\,\dot{\A}} \right] = \D_b^{~\,c}\bar{\mathbb{S}}^a_{~\,\dot{\A}} - {1\over 4}\D_b^{~\,a}\bar{\mathbb{S}}^c_{~\,\dot{\A}}
\el
\left\{\mathbb{Q}^a_{~\,\A},\bar{\mathbb{Q}}_{b\,\dot{\A}}\right\} = \D_b^{~\,a} \mb{P}_{\A \dot{\A}} \qquad \left\{\mathbb{S}_{a\,\A},\bar{\mathbb{S}}^b_{~\,\dot{\A}}\right\} = \D_a^{~\,b} \mb{K}_{\A \dot{\A}}
\el
\left\{\mathbb{S}_{a\,\A},\mathbb{Q}^b_{~\,\B}\right\} = {1
\over 2}\D_a^{~\,b} \mb{M}_{\A \B} -\e_{\A \B} \mb{R}^b_{~\,a}+{1 \over 2}\e_{\A \B}\D_a^{~\,b}\left(\mb{D}+\mb{Z}\right)
\el
\left\{\bar{\mathbb{S}}^a_{~\,\dot{\A}},\bar{\mathbb{Q}}_{b\,\dot{\B}}\right\} = {1
\over 2}\D_b^{~\,a} \bar{\mb{M}}_{\dot{\A} \dot{\B}} +\e_{\dot{\A} \dot{\B}} \mb{R}^a_{~\,b}+{1 \over 2}\e_{\dot{\A} \dot{\B}}\D_b^{~\,a}\left(\mb{D}-\mb{Z}\right)
\label{commalg}
\eea
\cme
Our focus in this work is on the differential operator representation of the above superalgebra on on-shell superspace (discussed in Subsection \ref{DSI}). For a supermatrix representation of the $\Nsym$ superconformal algebra we refer the interested reader to~\cite{DHPYangian1}.  The representation that we present acts on the on-shell superspace of Subsection \ref{gendisonshell}:
\cmb{-1.0 in}{0 in}
\bea
P_{\A \dot{\A}} &=& \sum_{i = 1}^n \lambda_{i\,\A} \tilde{\lambda}_{i \,\dot{\A}} \qquad \qquad \qquad K_{\A \dot{\A}} = \sum_{i = 1}^n {\partial \over \partial \lambda_i^{~\,\A}}{\partial \over \partial \tilde{\lambda}_i^{~\,\dot{\alpha}}} 
\nonumber \\
M_{\A \B} &=&  \sum_{i = 1}^n \left(\lambda_{i\,\alpha}{\partial \over \partial \lambda_i^{~\,\B}}+\lambda_{i\,\B}{\partial \over \partial \lambda_i^{~\,\A}}\right) \qquad \bar{M}_{\dot{\A} \dot{\B}} =  \sum_{i = 1}^n \left(\tilde{\lambda}_{i\,\dot{\A}}{\partial \over \partial \tilde{\lambda}_i^{~\,\dot{\B}}}+\tilde{\lambda}_{i\,\dot{\B}}{\partial \over \partial \tilde{\lambda}_i^{~\,\dot{\alpha}}}\right)
\nonumber \\
D &=&  \sum_{i = 1}^n \left({1 \over 2}\lambda_i^{~\,\alpha}{\partial \over \partial \lambda_i^{~\,\A}}+{1 \over 2}\tilde{\lambda}_i^{~\,\dot{\A}}{\partial \over \partial \tilde{\lambda}_i^{~\,\dot{\alpha}}}+1\right)\qquad R^a_{~\,b} = \sum_{i = 1}^n \left(\eta_i^a{\partial \over \partial \eta_{i}^b}-{1 \over 4}\D_b^{\,~a}\eta^c_i{\partial \over \partial \eta_{i}^c}\right)
\nonumber \\
Q^a_{~\,\alpha}&=&  \sum_{i = 1}^n \lambda_{i\,\alpha}\eta_i^a \qquad \bar{Q}_{a\,\dot{\A}} = \sum_{i = 1}^n \tilde{\lambda}_{i\,\dot{\alpha}} {\partial \over \partial \eta_{i}^a}\qquad S_{a\,\A} = \sum_{i = 1}^n {\partial \over \partial \lambda_i^{~\,\A}} {\partial \over \partial \eta_{i}^a} \qquad \bar{S}^a_{~\,\dot{\alpha}} =  \sum_{i = 1}^n \eta^a_i {\partial \over \partial \tilde{\lambda}_i^{~\,\dot{\alpha}}}
\nonumber \\
Z &=& \sum_{i = 1}^n \left(1 + {1\over 2}\lambda_i^{~\,\A}{\partial \over \partial \lambda_i^{~\,\A}}-{1\over 2}\tilde{\lambda}_i^{~\,\dot{\alpha}}{\partial \over \partial \tilde{\lambda}_i^{~\,\dot{\alpha}}}-{1\over 2}\eta_i^a{\partial \over \partial \eta_i^a} \right)
\eea
\cme
We have painstakingly checked (using the conventions of Penrose and Rindler~\cite{PenroseRindler} for raising and lowering spinor indices) that this representation satisfies the above (anti)commutation relations. 

It's important to note that the above superalgebra is not appropriate for the dual superconformal symmetry discussed in Subsection \ref{DSI}; to write down the dual superconformal algebra one should take (\ref{commalg}), swap the $SU(4)_R$ chiralities of all operators ({\it e.g.} $\mathbb{Q}^a_{~\,\A}$ becomes $\mathbb{Q}_{a\,\A}$), and then appropriately adjust the (anti)commutation relations involving $\mb{R}^a_{~\,b}$. Using the explicit expressions given in Subsection \ref{DSI}, we find:
\cmb{-.5 in}{0 in}
\bea
&&\left[ \mathbb{D}, \mathbb{P}_{\alpha \dot{\alpha}} \right] = \mathbb{P}_{\alpha \dot{\alpha}} \qquad ~~~~~\left[ \mathbb{D}, \mathbb{K}_{\alpha \dot{\alpha}} \right] = -\mathbb{K}_{\alpha \dot{\alpha}} 
\el
\left[ \mathbb{D}, \mathbb{Q}_{a\,\alpha} \right] = {1\over 2} \mathbb{Q}_{a\,\alpha} \qquad ~~\left[ \mathbb{D}, \bar{\mathbb{Q}}^a_{~\,\dot{\alpha}} \right] = {1\over 2}\bar{\mathbb{Q}}^a_{~\,\dot{\alpha}}
\el
\left[ \mathbb{D}, \mathbb{S}^a_{~\,\alpha} \right] = -{1\over 2} \mathbb{S}^a_{~\,\alpha} \qquad ~\left[ \mathbb{D}, \bar{\mathbb{S}}_{a\,\dot{\alpha}} \right] = -{1\over 2}\bar{\mathbb{S}}_{a\,\dot{\alpha}}
\el
\left[ \mathbb{K}_{\alpha \dot{\alpha}}, \mathbb{Q}_{a\,\beta} \right] = \e_{\A \B}\bar{\mathbb{S}}_{a\,\dot{\alpha}} \qquad ~\left[ \mathbb{K}_{\alpha \dot{\alpha}}, \bar{\mathbb{Q}}^a_{~\,\dot{\beta}} \right] = \e_{\dot{\A} \dot{\B}} \mathbb{S}^a_{~\,\alpha}
\el
\left[ \mathbb{K}_{\alpha \dot{\alpha}}, \mathbb{P}_{\beta \dot{\beta}} \right] = \e_{\A \B}\e_{\dot{\A}\dot{\B}} \mb{D}+{1
\over 2}\e_{\dot{\A}\dot{\B}}\mb{M}_{\A \B}+{1
\over 2}\e_{\A \B}\bar{\mb{M}}_{\dot{\A}\dot{\B}}
\el
\left[ \mathbb{P}_{\A \dot{\A}}, \mathbb{S}^a_{~\,\B} \right] = \e_{\A \B} \bar{\mathbb{Q}}^a_{~\,\dot{\alpha}} \qquad ~~\left[ \mathbb{P}_{\A \dot{\A}}, \bar{\mathbb{S}}_{a\,\dot{\B}} \right] = \e_{\dot{\A} \dot{\B}}\mathbb{Q}_{a\,\alpha}
\el
\left[ \mathbb{M}_{\A \B}, \mathbb{M}_{\g \D} \right] = \e_{\A \g} \mathbb{M}_{\B \D} + \e_{\B \g} \mathbb{M}_{\A \D}+\e_{\A \D} \mathbb{M}_{\B \g}+\e_{\B \D} \mathbb{M}_{\A \g}
\el
\left[ \bar{\mathbb{M}}_{\dot{\A} \dot{\B}}, \bar{\mathbb{M}}_{\dot{\g} \dot{\D}} \right] = \e_{\dot{\A} \dot{\g}} \bar{\mathbb{M}}_{\dot{\B} \dot{\D}} + \e_{\dot{\B} \dot{\g}} \bar{\mathbb{M}}_{\dot{\A} \dot{\D}}+\e_{\dot{\A} \dot{\D}} \bar{\mathbb{M}}_{\dot{\B} \dot{\g}}+\e_{\dot{\B} \dot{\D}} \bar{\mathbb{M}}_{\dot{\A} \dot{\g}}
\el
\left[ \mathbb{M}_{\A \B}, \mathbb{S}^a_{~\,\g} \right] = \e_{\B \g} \mathbb{S}^a_{~\,\A}+\e_{\A \g} \mathbb{S}^a_{~\,\B} \qquad ~~\left[ \bar{\mathbb{M}}_{\dot{\A} \dot{\B}}, \bar{\mathbb{S}}_{a\,\dot{\g}} \right] = \e_{\dot{\A} \dot{\g}}\bar{\mathbb{S}}_{a\,\dot{\B}}+\e_{\dot{\B} \dot{\g}}\bar{\mathbb{S}}_{a\,\dot{\A}}
\el
\left[ \mathbb{M}_{\A \B}, \mathbb{Q}_{a\,\g} \right] = \e_{\B \g} \mathbb{Q}_{a\,\A}+\e_{\A \g} \mathbb{Q}_{a\,\B} \qquad \left[ \bar{\mathbb{M}}_{\dot{\A} \dot{\B}}, \bar{\mathbb{Q}}^a_{~\,\dot{\g}} \right] = \e_{\dot{\A} \dot{\g}}\bar{\mathbb{Q}}^a_{~\,\dot{\B}}+\e_{\dot{\B} \dot{\g}}\bar{\mathbb{Q}}^a_{~\,\dot{\A}}
\el
\left[ \mathbb{M}_{\A \B},\mathbb{K}_{\g \dot{\g}}\right] = \e_{\B \g} \mathbb{K}_{\A \dot{\g}}+\e_{\A \g} \mathbb{K}_{\B \dot{\g}} \qquad ~\left[ \bar{\mathbb{M}}_{\dot{\A} \dot{\B}}, \mathbb{K}_{\g \dot{\g}} \right] = \e_{\dot{\A} \dot{\g}}\mathbb{K}_{\g \dot{\B}}+\e_{\dot{\B} \dot{\g}}\mathbb{K}_{\g \dot{\A}}
\el
\left[ \mathbb{M}_{\A \B},\mathbb{P}_{\g \dot{\g}}\right] = \e_{\B \g} \mathbb{P}_{\A \dot{\g}}+\e_{\A \g} \mathbb{P}_{\B \dot{\g}} \qquad ~~~\left[ \bar{\mathbb{M}}_{\dot{\A} \dot{\B}}, \mathbb{P}_{\g \dot{\g}} \right] = \e_{\dot{\A} \dot{\g}}\mathbb{P}_{\g \dot{\B}}+\e_{\dot{\B} \dot{\g}}\mathbb{P}_{\g \dot{\A}}
\el
\left[ \mb{R}^a_{~\,b},\mb{R}^c_{~\,d} \right] = \D_b^{~\,c}\mb{R}^a_{~\,d}-\D_d^{~\,a}\mb{R}^c_{~\,b}
\el
\left[ \mb{R}^a_{~\,b},\mathbb{Q}_{c\,\A} \right] = -\left(\D_c^{~\,a}\mathbb{Q}_{b\,\A} - {1\over 4}\D_b^{~\,a}\mathbb{Q}_{c\,\A}\right) \qquad \left[ \mb{R}^a_{~\,b},\bar{\mathbb{Q}}^c_{~\,\dot{\A}} \right] = \D_b^{~\,c}\bar{\mathbb{Q}}^a_{~\,\dot{\A}} - {1\over 4}\D_b^{~\,a}\bar{\mathbb{Q}}^c_{~\,\dot{\A}}
\el
\left[ \mb{R}^a_{~\,b},\mathbb{S}^c_{~\,\A} \right] = \D_b^{~\,c}\mathbb{S}^a_{~\,\A} - {1\over 4}\D_b^{~\,a}\mathbb{S}^c_{~\,\A} \qquad \left[ \mb{R}^a_{~\,b},\bar{\mathbb{S}}_{c\,\dot{\A}} \right] = -\left(\D_c^{~\,a}\bar{\mathbb{S}}_{b\,\dot{\A}} - {1\over 4}\D_b^{~\,a}\bar{\mathbb{S}}_{c\,\dot{\A}}\right)
\el
\left\{\mathbb{Q}_{a\,\A},\bar{\mathbb{Q}}^b_{~\,\dot{\A}}\right\} = \D_a^{~\,b} \mb{P}_{\A \dot{\A}} \qquad \left\{\mathbb{S}^a_{~\,\A},\bar{\mathbb{S}}_{b\,\dot{\A}}\right\} = \D_b^{~\,a} \mb{K}_{\A \dot{\A}}
\el
\left\{\mathbb{S}^a_{~\,\A},\mathbb{Q}_{b\,\B}\right\} = {1
\over 2}\D_b^{~\,a} \mb{M}_{\A \B} +\e_{\A \B} \mb{R}^a_{~\,b}+{1 \over 2}\e_{\A \B}\D_b^{~\,a}\left(\mb{D}+\mb{Z}\right)
\el
\left\{\bar{\mathbb{S}}_{a\,\dot{\A}},\bar{\mathbb{Q}}^b_{~\,\dot{\B}}\right\} = {1
\over 2}\D_a^{~\,b} \bar{\mb{M}}_{\dot{\A} \dot{\B}} - \e_{\dot{\A} \dot{\B}} \mb{R}^b_{~\,a}+{1 \over 2}\e_{\dot{\A} \dot{\B}}\D_a^{~\,b}\left(\mb{D}-\mb{Z}\right) \,.
\eea
\cme
For the sake of completeness, we collect the dual superconformal generators here as well:
\cmb{-1.4 in}{0 in}
\bea
\mc{Q}_{a\,\alpha} &=& \sum_{i = 1}^n {\partial\over \partial \T_{i}^{a\,\A}} ~~~ \mathcal{P}_{\alpha \dot{\alpha}} = \sum_{i = 1}^n {\partial\over \partial x_{i}^{~\,\A\dot{\A}}} \nn
\bar{\mc{S}}_{b\,\dot{\alpha}} &=& \sum_{i = 1}^n \left(x_{i\,\dot{\alpha}}^{\,~~\beta}{\partial\over \partial \T_{i}^{b\,\B}}+\tilde{\lambda}_{i\,\dot{\alpha}}{\partial\over \partial \eta^b_{i}}\right) ~~~ \bar{\mc{Q}}^b_{~\dot{\alpha}} = \sum_{i = 1}^n \left(\theta^{b\,\alpha}_{i}{\partial\over \partial x_{i}^{~\,\A\dot{\A}}}+\eta_{i}^{b}{\partial\over \partial \tilde{\lambda}_{i}^{~\,\dot{\A}}}\right)\nn
\mc{S}^a_{~\alpha} &=& \sum_{i = 1}^n \left(-\theta^b_{i\,\alpha}\theta^{a\,\beta}_{i}{\partial\over \partial \T_{i}^{b\,\B}}+x_{i\,\alpha}^{~~\,\dot{\beta}}\theta^{a\,\beta}_{i}{\partial\over \partial x_{i}^{~\,\B \dot{\B}}}+\lambda_{i\,\alpha}\theta^{a\,\gamma}_{i}{\partial\over \partial \lambda_{i}^{\,~\g}}+x_{i+1\,\alpha}^{\,~~~~~\dot{\beta}}\eta^a_i{\partial\over \partial \tilde{\lambda}_{i}^{~\,\dot{\B}}}-\theta^b_{i+1\,\alpha}\eta^a_i{\partial\over \partial \eta^b_i}\right)  \nonumber \\
\mc{K}_{\alpha \dot{\alpha}} &=& \sum_{i = 1}^n \left(x_{i\,\alpha}^{~~\,\dot{\beta}}x_{i\,\dot{\alpha}}^{~~\,\beta}{\partial\over \partial x_{i}^{~\,\B \dot{\B}}}+x_{i\,\dot{\alpha}}^{~~\,\beta}\theta_{i\,\alpha}^b{\partial\over \partial \T_{i}^{b\,\B}}+x_{i\,\dot{\alpha}}^{\,~~\beta}\lambda_{i\,\alpha}{\partial\over \partial \lambda_{i}^{\,~\B}}+x_{i+1\,\alpha}^{~~~~~\,\dot{\beta}}\tilde{\lambda}_{i\,\dot{\alpha}}{\partial\over \partial \tilde{\lambda}_{i}^{~\,\dot{\B}}}+\tilde{\lambda}_{i\,\dot{\alpha}}\theta^b_{i+1\,\alpha}{\partial\over \partial \eta_{i}^b}\right)\nn
\mc{M}_{\A \B} &=& \sum_{i = 1}^n \left(x_{i\,\alpha}^{~~\,\dot{\alpha}}{\partial\over \partial x_{i}^{~\,\B \dot{\A}}}+x_{i\,\beta}^{~~\,\dot{\alpha}}{\partial\over \partial x_{i}^{~\,\alpha \dot{\A}}}+\theta_{i\,\alpha}^{a}{\partial\over \partial \T_{i}^{a\,\B}}+\theta_{i\,\beta}^{a}{\partial\over \partial \T_{i}^{a\,\alpha}}+\lambda_{i\,\alpha}{\partial\over \partial \lambda_{i}^{\,~\B}}+\lambda_{i\,\beta}{\partial\over \partial \lambda_{i}^{\,~\alpha}}\right)\nonumber \\
\bar{\mc{M}}_{\dot{\A}\dot{\B}} &=& \sum_{i = 1}^n \left(x_{i\,~~\dot{\alpha}}^{~\,\alpha}{\partial\over \partial x_{i}^{~\,\alpha \dot{\B}}}+x_{i\,~~\dot{\beta}}^{~\,\alpha}{\partial\over \partial x_{i}^{~\,\alpha \dot{\alpha}}}+\tilde{\lambda}_{i\,\dot{\alpha}}{\partial\over \partial \tilde{\lambda}_{i}^{~\,\dot{\B}}}+\tilde{\lambda}_{i\,\dot{\beta}}{\partial\over \partial \tilde{\lambda}_{i}^{~\,\dot{\alpha}}}\right)\nonumber \\
\mc{R}^a_{~b} &=& \sum_{i = 1}^n \left(\T_i^{a\,\alpha}{\partial\over \partial \T_{i}^{a\,\A}}+\eta_i^a {\partial\over \partial \eta^b_{i}}-{1\over 4}\D_{b}^{~\,a}\T_{i}^{c\,\alpha}{\partial\over \partial \T_{i}^{c\,\A}}-{1\over 4}\D_{b}^{~\,a}\eta_i^c{\partial\over \partial \eta^c_i}\right)\nonumber \\
\mc{D} &=& -\sum_{i = 1}^n \left(x_{i}^{~\,\alpha \dot{\alpha}}{\partial\over \partial x_{i}^{~\,\alpha \dot{\alpha}}}+{1\over 2}\T_{i}^{a\,\alpha}{\partial\over \partial \T_{i}^{a\,\alpha}}+{1\over 2}\lambda_{i}^{\,~\alpha}{\partial\over \partial \lambda_{i}^{\,~\alpha}}+{1\over 2}\tilde{\lambda}_{i}^{~\,\dot{\alpha}}{\partial\over \partial \tilde{\lambda}_{i}^{~\,\dot{\alpha}}}\right)\nonumber \\
\mc{Z} &=& -{1\over 2}\sum_{i = 1}^n \left(\lambda_{i}^{\,~\alpha}{\partial\over \partial \lambda_{i}^{\,~\alpha}}-\tilde{\lambda}_{i}^{\,~\dot{\alpha}}{\partial\over \partial \tilde{\lambda}_{i}^{\,~\dot{\alpha}}}-\eta^a_i{\partial\over \partial \eta_{i}^a}\right)\,.
\eea
\cme
\newpage

\section*{References}

\bibliographystyle{unsrt}
\bibliography{JHEPexample2}
\end{document}